\documentclass[usenatbib]{mn2e}

\usepackage{amsmath}
\usepackage{natbib}
\usepackage{epsfig}
\usepackage{txfonts}

\newcommand{\id}{{\,\rm d}}

\newcommand{\beq}{\begin{equation}}   %

\newcommand{\eeq}{\end{equation}}   %

\newcommand{\beqa}{\begin{eqnarray}}   %

\newcommand{\eeqa}{\end{eqnarray}}   %

\newcommand{\beal}{\begin{align}}

\newcommand{\enal}{\end{align}}

\newcommand{\bspl}{\begin{split}}

\newcommand{\espl}{\end{split}}

\newcommand{\bsub}{\begin{subequations}}

\newcommand{\esub}{\end{subequations}}

\newcommand{\bmulti}{\begin{multline}}   %

\newcommand{\beqm}{\begin{mathletters}}   %

\newcommand{\eeqm}{\end{mathletters}}   %

\newcommand{\kB}{k_{\rm B}}

\newcommand{\me}{m_{\rm e}}

\newcommand{\Te}{T_{\rm e}}

\newcommand{\Tg}{T_{\gamma}}

\newcommand{\pd}{\partial}

\newcommand{\pAb}[2]{\frac{\displaystyle\pd #1}{\displaystyle\pd #2}}

\newcommand{\Abl}[2]{\frac{{\rm d} #1}{{\rm d} #2}}

\newcommand{\pot}[2]{#1 \times 10^{#2}}

\newcommand{\ion}[2]{{\text{{\sc #1}\,{\sc #2}}}}
\newcommand{\HeIlevel}[4]{{#1^{#2} {\rm #3}_{#4}}}   

\newcommand{\xD}{{{x_{\rm D}}}}

\usepackage{color}
\newcommand{\changeI}[1]{{#1}}
\newcommand{\changeII}[1]{{#1}}
\newcommand{\changeIII}[1]{{#1}}

\title[Cosmological recombination: feedback of helium and its effect on the recombination spectrum]{Cosmological recombination: feedback of helium photons and its effect on the recombination spectrum}

\author[Chluba and Sunyaev]{J. Chluba$^{1,2}$\thanks{E-mail:
  jchluba@cita.utoronto.ca} and R.~A. Sunyaev$^{2,3}$ 
  \\
$^{1}$ Canadian Institute for Theoretical Astrophysics, 60 St. George Street,
Toronto, ON M5S 3H8, Canada\\
$^{2}$ Max-Planck Institut f\"ur Astrophysik, Karl-Schwarzschild-Str. 1,
D-85740 Garching, Germany\\
$^{3}$ Space Research Institute, Russian Academy of Sciences, Profsoyuznaya 84/32,
117997 Moscow, Russia
}

\begin{document}

\date{Received **insert**; Accepted **insert**}

\maketitle

\begin{abstract}
In this paper we consider the {\it re-processing} of high frequency photons emitted by \changeIII{\ion{He}{ii} and \ion{He}{i}} during the epoch of cosmological recombination \changeIII{by \ion{He}{i} and  \ion{H}{i}}. We demonstrate that, in comparison to \changeI{computations which {\it neglect}} all feedback processes, the \changeI{number of cosmological recombination photons that are} related to the presence of helium in the early Universe could be increased by $\sim 40\%-70\%$. \changeI{Our computations imply that per helium nucleus $\sim 3-6$ {\it additional} photons could be produced. Therefore, a total of $\sim12-14$ helium-related photons are emitted during cosmological recombination.}
This is an important addition to cosmological recombination spectrum \changeI{which in the future} may render it slightly easier to determine the primordial abundance of helium using differential measurements of the CMB energy spectrum. Also, since these photons are the only witnesses of the feedback process at high redshift, observing them in principle offers a way to check our understanding of the recombination physics. 
Here most interestingly, the feedback of \ion{He}{ii} photons on \ion{He}{i} leads to the appearance of several additional, rather {\it narrow spectral features} in the \ion{He}{i} recombination spectrum at low frequencies. Consequently, the signatures of helium-related features in the CMB  spectral distortion \changeI{due to cosmological recombination} at some given frequency can exceed the average level of $\sim 17\%$ several times. We find that in particular the bands around $\nu\sim 10 \,$GHz, $\sim 35\,$GHz, $\sim 80\,$GHz, and $\sim 200\,$GHz seem to be affected strongly.
In addition, we computed the changes in the cosmological ionization history, finding that only  the feedback of \changeI{primary} $\ion{He}{i}$ photons on the dynamics of $\ion{He}{ii}\rightarrow \ion{He}{i}$ recombination has an effect, \changeI{producing} a change of $\Delta N_{\rm e}/N_{\rm e}\sim +0.17\%$ at $z\sim 2300$. This result seems to be $\sim 2-3$ times smaller than the one obtained in earlier computations for this process, however, the difference will not be very important for the analysis of future CMB data. 

\end{abstract}


\section{Introduction}
\label{sec:Intro}
It is well known that cosmological recombination of hydrogen and helium in the Universe
leads to the emission of several photons per baryon, modifying the cosmic microwave background (CMB) energy
spectrum \citep{Zeldovich68, Peebles68, Dubrovich1975, Dubrovich1997}.  
Recently, \changeI{detailed computations of the cosmological recombination spectrum were carried out  \citep[e.g.][]{Jose2006, Chluba2006, Jose2008}}, showing that the recombinations of hydrogen and helium lead to relatively {\it narrow spectral features} in the CMB energy spectrum. These features were created at redshift $z\sim 1400$, $\sim 2500$ and $\sim 6000$, corresponding to the times of \ion{H}{i}, \ion{He}{i} and \ion{He}{ii} recombination, and, due to redshifting, today should still be visible at \changeI{mm, cm and dm} wavelength. Observing these signatures from cosmological recombination may offer an independent way to determine some of the key cosmological parameters, such as the {\it primordial helium abundance}, the {\it number density of baryons} and the CMB {\it monopole temperature} at recombination \citep[for overview e.g. see][]{Sunyaev2009}.

However, to \changeIII{approach} this observationally challenging task it is important to theoretically understand all the possible contributions to the cosmological recombination spectrum and how they may be affected by physical processes occurring at high redshift.
In this paper we investigate the re-processing of {\it energetic photons} initially released by \ion{He}{i} and \ion{He}{ii} during cosmological recombination \changeI{by neutral hydrogen and helium.}
For this problem in particular quanta \changeI{emitted in} the \ion{He}{ii} Lyman $\alpha$ (transition energy $E\sim 40.8\,$eV), $\ion{He}{i} \;\HeIlevel{2}{1}{P}{1}-\HeIlevel{1}{1}{S}{0}$ ($E\sim 21.2\,$eV), $\ion{He}{i} \;\HeIlevel{2}{3}{P}{1}-\HeIlevel{1}{1}{S}{0}$ ($E\sim 21.0\,$eV) dipole transitions are important, but e.g. also the higher $\ion{He}{i} \;\HeIlevel{n}{1}{P}{1}-\HeIlevel{1}{1}{S}{0}$ series and \ion{He}{ii} two-photon continuum do play some role\footnote{We give a detailed inventory of possible primary \ion{He}{i} and \ion{He}{ii} feedback photons in Sect.~\ref{sec:feed_num_HeI} and \ref{sec:feed_num_HeII}.}.
In contrast to the CMB spectral distortions created by helium at low frequencies, these photons lead to a {\it large} deviation of the CMB spectrum from the one of a pure blackbody, so that, after some (significant) redshifting, they are able to {\it re-excite} energetically lower-lying atomic transitions in  \ion{H}{i} and \ion{He}{i}, significantly affecting the net rates of resonant and continuum transitions starting from the ground-state.
This can lead to both changes in the {\it cosmological recombination spectrum} and the {\it cosmological ionization history}, and, as we demonstrate here, in particular the total contribution of photons related to the presence of helium in the early Universe is increased by $\sim 40\%-70\%$ in comparison to computations that do not include the {\it feedback} processes under discussion here.
Such a large addition to the cosmological recombination spectrum is very important, since \changeI{in the future} it may render a determination of the primordial helium abundance using differential measurements of the CMB energy spectrum slightly easier.

To understand the physics behind this problem, we distinguish between {\it two} main types of feedback: (i) the {\it self-feedback} or {\it intra-species feedback}; and (ii) the {\it inter-species feedback}.
The first type of feedback is related to photons that are emitted by some atomic species (e.g. \ion{He}{i}) and then affect the lower-lying transitions of the {\it same} atomic species. 
Since the difference between the time of emission and feedback is connected with the redshift interval it takes to cross the energy distance between the resonances that are affected, for \ion{H}{i} and \ion{He}{ii} one therefore typically expects a delay of $\Delta z/z\sim 1\%-20\%$.
In contrast to this, for the inter-species feedback energetic photons are emitted by some atomic species (e.g. \ion{He}{ii}) and then feedback on lower-lying transitions of some {\it other} atomic species (e.g. \ion{He}{i}).
Here the typical delay between emission and feedback is \changeI{significantly} larger (e.g. reaching $\Delta z/z\sim 40\%$ in the case of \ion{He}{ii} Lyman $\alpha$ to $\ion{He}{i} \;\HeIlevel{1}{1}{S}{0}$ continuum feedback as shown in Sect.~\ref{sec:time-HeII-feed}), since the possible energy differences are much bigger.

For hydrogen the self-feedback problem was already studied earlier \citep{Chluba2007b, Switzer2007I} in connection with the \ion{H}{i} Lyman series.
There, for example, photons escaping from the \ion{H}{i} Lyman $\beta$ resonance, after redshifting by $\Delta z/z\sim 16\%$, will feedback on the \ion{H}{i} Lyman $\alpha$ line, leading to a small {\it inhibition} of hydrogen recombination.
This occurs because the \ion{H}{i} Lyman $\beta$ feedback adds photons to the phase space density around the \ion{H}{i} Lyman $ 
\alpha$ resonance and hence increases the population of the 2p-state.
These additional photons have to reach the very distant red wing of the \ion{H}{i} Lyman $\alpha$ line (e.g. via redshifting or by some chain of transitions towards higher levels or the continuum\footnote{\changeII{Excluding the Lyman $\alpha$ resonance itself,} the main way out of the 2p-level is via a transition to the 3d state (occurring in $\sim 90\%$ of the cases), while a transition to the continuum happens with $\lesssim 0.5\%$ probability \citep{Chluba2009}} that eventually leads to an emission in the 2s-1s two-photon channel) before the initial re-excitation of the electron by the feedback is reversed.
%
%
Similarly, photons emitted in the $\ion{He}{i} \;\HeIlevel{2}{1}{P}{1}-\HeIlevel{1}{1}{S}{0}$ resonance will feedback on the $\ion{He}{i} \;\HeIlevel{2}{3}{P}{1}-\HeIlevel{1}{1}{S}{0}$ line, but here the difference between the emission and feedback redshift is only $\Delta z/z\sim 1\%$, due to the smaller energy distance between these resonances.
The latter problem has as well been addressed in the literature \citep{Switzer2007I}, also including the fact that on their way from one resonance to the other some of the \ion{He}{i} photons are absorbed by the small amount of neutral hydrogen that is already present at redshift $z\sim 2500$ before they can actually feed back.

In this context, it was also shown that the intra-species feedback of \ion{H}{i} and \ion{He}{i} photons leads to some small correction to the cosmological ionization history, reaching $\Delta N_{\rm e}/N_{\rm e}\sim 0.2\%-0.3\%$ at $z\sim 1100$ for hydrogen \citep{Chluba2007b} and 
$\Delta N_{\rm e}/N_{\rm e}\sim 0.4\%-0.5\%$ at $z\sim 2000$ for helium \citep{Switzer2007I}.
This level of precision in our understanding of the dynamics of cosmological recombination will be important for the analysis of upcoming CMB data from the {\sc Planck} Surveyor, which was successfully launched in May this year.
In particular, our ability to precisely measure the spectral index of CMB fluctuations may be compromised by the neglect of physical processes that can affect the ionization history at the level of $\sim 0.1\%$ close to the maximum of the Thomson visibility function \citep{Sunyaev1970} at $z\sim 1100$. Over the past few years many such processes have already been identified \citep[for overview e.g. see][]{Fendt2009, Sunyaev2009} by several independent groups \citep[e.g.][]{Dubrovich2005, Chluba2006, Kholu2006, Jose2006, Wong2007, Switzer2007II, Karshenboim2008, Labzowsky2009, Jentschura2009}, also emphasizing that in principle {\it all} these processes do directly change the cosmological recombination spectrum \citep[e.g. see][]{Chluba2009b}.

\changeII{Although the  intra-species feedback affects the dynamics of recombination, it does not lead} to any significant change in the cosmological recombination spectrum, and in particular the total number of (low-frequency) photons released during recombination.
This is mainly because the total number of available feedback photons is small\footnote{e.g. the number of \ion{H}{i} Lyman $\beta$ relative to \ion{H}{i} Lyman $\alpha$ photons is about $\sim 0.5\%$ \citep{Chluba2007b}}, but \changeII{as we explain here (see Sect.~\ref{sec:direct}),} also the branching ratios to other levels play some important role.
In terms of photon production the {\it inter-species feedback} is much more interesting, and was not taken into account at full depth so far. 
The simplest example is connected with the feedback of primary $\ion{He}{i} \;\HeIlevel{2}{3}{P}{1}-\HeIlevel{1}{1}{S}{0}$ photons on \ion{H}{i}.
As we show here, practically all these photons never reach frequencies below the \ion{H}{i} Lyman continuum. Furthermore, we find that they feedback during the {\it pre-recombinational} epoch of \ion{H}{i}.
At that time the degree of ionization for hydrogen is still very close to the equilibrium Saha ionization \changeIII{(even when including the feedback)}.  
Obviously, the presence of additional non-equilibrium ionizing photons tends to increase the degree of 
ionization, but the recombination rate is high, so that this feedback eventually only leads to additional features 
in the cosmological recombination radiation, but no significant correction to the ionization history.

Here the most important aspect is that the electron which is liberated by the \ion{He}{i} feedback on the \ion{H}{i} Lyman continuum afterward has the possibility to recombine to some highly excited state and then emit several photons on its way towards the ground-state.
This is in stark contrast to the intra-species feedback, where it is unlikely to reach very highly excited levels.
The inter-species feedback therefore leads to {\it loops of atomic transitions} in the non-equilibrium ambient radiation field of the CMB \citep{Liubarskii83}, which tend to erase the high frequency spectral distortion introduced by \ion{He}{i}. In these loops {\it one} energetic photon is absorbed, while {\it several} low-frequency photons can be emitted. A similar process was studied in connection with the release of energy by some decaying or annihilating particles prior to cosmological recombination \citep{Chluba2008c}.
As we explain here, per primary \ion{He}{i} feedback photon about $2.6\,\gamma$ are produced by \ion{H}{i} in addition to one \ion{H}{i} Lyman $\alpha$ photon replacing the \ion{He}{i} feedback photon (see Sect.~\ref{sec:HeI_total_num} for more explanation).
We also study the feedback of \ion{He}{ii} photons on both \ion{He}{i} and \ion{H}{i}, in detail explaining all the important physical aspects of this problem (Sect.~\ref{sec:HeIII_feedback}), \changeII{and computing the effect of \ion{He}{ii} Lyman $\alpha$ feedback (Sect.~\ref{sec:spec_HeII}-\ref{sec:number_HeII}) on the cosmological recombination spectrum. Our main results for this problem are shown in Fig.~\ref{fig:DI.HeI.20} and \ref{fig:DI.total}}.

The paper is structured as follows: in Sect.~\ref{sec:TransEq} we give the formulation and solution of several photons transfer equations which are important in the context of the \changeII{$\ion{He}{i}$ intra-species} feedback problem. This Section is rather technical, and is only addressed to the interested reader.
In Sect.~\ref{sec:sol_F} and \ref{sec:Pesc_corr} we apply the results of Sect.~\ref{sec:TransEq} to the problem of the $\ion{He}{i} \;\HeIlevel{2}{1}{P}{1}-\HeIlevel{1}{1}{S}{0}$ and $\ion{He}{i} \;\HeIlevel{2}{3}{P}{1}-\HeIlevel{1}{1}{S}{0}$ resonances including the \ion{H}{i} Lyman continuum, solving them numerically. This Section provides some intuition for the important aspects of the problem, which then lead to the analytic approximations for the net rates and escape probabilities derived in Sect.~\ref{sec:Pesc_corr_ana}.
In Sect.~\ref{sec:DNe_Ne} we compute the corrections to the \changeII{ionization} history and in Sect.~\ref{sec:pre-rec}-\ref{sec:HeIII_feedback} we discuss the changes in the cosmological recombination \changeII{spectrum}.

\section{Different transfer equations in the no scattering approximation and their solutions}
\label{sec:TransEq}
In this Section we give the transfer equations and their solutions including different combinations of emission and absorption processes. 
In particular we are interested in the combined problem of the $\ion{He}{i} \;\HeIlevel{2}{1}{P}{1}-\HeIlevel{1}{1}{S}{0}$ and $\ion{He}{i} \;\HeIlevel{2}{3}{P}{1}-\HeIlevel{1}{1}{S}{0}$ resonances and the \ion{H}{i} Lyman continuum. However, the equations and solutions given here can be easily applied to the other resonances of neutral helium and may also be useful for further studies related to the CMB spectral distortions generated by helium.

We assume that the modifications to the solutions caused by partial frequency redistribution can be neglected, and work in the {\it no line-scattering} approximation. This procedure has been used in several recent studies \citep[e.g.][]{Switzer2007I, Chluba2008a, Chluba2008b} and seems to be well justified given that the additional corrections due to partial frequency redistribution are small \citep[e.g.][]{Switzer2007I, Jose2008, Hirata2009, Chluba2009b}.
\changeI{However, as we comment here (see Sect.~\ref{sec:feedback_higher}), when the considered resonances are very close to each other, additional corrections are expected.}

We will then use the results of this section to compute the CMB spectral distortion due to the $\ion{He}{i} \;\HeIlevel{2}{1}{P}{1}-\HeIlevel{1}{1}{S}{0}$ and $\ion{He}{i} \;\HeIlevel{2}{3}{P}{1}-\HeIlevel{1}{1}{S}{0}$ resonances at different stages of $\ion{He}{ii}\rightarrow \ion{He}{i}$ recombination (Sect.~\ref{sec:sol_F}). This also allows us to check for additional corrections (i.e. due to time-dependence, the thermodynamic correction factor and line cross-talk including feedback) to the effective escape probability of these resonances with respect to the normal quasi-stationary approximation (Sect.~\ref{sec:Pesc_corr}).

\subsection{Isolated helium resonances}
\label{sec:TransEqS}
Following the procedure described in \citet{Chluba2009}, in the no scattering approximation the transfer equation for photons in an {\it isolated helium resonance} connected with the transition from an excited level $i$ to the \ion{He}{i} ground state can be cast into the form
\bsub
\label{eq:real_em_abs_simp}
\beal
\label{eq:real_em_abs_simp_a}
\frac{1}{c}\left.\Abl{N_{\nu}}{t}\right|_{\rm res, \it i}
&\!=\!
p^{i}_{\rm d}\,\sigma^{i}_{\rm r}\,N^{\ion{He}{i}}_{\rm 1s}\,\phi^{i}_{\rm abs}(\nu, z)
\left\{N^{\rm em, \it i}_\nu-N_{\nu}\right\}.
\end{align}
Here $p^{i}_{\rm d}$ is the one photon death probability of the resonance; $\sigma^{i}_{\rm r}=\frac{h\nu_i}{4\pi}\,\frac{B_{\rm 1s\it i}}{\Delta\nu_{\rm D}^i}$ is the averaged one photon cross-section of the resonance, where $B_{\rm 1s\it i}$ is Einstein-B-coefficient of the line and $\Delta\nu_{\rm D}^i$ its Doppler width; $N^{\ion{He}{i}}_{\rm 1s}$ is the population of the helium ground state; $\phi^{i}_{\rm abs}(\nu, z)$ denotes the effective absorption profile; and $N^{\rm em, \it i}_\nu$ describes the production of photons in the line.

If we {\it neglect two-photon corrections} to the shapes of the profiles in the different absorption channels contributing to the effective absorption profile of the resonance, and if we assume that all are given by the normal Voigt profile $\phi^{i}_{\rm V}(\nu, z)$ for the considered resonance, then it is clear that 
\beal
\label{eq:real_em_abs_simp_b}
\phi^{i}_{\rm abs}(\nu, z)&=\phi^{i}_{\rm V}(\nu, z)\,f^{i}_\nu
\\
f^{i}_\nu&=\frac{\nu_i^2}{\nu^2}e^{h[\nu-\nu_i]/k\Tg}
\end{align}
where $f^{i}_\nu$ denotes the frequency-dependent thermodynamic correction factor, which was discussed earlier in connection with the hydrogen Lyman $\alpha$ transfer problem \citep{Chluba2008b, Chluba2009}.
\changeII{There it was shown that this factor follows from the detailed balance principle and is required in order to fully conserve a blackbody spectrum in the distant wings of the line profile for the case of thermodynamic equilibrium.}
Furthermore, one can directly write
\beal
\label{eq:real_em_abs_simp_c}
N^{\rm em, \it i}_\nu
&\!=\!\frac{2\nu_{i}^2}{c^2}\frac{g_{\rm 1s}}{g_{i}}
\frac{R^{+}_{i}}{R^{-}_{i}N^{\ion{He}{i}}_{\rm 1s}}\times\frac{1}{f^{i}_\nu}
\equiv \frac{N^{i}_{\rm em}}{f^{i}_\nu},
\end{align}
\esub
where $N^{i}_{\rm em}$ is only redshift dependent.
Here $\nu_{i}$ is the transition frequency of the resonance, $g_{\rm 1s}$ and $g_i$ are the statistical weights of the helium ground state and level $i$, respectively. In addition, $R^{+}_{i}$ and $R^{-}_{i}$ denote the one photon rates at which electrons enter and exits the level $i$ via all possible channels excluding the resonance itself.

If we now introduce the \changeIII{{\it effective}} absorption optical depth in the resonance as
\beal
\label{sec:tau_abs_all}
\tau^{i}_{\rm abs}(\nu, z', z)
&=\int_{z}^{z'}
p^{i}_{\rm d}\,\frac{c\,\sigma^{i}_{\rm r}\,N^{\ion{He}{i}}_{\rm 1s}}{H (1+\tilde{z})}\,
\phi^{i}_{\rm abs}(x[1+\tilde{z}], \tilde{z})\id \tilde{z}
\end{align}
where we defined the dimensionless frequency $x=\nu/(1+z)$, then we can write Eq.~\eqref{eq:real_em_abs_simp} as
\beal
\label{eq:real_em_abs_simp_fin}
\left.\Abl{N_{\nu}}{t}\right|_{\rm res, \it i}
&\!=\!
\dot{\tau}^{i}_{\rm abs} \left\{N^{\rm em, \it i}_\nu-N_{\nu}\right\},
\end{align}
where $\dot{\tau}^{i}_{\rm abs}=\partial_t \tau^{i}_{\rm abs}$.
The solution of this equation in the expanding Universe was already discussed earlier \citep{Chluba2009}. With the notation used here it can be written as
\bsub
\label{app:kin_abs_em_Sol_phys_asym}
\beal
\label{app:kin_abs_em_Sol_phys_asym_a}
\Delta N^{i}_{\nu}(z)
&=[N^{i}_{\rm em}(z)-N^{\rm pl}_{\nu_{i}}]\times F^{\rm S, \it i}_\nu(z).
\end{align}
where the function $F_\nu$ represents the frequency dependent part of the solution for the spectral distortion, which is given by
\beal
\label{app:kin_abs_em_Sol_phys_asym_F}
F^{\rm S, \it i}_\nu(z)&=\int_{z_{\rm s}}^z \,\Theta^{i}_{\rm a}(z, z')\,\partial_{z'}
\,
e^{-\tau^i_{\rm abs}(\nu, z', z)}\id z'
\\
\label{app:kin_abs_em_Sol_phys_asym_b}
\Theta^i_{\rm a}(z, z')
&=\frac{
\tilde{N}^i_{\rm em}(z')-\tilde{N}^{\rm pl}_{x'_i}}{\tilde{N}^i_{\rm em}(z)-\tilde{N}^{\rm pl}_{x_{i}}}
\times\frac{1}{f^i_{\nu'}(z')}
\equiv
\frac{\Theta^i_{\rm t}(z, z')}{f^i_{\nu'}(z')},
\end{align}
\esub
where $\Delta N_{\nu}=N_{\nu}-N^{\rm pl}_\nu$, $\nu'=x[1+z']$ and at $z>z_{\rm s}$ the CMB spectrum is assumed to be given by a pure blackbody spectrum $N^{\rm pl}_\nu$.
Furthermore, $\tilde{N}^i_{\rm em}(z)=N^i_{\rm em}(z)/[1+z]^2$, $x_{i}=\nu_{i}/[1+z]$, $x'_{i}=\nu_{i}/[1+z']$, and $\tilde{N}^{\rm pl}_{x}=\frac{2}{c^2}\frac{x^2}{e^{hx/kT_0}-1}$, with $T_0=2.725\,$K \citep{Fixsen2002}.
Note that $\tilde{N}^{\rm pl}_{x}$ does not explicitly depend on redshift.
Also we have used that $f_{\nu'}(z') \tilde{N}^{\rm pl}_x\equiv \tilde{N}^{\rm pl}_{x'_{21}}$.

For numerical computations it is important to analytically separate the main contribution to the solution Eq.~\eqref{app:kin_abs_em_Sol_phys_asym_F}. It can be obtained using the quasi-stationary assumption for the photon emission rate (i.e. setting $\Theta^i_{\rm t}(z, z')=1$) and neglecting the thermodynamic correction factor in the definition of $\Theta^i_{\rm a}(z, z')$, \changeII{which yields}
\bsub
\label{app:kin_abs_em_Sol_phys_asym_F_numerically}
\beal
\label{app:kin_abs_em_Sol_phys_asym_F_numerically_a}
F^{\rm S, \it i}_\nu(z)&=F^{i, 1}_\nu(z)+\Delta F^{\rm S, \it i}_\nu(z)
\\
\label{app:kin_abs_em_Sol_phys_asym_F_numerically_b}
F^{i, 1}_\nu(z)&=1-e^{-\tau^i_{\rm abs}(\nu, z_{\rm s}, z)}
\\
\label{app:kin_abs_em_Sol_phys_asym_F_numerically_c}
\Delta F^{\rm S, \it i}_\nu(z)&=
\int_{z_{\rm s}}^z \,\left[\Theta^{i}_{\rm a}(z, z')-1\right]\,\partial_{z'}
\,
e^{-\tau^i_{\rm abs}(\nu, z', z)}\id z',
\end{align}
\esub
where $\Theta^{i}_{\rm a}(z, z')$ is defined by Eq.~\eqref{app:kin_abs_em_Sol_phys_asym_b}.
This solution describes the time-dependent evolution of the spectral distortion due to the emission and absorption of photons in the resonance. 
The properties of this solution are very similar to the case of hydrogen, when neglecting two-photon corrections to the shape of the absorption profile \citep{Chluba2009}. 
We will discuss this solution in more detail in Sect.~\ref{sec:sol_F}.

\subsection{Hydrogen Lyman continuum channel alone}
\label{sec:TransEqc}
If we only consider the evolution of the photon distribution in the 1s-continuum channel of the hydrogen atom then we can start from the equation describing the emission and absorption of photons due to direct recombinations to or ionizations from the ground state \citep[e.g. see][]{Chluba2006b}
\begin{equation}
\label{eq:DNic}
\frac{1}{c}\left.\pAb{N_\nu}{t}\right|^{\rm rec}_{\rm 1s} 
=N_{\rm e} N_{\rm p}\,f_{\rm 1s}(\Te)\,
\sigma_{{\rm 1sc}} 
\frac{2 \nu^2}{c^2} e^{-\frac{h\nu}{\kB\Te}}
-N^{\ion{H}{i}}_{\rm 1s}\,\sigma_{{\rm 1s c}} N_\nu. 
\end{equation}
Here $\sigma_{{\rm 1s c}}$ is the \ion{H}{i} 1s-photoionization cross section; $N^{\ion{H}{i}}_{\rm 1s}$ is the ground state population of the hydrogen atom; $\Te$ the electron temperature, which is
always very close to the radiation temperature $\Tg=T_0(1+z)$ with
$T_0=2.725\,$K; $N_{\rm e}$ and $N_{\rm p}$ are the free electron and proton number densities; 
and from the Saha-relation one has
$f_{\rm 1s}(\Te)=\left(\frac{N^{\ion{H}{i}}_{\rm 1s}}{N_{\rm e}\,N_{\rm p}}\right)^{\rm LTE}
\!\!=\left(\frac{h^2}{2\pi\,\me\kB \Te}\right)^{3/2}
e^{h\nu_{\rm c}/\kB \Te}$,
where \changeI{$h\nu_{\rm c}\sim 13.6\,$eV} is the ionization energy of the hydrogen ground state.
Note that in Eq.~\eqref{eq:DNic} we neglected the effect of stimulated recombination, since during recombination the photon occupation number $n_\nu$ around the Lyman continuum frequency is very small ($n_\nu\ll 1$). 

Comparing with Eq.~\eqref{eq:real_em_abs_simp_a} and Eq~\eqref{eq:real_em_abs_simp_fin} one can write
\bsub
\label{eq:continuum_channel_defs}
\beal
\label{eq:continuum_channel_defs_a}
\left.\Abl{N_{\nu}}{t}\right|_{\rm c}
&\!=\!
\dot{\tau}^{\rm c}_{\rm abs}
\left\{N^{\rm em, c}_\nu-N_{\nu}\right\}
\end{align}
with the definitions
\beal
\label{eq:continuum_channel_defs_b}
\tau^{\rm c}_{\rm abs}(\nu, z', z)
&=\int_{z}^{z'}
\frac{c\,\sigma_{\rm 1sc}(x[1+\tilde{z}])\,N^{\ion{H}{i}}_{\rm 1s}}{H (1+\tilde{z})}\id \tilde{z}
\\[1mm]
N^{\rm em, c}_\nu&=\frac{2\nu^2_{\rm c}}{c^2}
\frac{N_{\rm e}\,N_{\rm p} \,\tilde{f}_{\rm 1s}(\Tg)}{N^{\ion{H}{i}}_{\rm 1s}}
\times\frac{1}{f^{\rm c}_\nu}=\frac{N^{\rm c}_{\rm em}}{f^{\rm c}_\nu},
\end{align}
\esub
where $\tilde{f}_{\rm 1s}(\Tg)=\left(\frac{h^2}{2\pi\,\me\kB \Tg}\right)^{3/2}$, and the thermodynamic factor $f^{\rm c}_\nu=\frac{\nu^2_{\rm c}}{\nu^2}e^{h[\nu-\nu_{\rm c}]/k\Tg}$.
Note that here we have assumed $\Te\equiv\Tg$. This assumption is very well justified at redshift $z\gtrsim 1000$ \citep[e.g. see][]{Seager2000}.
Now it is clear that the solution of Eq.~\eqref{eq:continuum_channel_defs} is given by Eq.~\eqref{app:kin_abs_em_Sol_phys_asym} and similarly that Eq.~\eqref{app:kin_abs_em_Sol_phys_asym_F_numerically} is applicable.

\subsection{Cross-talk of the $\ion{He}{i} \;\HeIlevel{2}{1}{P}{1}-\HeIlevel{1}{1}{S}{0}$ and $\ion{He}{i} \;\HeIlevel{2}{3}{P}{1}-\HeIlevel{1}{1}{S}{0}$ resonances}
\label{sec:TransEqX}
If we now consider the problem for the simultaneous evolution of photons in the $\ion{He}{i} \;\HeIlevel{2}{1}{P}{1}-\HeIlevel{1}{1}{S}{0}$ and $\ion{He}{i} \;\HeIlevel{2}{3}{P}{1}-\HeIlevel{1}{1}{S}{0}$ resonances, then the transfer equation reads
\beal
\label{eq:tran_Xtalk}
\left.\Abl{N_{\nu}}{t}\right|_{\rm X-talk}
&\!=\!
\dot{\tau}^{\rm a}_{\rm abs} \left\{N^{\rm em, a}_\nu-N_{\nu}\right\}
+\dot{\tau}^{\rm b}_{\rm abs} \left\{N^{\rm em, b}_\nu-N_{\nu}\right\},
\end{align}
where the superscript 'a' is related to the $\ion{He}{i} \;\HeIlevel{2}{1}{P}{1}-\HeIlevel{1}{1}{S}{0}$ transition and 'b' to the $\ion{He}{i} \;\HeIlevel{2}{3}{P}{1}-\HeIlevel{1}{1}{S}{0}$ transition.
The solution of this equation can again be found using the same procedure as in \citet{Chluba2008b}, yielding
\bsub
\label{app:kin_abs_em_Sol_phys_asym_Xtalk}
\beal
\label{app:kin_abs_em_Sol_phys_asym_Xtalk_a}
\Delta N^{\rm X}_{\nu}(z)&=\Delta N^{\rm X,a}_{\nu}(z)+\Delta N^{\rm X,b}_{\nu}(z)
\\
\Delta N^{\rm X,\it i}_{\nu}(z)&=[N^{i}_{\rm em}(z)-N^{\rm pl}_{\nu_{i}}]\times F^{\rm X,\it i}_\nu(z).
\end{align}
where the function $F^{\rm X,\it i}_\nu$ is given by
\beal
\label{app:kin_abs_em_Sol_phys_asym_F_X_talk}
F^{\rm X, \it i}_\nu(z)&=\int^{z_{\rm s}}_z \,\Theta^{i}_{\rm a}(z, z')\,
\pAb{\tau^{i}_{\rm abs}}{z'}\,e^{-\tau^{\rm a+b}_{\rm abs}(\nu, z', z)}\id z',
\end{align}
\esub
with $\Theta^{i}_{\rm a}(z, z')$ defined by Eq.~\eqref{app:kin_abs_em_Sol_phys_asym_b} and $\tau^{\rm a+b}_{\rm abs}=\tau^{\rm a}_{\rm abs}+\tau^{\rm b}_{\rm abs}$.

For numerical computations it is again better to analytically separate the main term in $F^{\rm X, \it i}_\nu(z)$. This \changeII{results in}
\bsub
\label{app:kin_abs_em_Sol_phys_asym_F_X_talk_numerically}
\beal
\label{app:kin_abs_em_Sol_phys_asym_F_X_talk_numerically_a}
F^{\rm X, \it i}_\nu(z)&=F^{i,1}_\nu(z)+\Delta F^{\rm X, \it i}_\nu(z)
\\
\label{app:kin_abs_em_Sol_phys_asym_F_X_talk_numerically_b}
\Delta F^{\rm X, \it i}_\nu(z)&=
\!\int_{z_{\rm s}}^z \!\left[\Theta^{i}_{\rm a}(z, z')e^{-\tau^{\rm a+b-\it i}_{\rm abs}(\nu, z', z)}-1\right]\partial_{z'}
e^{-\tau^{i}_{\rm abs}(\nu, z', z)}\id z',
\end{align}
\esub
where $F^{i,1}_\nu(z)$ is given by Eq.~\eqref{app:kin_abs_em_Sol_phys_asym_F_numerically_b} and $\tau^{\rm a+b-\it i}_{\rm abs}=\tau^{\rm a+b}_{\rm abs}-\tau^{i}_{\rm abs}$.

Comparing Eq.~\eqref{app:kin_abs_em_Sol_phys_asym_F_X_talk_numerically_b} with Eq.~\eqref{app:kin_abs_em_Sol_phys_asym_F_numerically_c}, one can see that the cross-talk between the lines simply leads to a {\it frequency}- and {\it time-dependent modulation} of the effective photon emission rate $\Theta^{i}_{\rm a}(z, z')$ related to the resonance $i$ by $e^{-\tau^{\rm a+b-\it i}_{\rm abs}(\nu, z', z)}$. Physically this just reflects the fact that with time some photons from the resonance $i$ will disappear from the photon distribution due to the absorption in the other resonance.

\subsection{$\ion{He}{i} \;\HeIlevel{2}{1}{P}{1}-\HeIlevel{1}{1}{S}{0}$ and $\ion{He}{i} \;\HeIlevel{2}{3}{P}{1}-\HeIlevel{1}{1}{S}{0}$ resonances with hydrogen continuum opacity}
\label{sec:TransEqrc}
With the results of Sect.~\ref{sec:TransEqX}, It is now straightforward  to write down the solution for transfer problem in one helium resonance including the hydrogen continuum. This yields
\bsub
\label{app:kin_abs_em_Sol_phys_asym_rc}
\beal
\label{app:kin_abs_em_Sol_phys_asym_rc_a}
\Delta N^{\rm rc}_{\nu}(z)&=\Delta N^{\rm rc,a/b}_{\nu}(z)+\Delta N^{\rm rc,c}_{\nu}(z)
\\
\Delta N^{\rm rc,\it i}_{\nu}(z)&=[N^{i}_{\rm em}(z)-N^{\rm pl}_{\nu_{i}}]\times F^{\rm rc,\it i}_\nu(z).
\end{align}
\esub
where the function $F^{\rm rc,\it i}_\nu$ can be written as
\bsub
\label{app:kin_abs_em_Sol_phys_asym_F_rc_numerically}
\beal
\label{app:kin_abs_em_Sol_phys_asym_F_rc_numerically_a}
F^{\rm rc, \it i}_\nu(z)&=F^{i,1}_\nu(z)+\Delta F^{\rm rc, \it i}_\nu(z)
\\
\label{app:kin_abs_em_Sol_phys_asym_F_rc_numerically_b}
\Delta F^{\rm rc, \it i}_\nu(z)&=
\!\!\int_{z_{\rm s}}^z \!\left[\Theta^{i}_{\rm a}(z, z')e^{-\tau^{\rm a/b+c-\it i}_{\rm abs}(\nu, z', z)}-1\right]
\!\partial_{z'}
e^{-\tau^{i}_{\rm abs}(\nu, z', z)}\id z'\!,
\end{align}
\esub
with $\Theta^{i}_{\rm a}(z, z')$ defined by Eq.~\eqref{app:kin_abs_em_Sol_phys_asym_b}; $\tau^{\rm a/b+c-\it i}_{\rm abs}=\tau^{\rm a/b}_{\rm abs}+\tau^{\rm c}_{\rm abs}-\tau^{i}_{\rm abs}$. \changeII{Here} we used the notation 'a/b' in the superscripts, which means either resonance 'a' or 'b'; and $F^{i,1}_\nu(z)$ as defined in Eq.~\eqref{app:kin_abs_em_Sol_phys_asym_F_numerically_b}.

Furthermore, when simultaneously including both resonances and the \ion{H}{i} continuum one finds
\bsub
\label{app:kin_abs_em_Sol_phys_asym_all}
\beal
\label{app:kin_abs_em_Sol_phys_asym_all_a}
\Delta N^{\rm all}_{\nu}(z)
&=\Delta N^{\rm all,a}_{\nu}(z)+\Delta N^{\rm all,b}_{\nu}(z)+\Delta N^{\rm all,c}_{\nu}(z)
\\
\Delta N^{\rm all,\it i}_{\nu}(z)&=[N^{i}_{\rm em}(z)-N^{\rm pl}_{\nu_{i}}]\times F^{\rm all,\it i}_\nu(z).
\end{align}
\esub
where the function $F^{\rm all,\it i}_\nu$ is given by
\bsub
\label{app:kin_abs_em_Sol_phys_asym_F_all_numerically}
\beal
\label{app:kin_abs_em_Sol_phys_asym_F_all_numerically_a}
F^{\rm all, \it i}_\nu(z)&=F^{i,1}_\nu(z)+\Delta F^{\rm all, \it i}_\nu(z)
\\
\label{app:kin_abs_em_Sol_phys_asym_F_all_numerically_b}
\Delta F^{\rm all, \it i}_\nu(z)&=
\!\!\int_{z_{\rm s}}^z \!\left[\Theta^{i}_{\rm a}(z, z')e^{-\tau^{\rm a+b+c-\it i}_{\rm abs}(\nu, z', z)}-1\!\right]
\!\partial_{z'}
e^{-\tau^{i}_{\rm abs}(\nu, z', z)}\!\id z'\!,
\end{align}
\esub
with $\Theta^{i}_{\rm a}(z, z')$ defined by Eq.~\eqref{app:kin_abs_em_Sol_phys_asym_b}; $\tau^{\rm a+b+c-\it i}_{\rm abs}=\tau^{\rm a}_{\rm abs}+\tau^{\rm b}_{\rm abs}+\tau^{\rm c}_{\rm abs}-\tau^{i}_{\rm abs}$; and $F^{i,1}_\nu(z)$ as defined in Eq.~\eqref{app:kin_abs_em_Sol_phys_asym_F_numerically_b}

\changeI{Note that it is very easy to extend the solution to $n$ different resonances with \ion{H}{i} continuum absorption between the lines}, \changeII{or other processes of emission and absorption (e.g. \ion{H}{i} 2s-1s two-photon emission). However, since in most cases resonances in hydrogen and helium are very distant to each other, for our purpose we neglect the case of cross talk between many resonances.}

\begin{figure*}
\centering 
\includegraphics[width=0.95\columnwidth]
{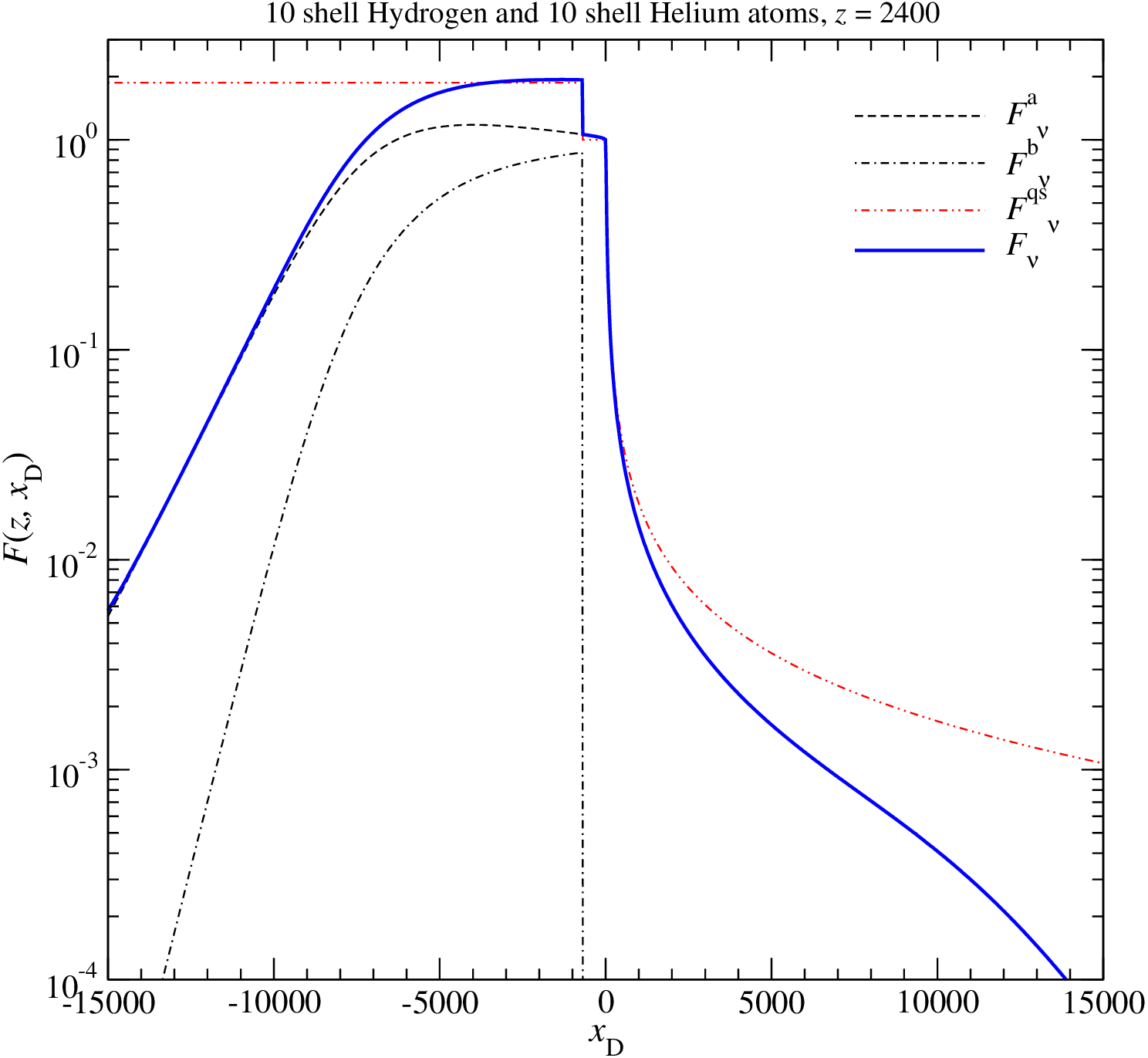}
\hspace{4mm}
\includegraphics[width=0.95\columnwidth]
{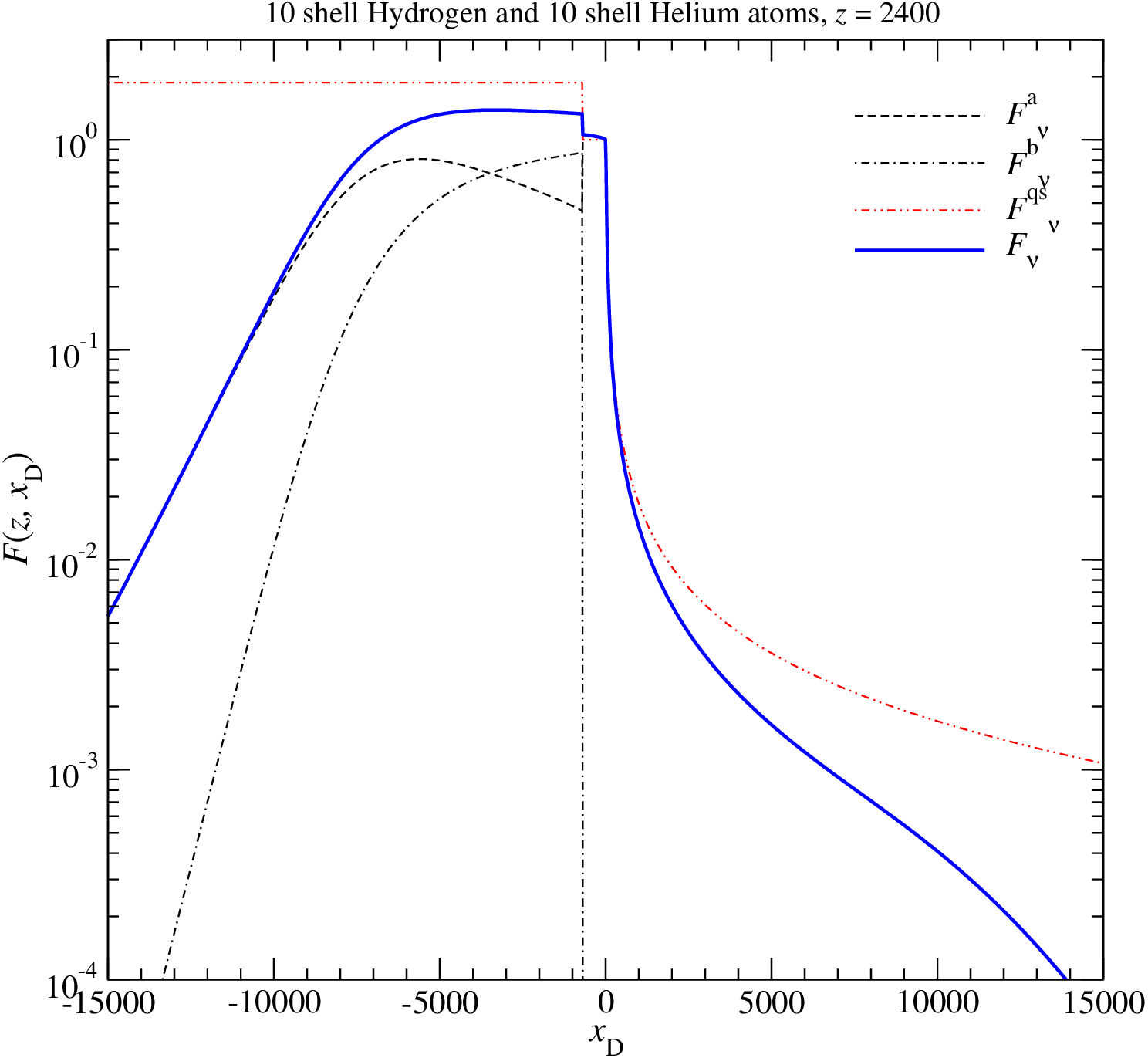}
\\[1mm]
\includegraphics[width=0.95\columnwidth]
{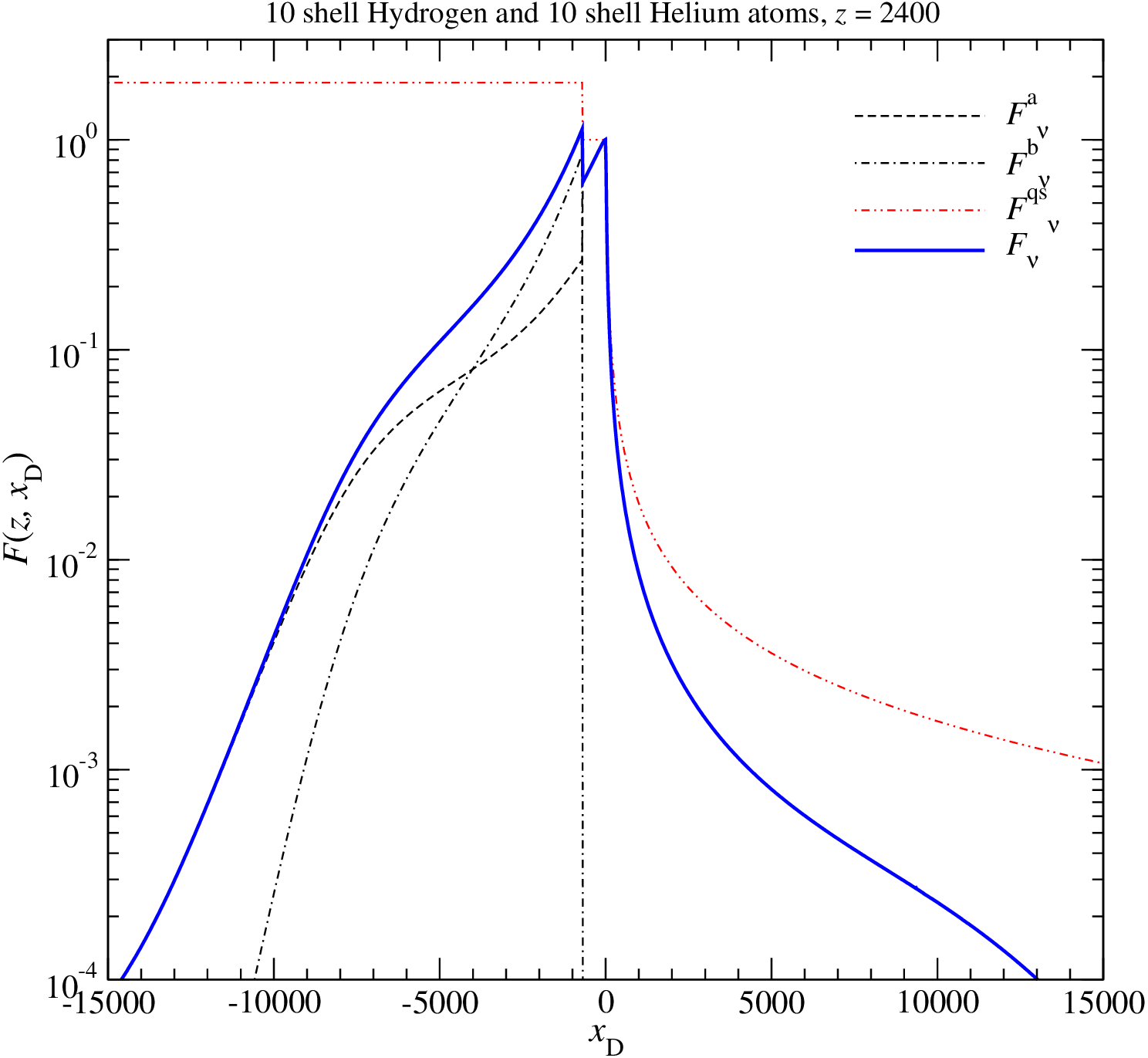}
\hspace{4mm}
\includegraphics[width=0.95\columnwidth]
{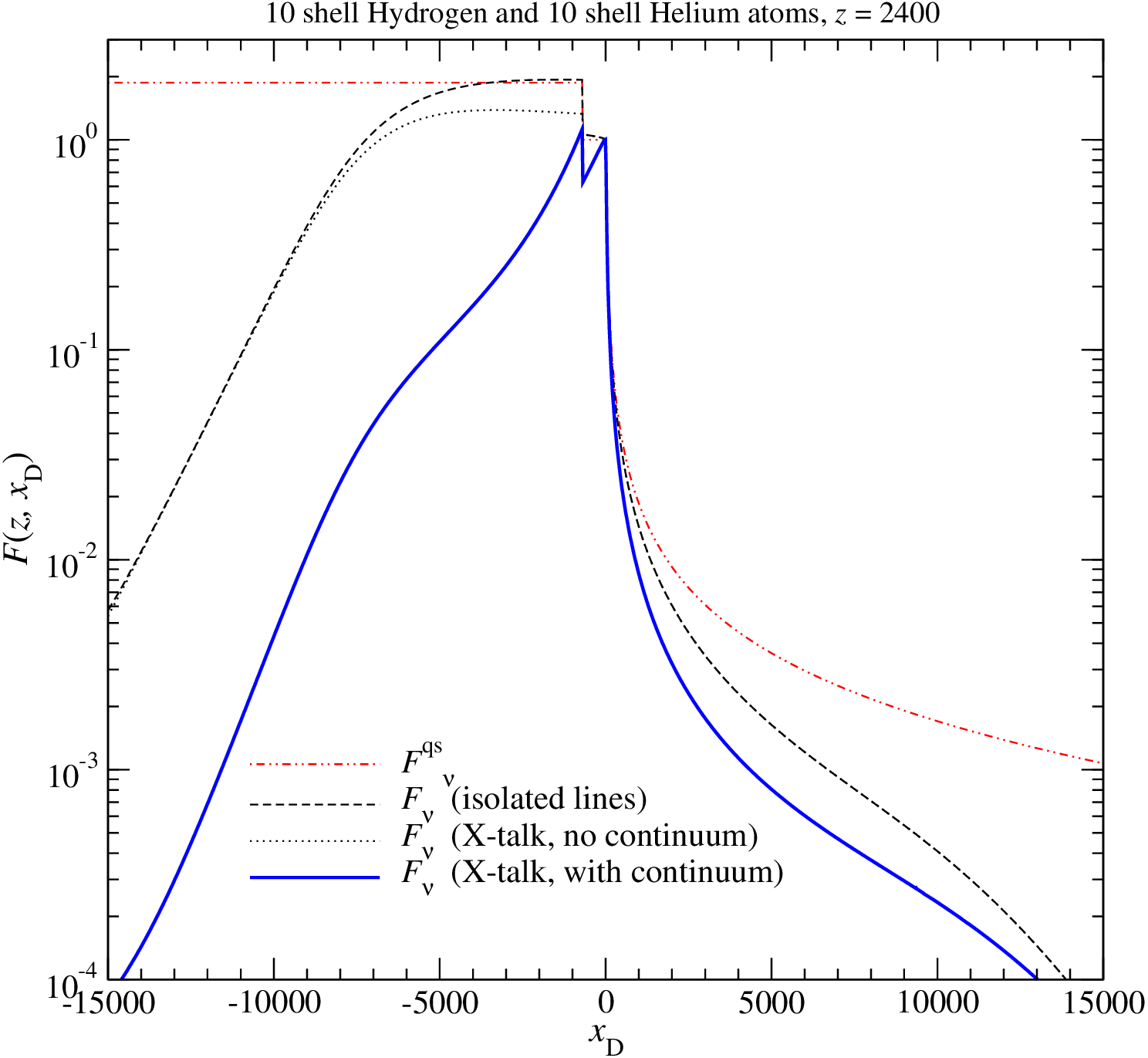}
\caption{Spectral distortion due to the $\ion{He}{i} \;\HeIlevel{2}{1}{P}{1}-\HeIlevel{1}{1}{S}{0}$ and $\ion{He}{i} \;\HeIlevel{2}{3}{P}{1}-\HeIlevel{1}{1}{S}{0}$ resonances at $z=2400$
-- upper left: isolated lines -- upper right: cross-talk without \ion{H}{i} continuum opacity -- lower left: cross-talk with \ion{H}{i} continuum opacity -- lower right: comparison of the total distortion. 
The frequency is given as the distance to the central frequency $\nu_{\rm a}$ of the  $\ion{He}{i} \;\HeIlevel{2}{1}{P}{1}-\HeIlevel{1}{1}{S}{0}$ resonance in Doppler units, $\xD=\frac{\nu-\nu_{\rm a}}{\Delta\nu^{\rm a}_{\rm D}}$.
In all panels we normalized the total distortion $\Delta N_\nu(z)$ by $\Delta N^{\rm a}_{\rm em}(z)=N_{\rm em}^{\rm a}(z)-N^{\rm pl}_{\nu_{\rm a}}$ for the $\ion{He}{i} \;\HeIlevel{2}{1}{P}{1}-\HeIlevel{1}{1}{S}{0}$ resonance, i.e. $F_\nu(z)=\Delta N_\nu(z)/\Delta N^{\rm a}_{\rm em}(z)$. Furthermore, $F_\nu(z)$ denotes the total distortion, $F_\nu^{\rm a}(z)$ the distortion due to the  $\ion{He}{i} \;\HeIlevel{2}{1}{P}{1}-\HeIlevel{1}{1}{S}{0}$ line, and $F_\nu^{\rm b}(z)$ the distortion due to the  $\ion{He}{i} \;\HeIlevel{2}{3}{P}{1}-\HeIlevel{1}{1}{S}{0}$ resonance.
For comparison we show the simple quasi-stationary solution $F^{\rm qs, ab}_\nu(z)$ according to Eq.~\eqref{eq:F_nu_QS_ab}.}
\label{fig:Fnu_2400}
\end{figure*}
\section{Numerical results for the high frequency CMB spectral distortion at different times}
\label{sec:sol_F}
Using the solutions given in the previous sections it is possible to compute the high frequency spectral distortions due to the emission of photons in the main helium resonances at different epochs.
To demonstrate the main aspects of the problem here we shall {\it only} consider the evolution of the photon field due to the $\ion{He}{i} \;\HeIlevel{2}{1}{P}{1}-\HeIlevel{1}{1}{S}{0}$ and $\ion{He}{i} \;\HeIlevel{2}{3}{P}{1}-\HeIlevel{1}{1}{S}{0}$ transitions.
These two resonances are separated by $\Delta\nu/\nu\sim 1\%$, so that photons from the $\ion{He}{i} \;\HeIlevel{2}{1}{P}{1}-\HeIlevel{1}{1}{S}{0}$ line will mainly feedback on the $\ion{He}{i} \;\HeIlevel{2}{3}{P}{1}-\HeIlevel{1}{1}{S}{0}$ intercombination line after $\Delta z/z\sim 1\%$ which corresponds to about $700$ Doppler width at $z\sim 2200$ (see Table~\ref{tab:feedback}).
In addition, it is expected that due to the presence of a small fraction of neutral hydrogen during helium recombination resonance photons will be absorbed in the \ion{H}{i} Lyman continuum \citep{Switzer2007I, Jose2008}. These photons will re-appear as pre-recombinational photons from hydrogen \citep{Jose2008}, at frequencies very far red-ward of the main helium resonances.
This leads to two effects: 
\begin{itemize}

\item[(i)] At a given time the main resonances will be supported by {\it fewer} photons than in the case without \ion{H}{i} continuum opacity. This leads to an {\it increase} in the effective escape probability which becomes very large at redshift $z\lesssim 2400$ \citep{Kholupenko2007, Switzer2007I, Jose2008}.

\item[(ii)] The feedback of photons from the $\ion{He}{i} \;\HeIlevel{2}{1}{P}{1}-\HeIlevel{1}{1}{S}{0}$ resonance on the $\ion{He}{i} \;\HeIlevel{2}{3}{P}{1}-\HeIlevel{1}{1}{S}{0}$ intercombination line will be reduced, since part of the photons released in the $\ion{He}{i} \;\HeIlevel{2}{1}{P}{1}-\HeIlevel{1}{1}{S}{0}$ line will disappear from the photon distribution before they can actually reach the $\ion{He}{i} \;\HeIlevel{2}{3}{P}{1}-\HeIlevel{1}{1}{S}{0}$ resonance \citep[see also explanation in][]{Switzer2007I}. 
As we demonstrate here due to this process the feedback of $\gamma(\ion{He}{i} \;\HeIlevel{2}{1}{P}{1}-\HeIlevel{1}{1}{S}{0})\rightarrow \gamma( \ion{He}{i} \;\HeIlevel{2}{3}{P}{1}-\HeIlevel{1}{1}{S}{0})$ practically stops at redshift $z\lesssim 2200-2400$ (see Sect.~\ref{sec:feedback_21P1_23P1}).

\end{itemize}
%

In order to compute the solution for the spectral distortion caused by the considered resonances one has to give the solutions for the population of the \ion{He}{i} levels as a function of time. For this one has to assume that the approximations used in the multi-level helium recombination code\footnote{We based this code on the works of \citet{Jose2006}, \citet{Chluba2007}, and \citet{Jose2008}. As similar code was already used for the first training of {\sc Rico} \citep{Fendt2009}.} already captures the main processes, and that the corrections due to the additional effects (e.g. time-dependence, thermodynamic correction factor, line feedback and line cross-talk) are small. 
It was already shown \citep{Kholupenko2007, Switzer2007I, Jose2008} that the main correction during helium recombination \changeI{in comparison with the standard computation \citep{Seager2000}} is due to the speed-up of the $\ion{He}{i} \;\HeIlevel{2}{1}{P}{1}-\HeIlevel{1}{1}{S}{0}$ and $\ion{He}{i} \;\HeIlevel{2}{3}{P}{1}-\HeIlevel{1}{1}{S}{0}$ channels caused by the \ion{H}{i} continuum opacity.
\changeI{To include this process in our computations of the} \ion{He}{i} populations we will follow the approach of \citet{Jose2008} using the 1D-integral approximation (see Eq.~(B.3) in their paper) to take the increase in the escape probability of the $\ion{He}{i}\;\HeIlevel{2}{1}{P}{1}-\HeIlevel{1}{1}{S}{0}$ and $\ion{He}{i}\;\HeIlevel{2}{3}{P}{1}-\HeIlevel{1}{1}{S}{0}$ transitions into account. 
In addition, for the $\ion{He}{i}\;\HeIlevel{2}{1}{P}{1}-\HeIlevel{1}{1}{S}{0}$ resonance, corrections due to {\it partial frequency redistribution} are important \citep{Switzer2007I, Jose2008} which we account for with the 'fudge'-function used in \citet{Jose2008}.
Henceforth we will refer to this model as our {\it reference model}. Consistent with our previous works \citep{Jose2006, Chluba2007, Jose2008, Chluba2009, Chluba2009b} we used the cosmological parameters $T_0=2.725\,$K, $Y_{\rm p}=0.24$, $h=0.71$, $\Omega_{\rm b}=0.0444$, $\Omega_{\rm m}=0.2678$, $\Omega_{\Lambda}=0.7322$, and $\Omega_{\rm k}=0$.

To understand the importance of the different corrections for the shape of the CMB spectral distortion introduced by the $\ion{He}{i} \;\HeIlevel{2}{1}{P}{1}-\HeIlevel{1}{1}{S}{0}$ and $\ion{He}{i} \;\HeIlevel{2}{3}{P}{1}-\HeIlevel{1}{1}{S}{0}$ transitions we now \changeI{discuss} the solution for two different representative stages (Sect.~\ref{sec:sol_F_2400} and \ref{sec:sol_F_2000}) during helium recombination.
We then also compute the present-day ($z=0$) CMB distortion at high frequencies including different processes. In particular we show that the absorption in the \ion{H}{i} continuum completely erases the high frequency spectral distortion from $\ion{He}{ii} \rightarrow \ion{He}{i} $ recombination, \changeII{a fact that was already suspected} earlier \citep{Chluba2007b, Jose2008}.

\subsection{High frequency spectral distortion at $z=2400$}
\label{sec:sol_F_2400}
%
In Fig.~\ref{fig:Fnu_2400} we show the results for $F_\nu$ at $z=2400$ for different combinations of the line emission and absorption processes.
We normalized all the curves to the central distortion in the $\ion{He}{i} \;\HeIlevel{2}{1}{P}{1}-\HeIlevel{1}{1}{S}{0}$ line, $\Delta N_{\nu_{\rm a}}^{\rm a}(z)=\Delta N_{\rm em}^{\rm a}(z)=N_{\rm em}^{\rm a}(z)-N^{\rm pl}_{\nu_{\rm a}}$.
In all cases we neglected the possible emission in the \ion{H}{i} continuum channel, since close to the resonances it contributes very little to the total distortion.
For comparison we also show the simple quasi-stationary solution
\beal
\label{eq:F_nu_QS_ab}
F^{\rm qs, ab}_\nu(z)
&=1-e^{-\tau^{\rm a}_{\rm d}[1-\chi^{\rm a}_{\nu}]}
+\frac{\Delta N_{\rm em}^{\rm b}(z)}{\Delta N_{\rm em}^{\rm a}(z)}
\left[1-e^{-\tau^{\rm b}_{\rm d}[1-\chi^{\rm b}_{\nu}]}\right],
\end{align}
\changeI{which neglects any time-dependence, cross talk, the \ion{H}{i} continuum opacity, or feedback of the lines.}
Here $\tau_{\rm d}^i=p_{\rm d}^i\,\tau_{\rm S}^i$, where $\tau_{\rm S}^i$ is the standard Sobolev optical depth for the resonance $i$, and $\Delta N_{\rm em}^{\rm b}(z)=\Delta N_{\nu_{\rm b}}^{\rm b}(z)$ is the central distortion in the $\ion{He}{i} \;\HeIlevel{2}{3}{P}{1}-\HeIlevel{1}{1}{S}{0}$ resonance.
Also we defined $\chi^i_\nu=\int_0^\nu \varphi^i(\nu')\id\nu'$ with $\varphi^i_{\rm V}(\nu, z)=\phi^i_{\rm V}(\nu, z)/\Delta\nu^i_{\rm D}$.

\changeI{Note that the \changeIII{effective} absorption optical depth, $\tau_{\rm d}^i$, takes into account that only a fraction $p_{\rm d}^i$ of interactions with the resonance really leads to an {\it absorption} or {\it death} of the photon. As an example, for the \ion{H}{i} Lyman $\alpha$ resonance this absorption is related to a transition of the electron towards higher levels or the continuum in a two-photon process \citep[e.g. see][]{Chluba2009}, however, for the higher \ion{H}{i} Lyman-series also spontaneous decays towards lower levels matter. All these contributions can be taken into account using the appropriate {\it branching ratios} for each excited level in hydrogen and helium, assuming that the ambient radiation field is given by the CMB blackbody spectrum.}

When considering the two resonances separately and neglecting the \ion{H}{i} continuum opacity (Fig.~\ref{fig:Fnu_2400}, upper left panel), $F_\nu$ is simply given by the sum of the distortions from each resonance.
At low frequencies one can clearly see the modulation of the distortion due to the time-dependence of the emission rate, like in the case of the \ion{H}{i} Lyman $\alpha$ distortion \citep{Chluba2008b}. In addition for the distortion caused by the $\ion{He}{i} \;\HeIlevel{2}{1}{P}{1}-\HeIlevel{1}{1}{S}{0}$ resonance one can also find the scaling $F^{\rm a}_\nu\sim 1/f^{\rm a}_\nu$, again in full analog to the  \ion{H}{i} Lyman $\alpha$ line \citep{Chluba2009}.
%

\begin{figure}
\centering 
\includegraphics[width=0.95\columnwidth]
{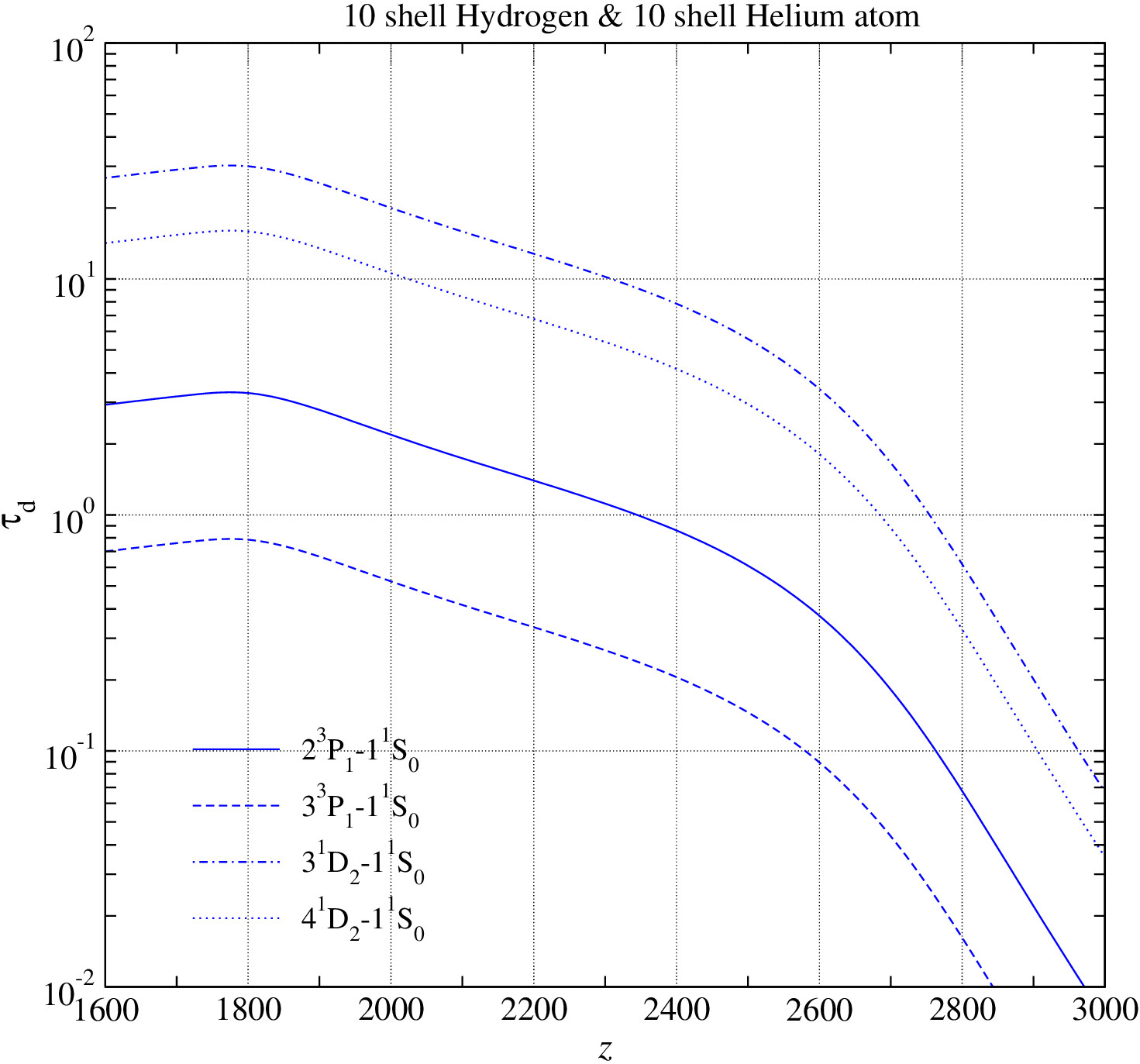}
\caption{ \changeIII{Effective} absorption optical depth, \changeI{$\tau^i_{\rm d}=p^i_{\rm d}\tau^i_{\rm S}$}, in the first two $\ion{He}{i}\;\HeIlevel{n}{3}{P}{1}-\HeIlevel{1}{1}{S}{0}$ intercombination and $\ion{He}{i}\;\HeIlevel{n}{1}{D}{2}-\HeIlevel{1}{1}{S}{0}$ quadrupole transitions. In all cases the death probabilities \changeI{$p^i_{\rm d}$} are very close to unity, so that $\tau^i_{\rm d}\approx\tau^i_{\rm S}$.}
\label{fig:tau_QT}
\end{figure}
If we now include the cross-talk between the lines (Fig.~\ref{fig:Fnu_2400}, upper right panel), then the main difference is only due to the fact that the photons from the $\ion{He}{i} \;\HeIlevel{2}{1}{P}{1}-\HeIlevel{1}{1}{S}{0}$ line have to pass through the $\ion{He}{i} \;\HeIlevel{2}{3}{P}{1}-\HeIlevel{1}{1}{S}{0}$ resonance, when they feedback on the  $\ion{He}{i} \;\HeIlevel{2}{3}{P}{1}-\HeIlevel{1}{1}{S}{0}$ transition and get partially re-absorbed. This leads to a drop of $F^{\rm a}_\nu$ by $e^{-\tau^{\rm b}_{\rm d}}$ in the vicinity of the $\ion{He}{i} \;\HeIlevel{2}{3}{P}{1}-\HeIlevel{1}{1}{S}{0}$ resonance (at around $\xD\sim -700$).
However, the additional absorption of photons from resonance $i$ in the (distant) damping wing of the line $j$ is negligible.
Since $\tau^{\rm b}_{\rm d}$ is a function of redshift, the amplitude of this drop depends on the redshift of line crossing. 
The farther one goes red-ward of the $\ion{He}{i} \;\HeIlevel{2}{3}{P}{1}-\HeIlevel{1}{1}{S}{0}$ line the smaller the absorption caused by the line crossing becomes. At $\xD\lesssim -10^4$ the distortion $F^{\rm a}_\nu$ again becomes comparable to $F^{\rm a}_\nu$ in the case without cross-talk (compare curves in the upper panels of Fig.~\ref{fig:Fnu_2400}). 
This is because those photons have passed through the $\ion{He}{i} \;\HeIlevel{2}{3}{P}{1}-\HeIlevel{1}{1}{S}{0}$ resonance at much earlier times, when the \changeIII{effective} $\ion{He}{i} \;\HeIlevel{2}{3}{P}{1}-\HeIlevel{1}{1}{S}{0}$ absorption optical depth was smaller (cf. Fig.~\ref{fig:tau_QT}).
At $z=2400$ for our reference model one has $\tau^{\rm b}_{\rm d}\sim 0.8$, so that the distortion due to the $\ion{He}{i} \;\HeIlevel{2}{3}{P}{1}-\HeIlevel{1}{1}{S}{0}$ line should drops by a factor of $\sim 2.2$ at $\xD\sim -700$. This is in good agreement with our computations (cf. Fig.~\ref{fig:Fnu_2400}, upper right panel).

Finally, when we also include the effect of the \ion{H}{i} continuum (Fig.~\ref{fig:Fnu_2400}, lower panels), the shape of the distortion changes drastically. In particular at large distances red-ward of the resonance the remaining distortion due to \ion{H}{i} absorption is strongly reduced.
From this Figure it is clear that basically {\it all} the photons from the $\ion{He}{i} \;\HeIlevel{2}{1}{P}{1}-\HeIlevel{1}{1}{S}{0}$ and $\ion{He}{i} \;\HeIlevel{2}{3}{P}{1}-\HeIlevel{1}{1}{S}{0}$ lines are re-absorbed in the  \ion{H}{i} continuum before they can actually reach frequencies below the \ion{H}{i} Lyman continuum threshold frequency $\nu_{\rm c}$, which at $z\sim 2400$ is at $\xD\sim -\pot{2.1}{4}$.
These absorbed photons should lead to additional ionizations of neutral hydrogen well {\it before} the actual epoch of hydrogen recombination ($z\lesssim 1600$). Since at those times hydrogen is still in very close equilibrium with the continuum, this process will not cause any important changes in the ionization history, but should lead to some {\it pre-recombinational emission} in the lines of hydrogen. 
However, until now this effect has not been fully taken into account, since only those absorptions causing the changes in the effective escape probabilities of the $\ion{He}{i} \;\HeIlevel{2}{1}{P}{1}-\HeIlevel{1}{1}{S}{0}$ and $\ion{He}{i} \;\HeIlevel{2}{3}{P}{1}-\HeIlevel{1}{1}{S}{0}$ resonances where accounted for \citep{Jose2008}.
We will discuss this problem in more detail below (Sect.~\ref{sec:pre-rec}).

\begin{figure}
\centering 
\includegraphics[width=0.95\columnwidth]{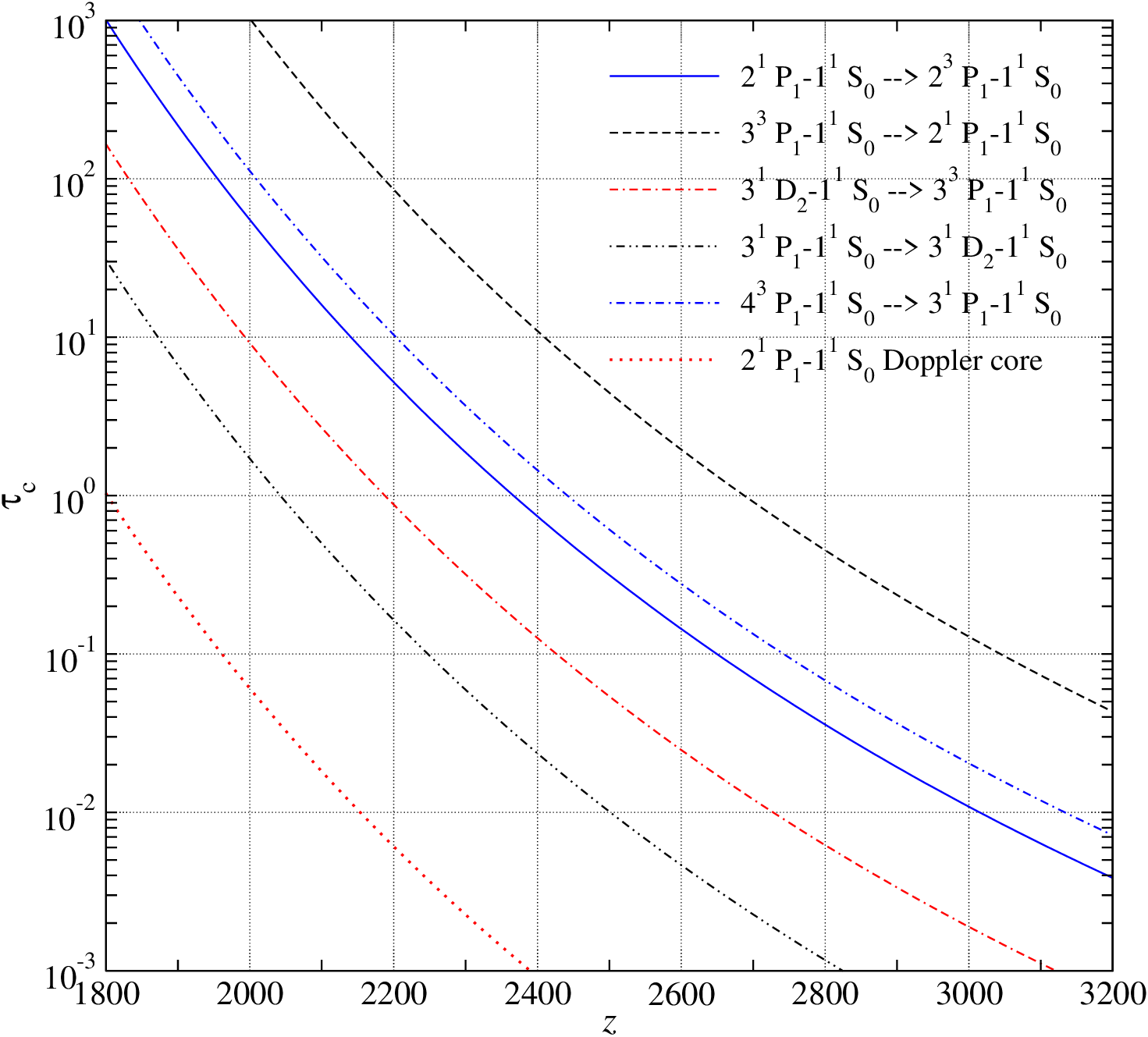}
\caption{Absorption optical depth, \changeI{$\tau_{\rm c}$,} in the \ion{H}{i} Lyman-continuum. \changeI{All curves can be computed using the definition Eq.~\eqref{eq:continuum_channel_defs_b} with the  \ion{H}{i} 1s-photoionization cross section, where the appropriated $\Delta z/z$ is defined by the distance between the considered resonances}. The dotted curve shows the optically depth when crossing the Doppler core of the $\ion{He}{i} \;\HeIlevel{2}{1}{P}{1}-\HeIlevel{1}{1}{S}{0}$ resonance.
%
%
The other curves show the total optical depth between the quoted resonances. These have been computed using the solution for $N^{\ion{H}{i}}_{\rm 1s}$ from our reference model \citep{Jose2008} using Eq.~\eqref{eq:continuum_channel_defs_b}.}
\label{fig:tauc}
\end{figure}
Again looking at the lower left panel in Fig.~\ref{fig:Fnu_2400}, one can also observe some modifications of the spectral distortion between the $\ion{He}{i} \;\HeIlevel{2}{1}{P}{1}-\HeIlevel{1}{1}{S}{0}$ and $\ion{He}{i} \;\HeIlevel{2}{3}{P}{1}-\HeIlevel{1}{1}{S}{0}$ resonances.
These changes are cause by the absorption of photons from the $\ion{He}{i} \;\HeIlevel{2}{1}{P}{1}-\HeIlevel{1}{1}{S}{0}$ line while they are on their way to the $\ion{He}{i} \;\HeIlevel{2}{3}{P}{1}-\HeIlevel{1}{1}{S}{0}$ intercombination line.
This will lead to a reduction of the feedback correction to the $\ion{He}{i} \;\HeIlevel{2}{3}{P}{1}-\HeIlevel{1}{1}{S}{0}$ transition, as we will explain in detail below (see Sect.~\ref{sec:feedback_21P1_23P1}).
One can estimate this reduction by simply computing the \ion{H}{i} absorption optical depth between the two lines (see solid line in Fig.~\ref{fig:tauc}). 
At $z\sim 2400$ we find $\tau_{\rm c}^{\rm a\rightarrow b}\sim 0.7$, so that, in good agreement with the result shown in Fig.~\ref{fig:Fnu_2400}, until the $\ion{He}{i} \;\HeIlevel{2}{1}{P}{1}-\HeIlevel{1}{1}{S}{0}$ photons have reached the $\ion{He}{i} \;\HeIlevel{2}{3}{P}{1}-\HeIlevel{1}{1}{S}{0}$ resonance due to \ion{H}{i} continuum absorption one expects an reduction of the spectral distortion by a factor of $\sim 2$.

\begin{figure*}
\centering 
\includegraphics[width=0.95\columnwidth]
{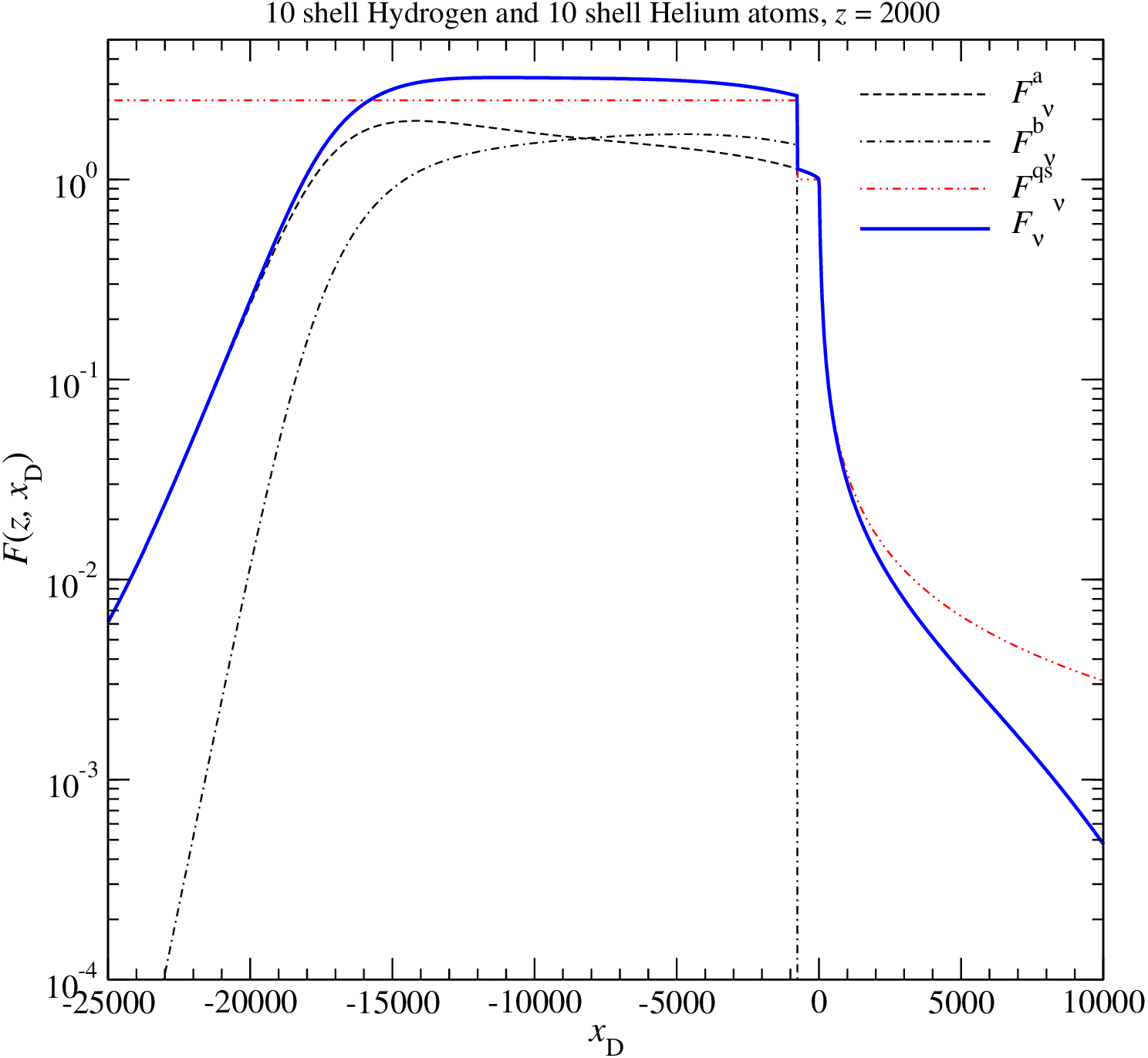}
\hspace{4mm}
\includegraphics[width=0.95\columnwidth]
{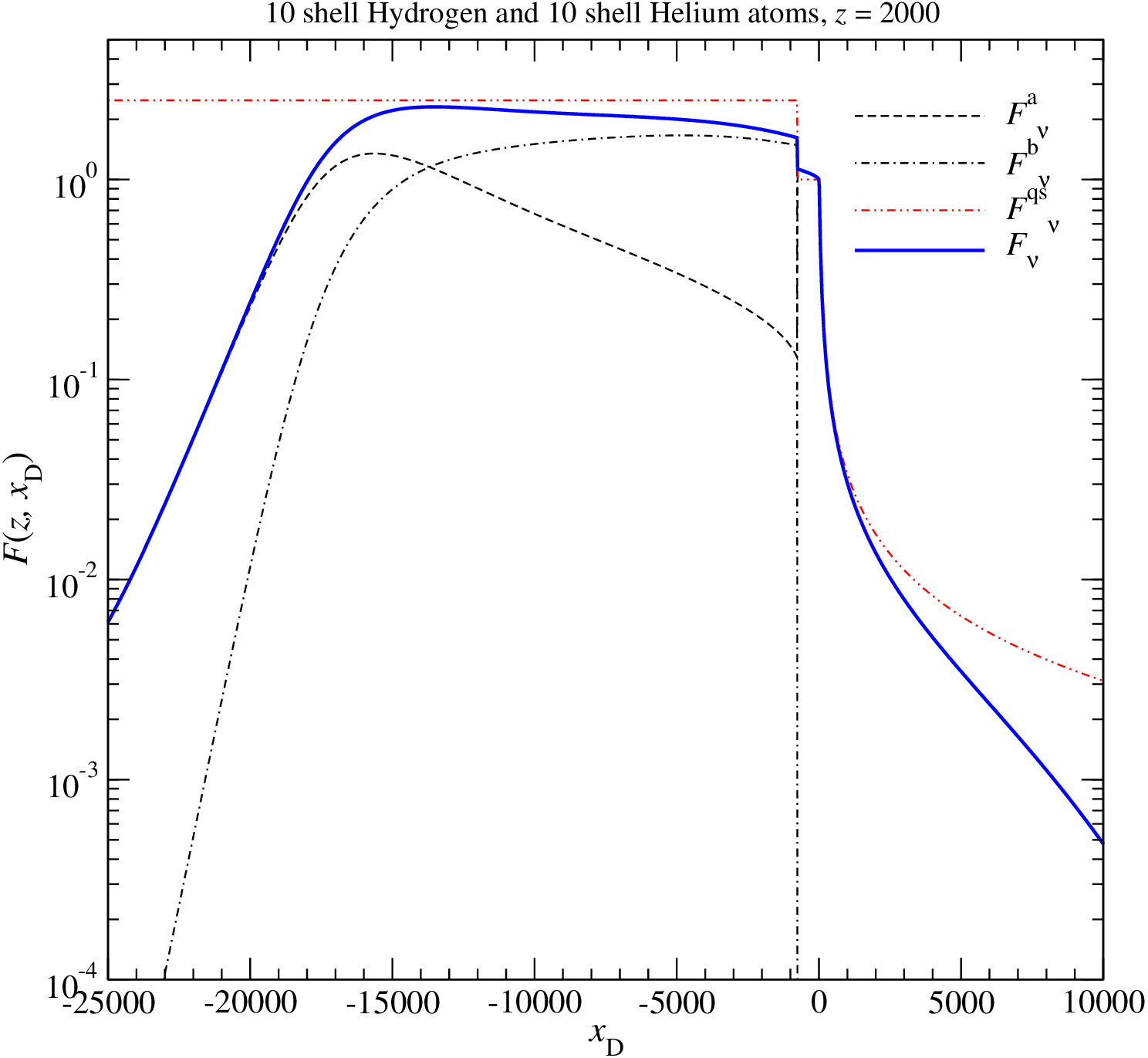}
\\[1mm]
\includegraphics[width=0.95\columnwidth]
{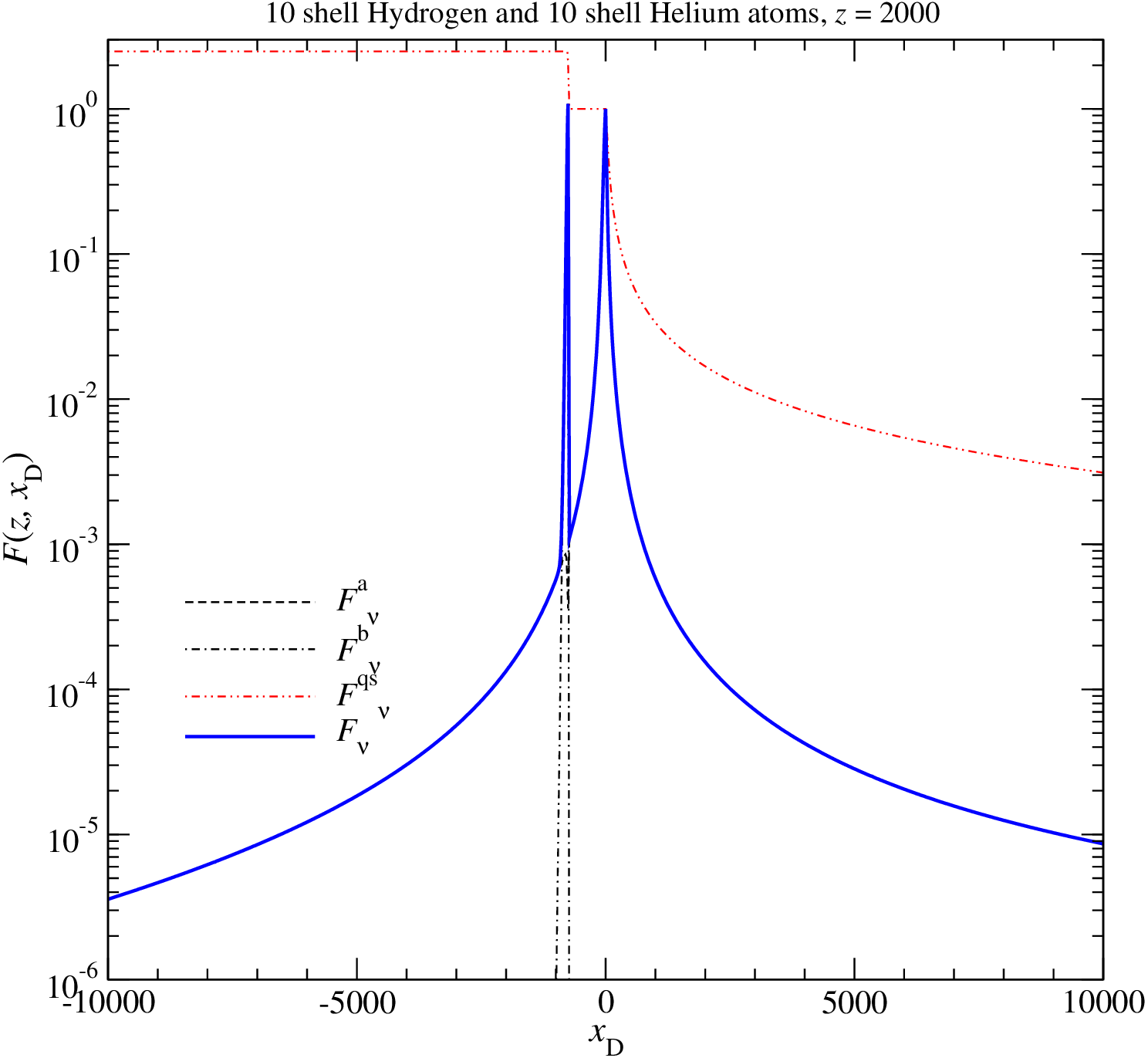}
\hspace{4mm}
\includegraphics[width=0.95\columnwidth]
{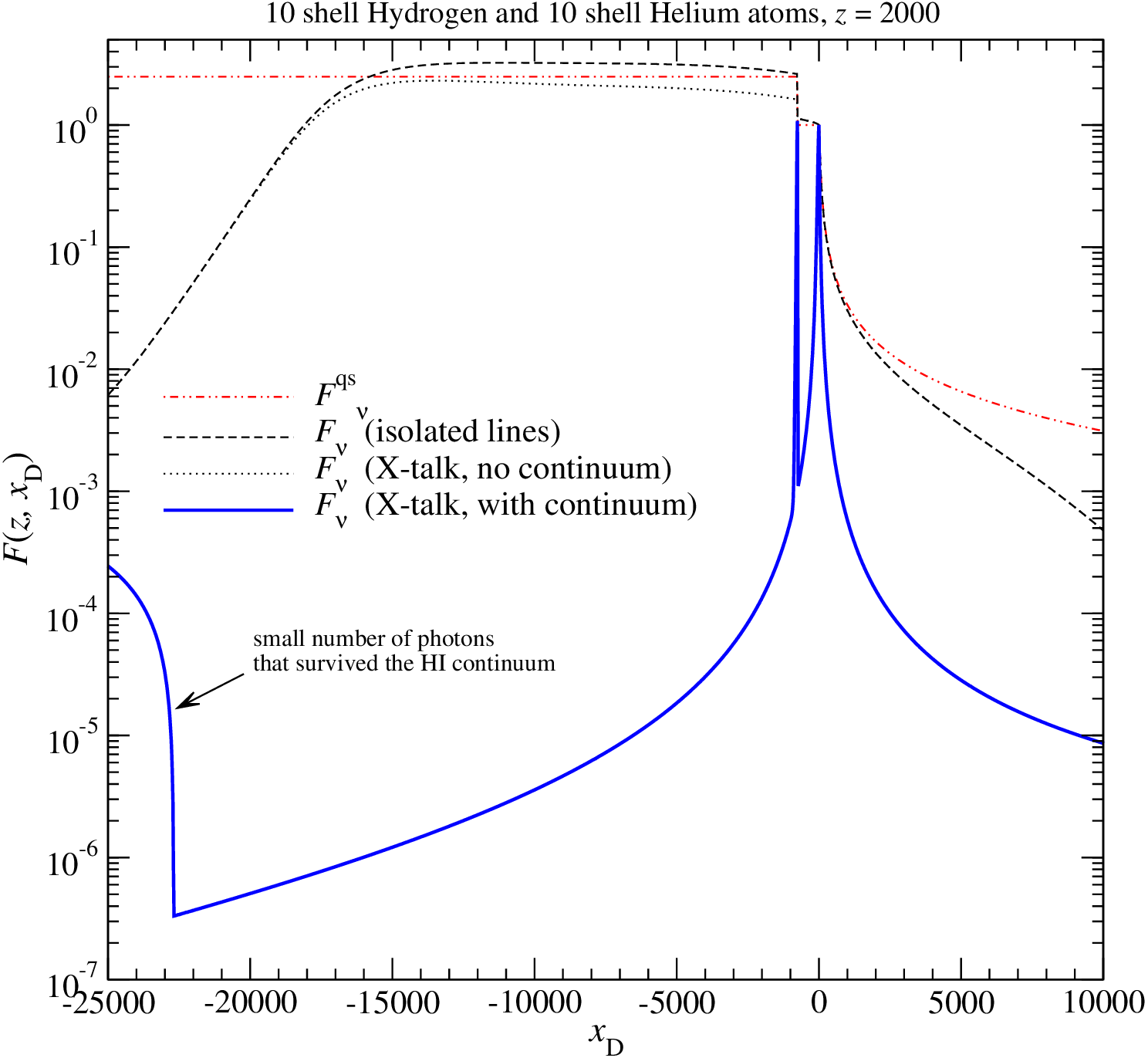}
\caption{Same as Fig.~\ref{fig:Fnu_2400} but at $z=2000$.}
\label{fig:Fnu_2000}
\end{figure*}
\subsection{High frequency spectral distortion at $z=2000$}
\label{sec:sol_F_2000}
Moving to $z\sim 2000$ one can see that the general behavior of the solution is very similar to the case we considered in the previous section.
However, for example, now the drop in the spectral distortion from $\ion{He}{i} \;\HeIlevel{2}{1}{P}{1}-\HeIlevel{1}{1}{S}{0}$ line the due to the feedback absorption in the $\ion{He}{i} \;\HeIlevel{2}{3}{P}{1}-\HeIlevel{1}{1}{S}{0}$ resonance has increased to about a factor of $\sim 8.5$ (cf. Fig.~\ref{fig:Fnu_2000}, upper right panel). Again looking at Fig.~\ref{fig:tau_QT} we find $\tau^{\rm b}_{\rm d}\sim 2.1$ at $z=2000$, which confirms this result.

Furthermore we can see that the influence of the hydrogen continuum opacity on the shape of the spectral distortion has become very drastic (Fig.~\ref{fig:Fnu_2000}, lower panels). The photon distribution is practically narrowed down to those photons appearing in the close vicinity of the two resonances. 
Practically none of the photons emitted in the $\ion{He}{i} \;\HeIlevel{2}{1}{P}{1}-\HeIlevel{1}{1}{S}{0}$ transition really reach the $\ion{He}{i} \;\HeIlevel{2}{3}{P}{1}-\HeIlevel{1}{1}{S}{0}$ intercombination line. 
%
%
This implies that the feedback correction to the $\ion{He}{i} \;\HeIlevel{2}{3}{P}{1}-\HeIlevel{1}{1}{S}{0}$ transition is already expected to be negligible.
In Sect.~\ref{sec:feedback_21P1_23P1} we will show that the $\ion{He}{i} \;\HeIlevel{2}{1}{P}{1}-\HeIlevel{1}{1}{S}{0}\rightarrow \ion{He}{i} \;\HeIlevel{2}{3}{P}{1}-\HeIlevel{1}{1}{S}{0}$ feedback practically stops at $z\lesssim 2000-2200$.

\begin{figure*}
\centering 
\includegraphics[width=0.95\columnwidth]
{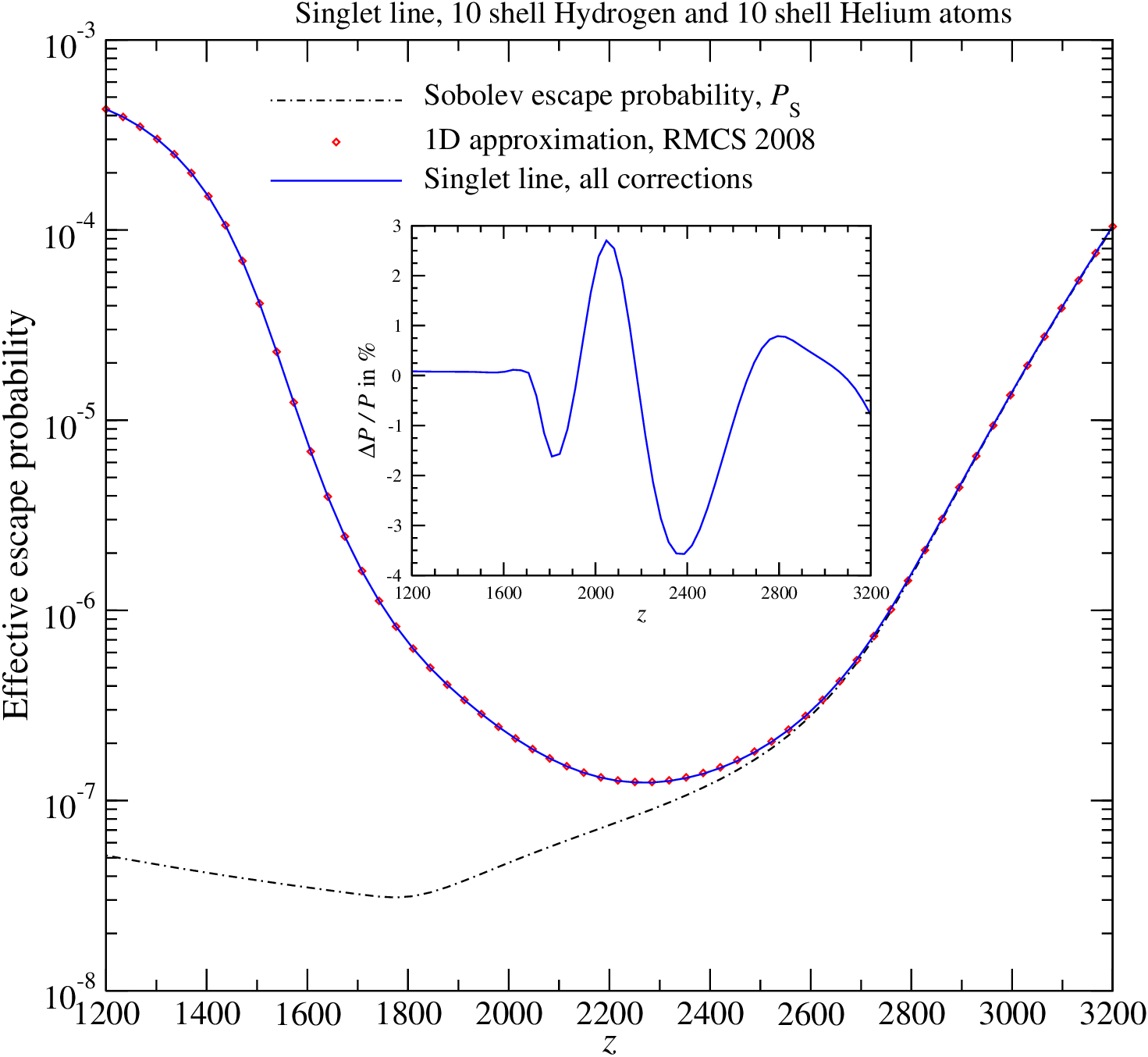}
\hspace{4mm}
\includegraphics[width=0.95\columnwidth]
{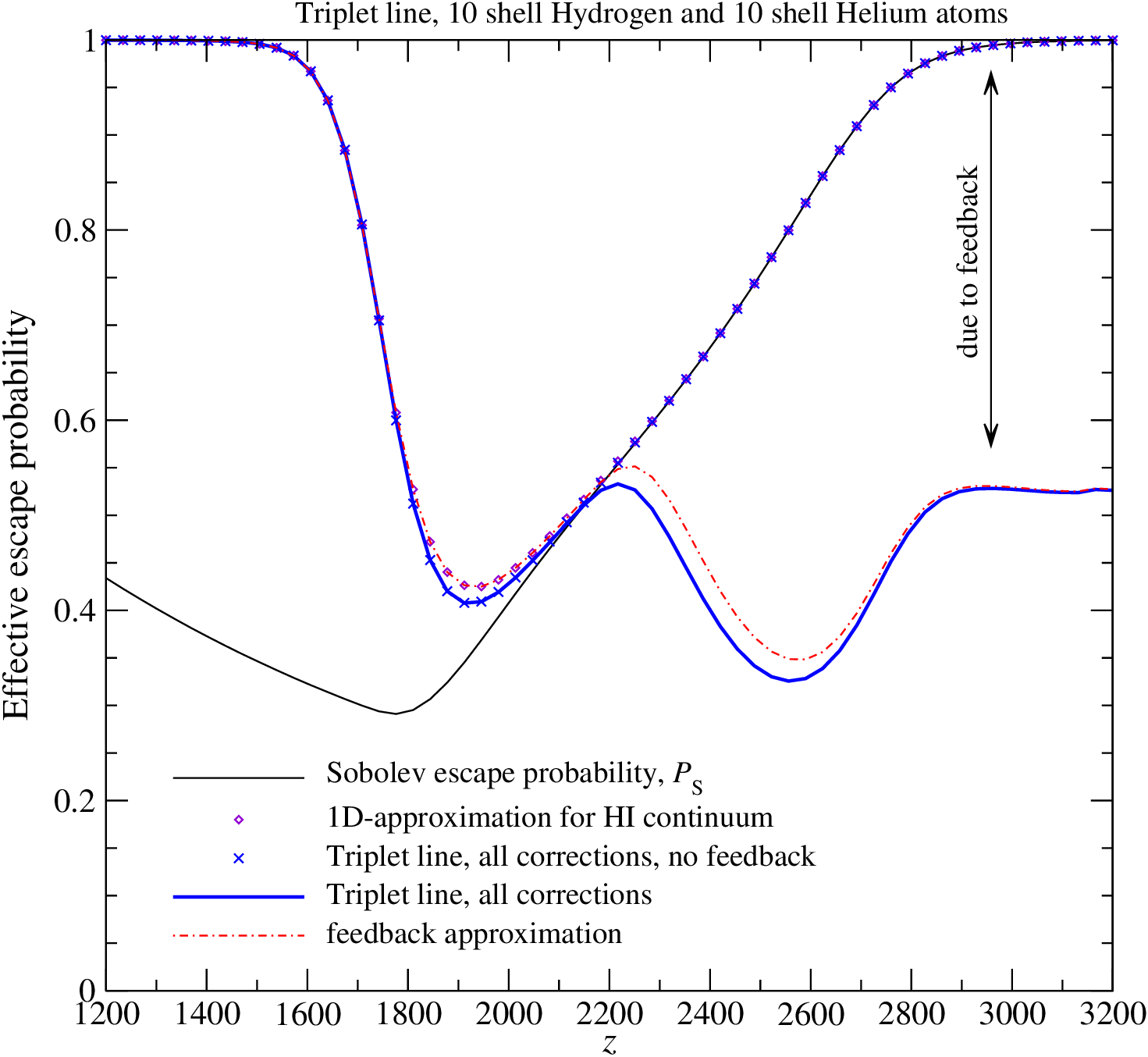}
\caption{Effective escape probabilities in the $\ion{He}{i}\;\HeIlevel{2}{1}{P}{1}-\HeIlevel{1}{1}{S}{0}$ (left panel) and $\ion{He}{i}\;\HeIlevel{2}{3}{P}{1}-\HeIlevel{1}{1}{S}{0}$ (right panel) resonance. We computed the effective escape probabilities using formula \eqref{eq:Peff} and the solution for the helium and hydrogen populations from our reference model. For comparison we also give the standard Sobolev escape probabilty and the result of different analytic approximations (see Sect.~\ref{sec:Pesc_corr_ana} for more details).
In the left panel we also show the relative difference of our numerical result with the 1D-approximation \citep{Jose2008} as inlay.
}
\label{fig:P.ST}
\end{figure*}

Given that photons supporting the flow of electrons to the $\ion{He}{i} \;\HeIlevel{2}{1}{P}{1}$ and $\ion{He}{i} \;\HeIlevel{2}{3}{P}{1}$ via the considered resonances are only present in a narrow range of frequencies around $\nu\sim \nu_{\rm a}$ and $\nu\sim \nu_{\rm b}$ it is also clear that time-dependent corrections and correction due to the thermodynamic factor $f^i_\nu$ cannot be very important.
This is in stark contrast to the Lyman $\alpha$ escape problem during hydrogen recombination, where a very large part of the total correction to the effective escape probability is caused by these processes \citep{Chluba2009}.  This is because during hydrogen recombination the \ion{H}{i} Lyman $\alpha$ spectral distortion is always very broad, so that even photons at $|\xD|\sim 10^3-10^4$ matter at the required level of precision.
However, during $\ion{He}{ii}\rightarrow\ion{He}{i}$ recombination these contributions turn out to be negligible (see Fig.~\ref{fig:P.ST} and Sect.~\ref{sec:Pesc_corr}).

In Fig.~\ref{fig:Fnu_2000} we also show the total spectral distortion at frequencies below the threshold of the \ion{H}{i} Lyman continuum. It is clear that only a very small amount of photons really reach below this frequency. In a complete treatment one should allow these additional photons to feedback on the transition in hydrogen during the pre-recombinational epoch. However, the total amount of photons created from this feedback is rather small in comparison to the emission produced from the absorption of helium photons before they pass the \ion{H}{i} Lyman continuum. 


\begin{figure*}
\centering 
\includegraphics[width=0.95\columnwidth]
{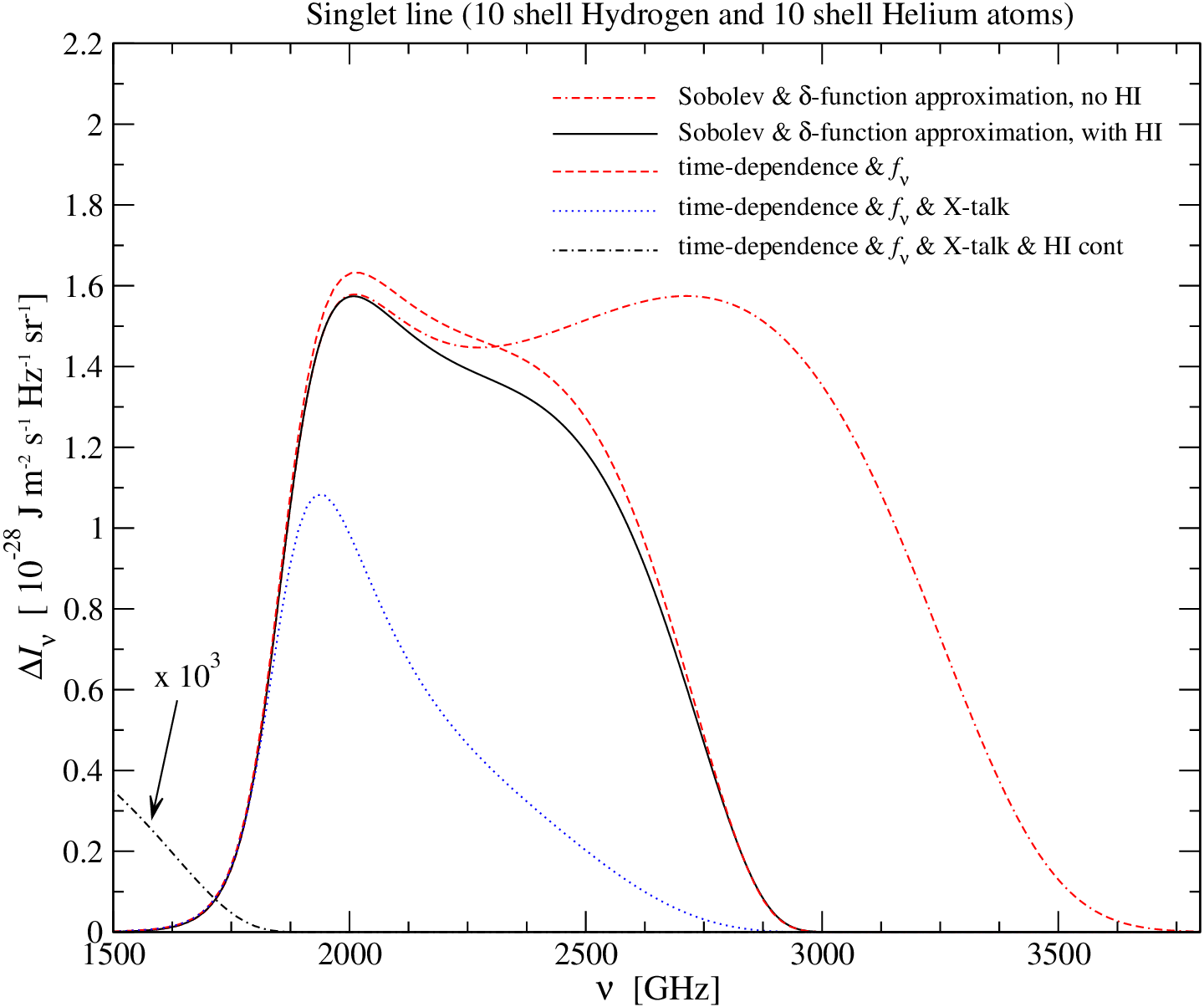}
\hspace{4mm}
\includegraphics[width=0.95\columnwidth]
{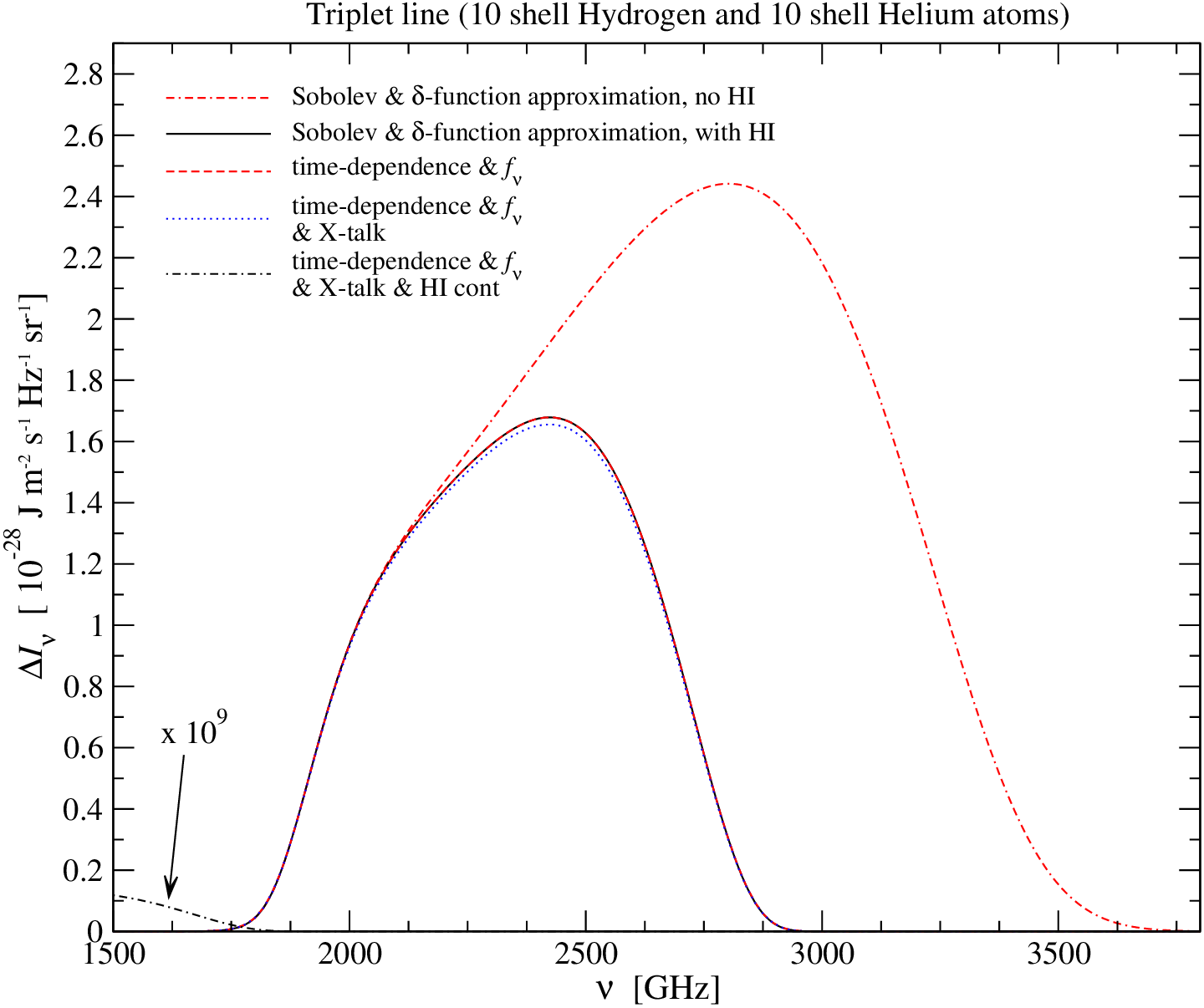}
\caption{CMB spectral distortion due to the $\ion{He}{i}\;\HeIlevel{2}{1}{P}{1}-\HeIlevel{1}{1}{S}{0}$ (left panel) and  $\ion{He}{i}\;\HeIlevel{2}{3}{P}{1}-\HeIlevel{1}{1}{S}{0}$ (right panel) transitions at $z=0$ for different cases (see Sect.~\ref{sec:DI_0} for details). In all computations we used the solution for the population from our reference model \citep{Jose2008}. However, for comparison we also considered the case without the speed-up of helium recombination caused by the hydrogen continuum opacity (dash-dash-dotted line).}
\label{fig:Fnu_0}
\end{figure*}
\subsection{CMB spectral distortion at redshift $z=0$}
\label{sec:DI_0}
In this Section we show the spectral distortion of the CMB caused by the emission and absorption in the $\ion{He}{i}\;\HeIlevel{2}{1}{P}{1}-\HeIlevel{1}{1}{S}{0}$ and  $\ion{He}{i}\;\HeIlevel{2}{3}{P}{1}-\HeIlevel{1}{1}{S}{0}$ lines as it would be visible to today for different assumptions on the considered processes.
However, it turns out that when including the absorption in the \ion{H}{i} Lyman continuum practically {\it all} the photons released by these transitions are erased before they can reach below the threshold frequency of the \ion{H}{i} Lyman continuum.
As we will explain in more detail below (Sect.~\ref{sec:pre-rec}), all these absorbed photons from helium lead to additional emission \changeII{by} hydrogen during its pre-recombinational epoch, which was neglected \changeII{in previous computations \citep{Jose2008}}.

In Figure~\ref{fig:Fnu_0} we give the present-day ($z=0$) CMB spectral distortion $\Delta I_\nu$ caused by the $\ion{He}{i}\;\HeIlevel{2}{1}{P}{1}-\HeIlevel{1}{1}{S}{0}$ (left panel) and  $\ion{He}{i}\;\HeIlevel{2}{3}{P}{1}-\HeIlevel{1}{1}{S}{0}$ (right panel) transition. In the computations we used the results for the populations from our reference model \citep{Jose2008} including 10 shells in hydrogen and 10 shells in helium.
The dash-dash-dotted lines give the distortion computed in the normal $\delta$-function approximation for the line-profile, when neglecting the speed-up of the helium recombination dynamics by the presence of neutral hydrogen.
When also including this process we obtain the solid curves, which were already presented in \citet{Jose2008}.
\changeI{One can clearly see how due to the speed-up of \ion{He}{i} recombination caused by \ion{H}{i} absorption in the latter case the high frequency spectral distortion from the considered lines narrows significantly in comparison to the former case \citep[compare][]{Jose2008}.}

If we now compute the distortion including the time-dependence of the emission process and the thermodynamic correction factor, we obtain the dashed curves. For the $\ion{He}{i}\;\HeIlevel{2}{1}{P}{1}-\HeIlevel{1}{1}{S}{0}$ resonance this leads to an increase of the spectral distortion in the frequency range $1800\,\text{GHz}\lesssim \nu \lesssim 3200\,\text{GHz}$ by about $10\%$, while the $\ion{He}{i}\;\HeIlevel{2}{3}{P}{1}-\HeIlevel{1}{1}{S}{0}$ line is hardly affected.
Like in the case of the \ion{H}{i} Lyman $\alpha$ line the wing contributions to the escaping photon distribution for the $\ion{He}{i}\;\HeIlevel{2}{1}{P}{1}-\HeIlevel{1}{1}{S}{0}$ are important, since in the  $\ion{He}{i}\;\HeIlevel{2}{1}{P}{1}-\HeIlevel{1}{1}{S}{0}$ Doppler core the \changeIII{effective} absorption optical depth is extremely large, \changeII{so that hardly any photon emitted there can survive \citep{Chluba2009b}}. Therefore in particular the scaling of the thermodynamic correction factor at frequencies below the line center ($f^{\rm a}_\nu\lesssim 1$) leads to additional leakage of photons, increasing the effective escape probability. This is completely analog to the case of the  \ion{H}{i} Lyman $\alpha$ line \citep{Chluba2008b, Chluba2009}.
In contrast to this, the wing contributions in the case of the $\ion{He}{i}\;\HeIlevel{2}{3}{P}{1}-\HeIlevel{1}{1}{S}{0}$ resonance are not very important, since the $\ion{He}{i}\;\HeIlevel{2}{3}{P}{1}-\HeIlevel{1}{1}{S}{0}$ Doppler core only becomes mildly optically thick \changeI{to absorption} during helium recombination (cf. Fig.~\ref{fig:tau_QT}). Therefore neither time-dependence nor $f^{\rm b}_\nu\neq 1$ can affect the escaping number of photons and hence the amplitude of the CMB spectral distortion very much.

If we now in addition allow for cross-talk between the lines, we can see that the feedback absorption during the passage of $\ion{He}{i}\;\HeIlevel{2}{1}{P}{1}-\HeIlevel{1}{1}{S}{0}$ photons through the $\ion{He}{i}\;\HeIlevel{2}{3}{P}{1}-\HeIlevel{1}{1}{S}{0}$ resonance, as expected, leads to an $e^{-\tau^{\rm b}_{\rm d}}$ suppression of the spectral distortion, which becomes more important toward higher frequencies, i.e. lower redshift of photon emission.
In the case of the $\ion{He}{i}\;\HeIlevel{2}{3}{P}{1}-\HeIlevel{1}{1}{S}{0}$ line one can also see a very small modification in the amplitude of the total distortion. This is due to the small amount of re-absorption of photons in the \changeII{distant} red damping wing of the $\ion{He}{i}\;\HeIlevel{2}{1}{P}{1}-\HeIlevel{1}{1}{S}{0}$ resonance, however this modification is negligible.
Note that here we have not yet included the change in the dynamics of helium recombination caused by the feedback of $\ion{He}{i}\;\HeIlevel{2}{1}{P}{1}-\HeIlevel{1}{1}{S}{0}$ photons in the $\ion{He}{i}\;\HeIlevel{2}{3}{P}{1}-\HeIlevel{1}{1}{S}{0}$ line, \changeI{but the final correction is small, so that the correction to correction can be neglected}. We will discuss this case in more detail below (Sect.~\ref{sec:feedback_21P1_23P1}).

If we finally also include the \ion{H}{i} continuum opacity in the computation of the spectral distortion, we can see that basically {\it all} photons disappear. Only a very small number of photons emitted a very early times during helium recombination can escape. This additional huge reduction of the spectral distortion due to \ion{H}{i} absorption was not yet taken into account, and it leads to a pre-recombinational feedback to hydrogen, which induces the emission of several additional \ion{H}{i} photons during helium recombination, as we will show in more detail below (Sect.~\ref{sec:pre-rec}).

\section{Corrections to the net rates and escape probabilities: numerical results}
\label{sec:Pesc_corr}
We now want to include the corrections in the solutions of the radiative transfer problem into the computations of the helium and hydrogen recombination history. 
Knowing the solution, $\Delta N_{\nu}(z)$, of the CMB spectral distortion as a function of time, and assuming that the differences in the net rates of the lines  introduced by the different corrections are small, one can incorporate the effect of these processes into the multi-level code by modifying the escape probabilities for the \ion{He}{i} resonances and the \ion{H}{i} continuum\footnote{This is only one possible approach, which is equivalent to directly including the relative changes in the net rates of the lines.}.
Here we will give the results of our numerical computations using the solutions of the photon transfer equation given in Sect.~\ref{sec:TransEq}. In Sect.~\ref{sec:Pesc_corr_ana} we will give simple analytic expressions which then allow to incorporate these corrections into the multi-level recombination code with sufficient precision.

\subsection{Corrections to the escape probabilities of the main helium resonances}
\label{sec:Pesc_corr_HeI}
If we start with the helium resonances \changeII{and neglect two-photon corrections as above} then it is clear that the net change in the population of level $i$ by transitions to the ground state can be cast into the form \citep[cf.][]{Chluba2009}
\beal
\label{eq:dNi_dt}
\left.\Abl{N_{i}}{t}\right|_{\rm res, \it i}&\equiv
- \int\frac{1}{c}\left.\Abl{N_{\nu}}{t}\right|_{\rm res, \it i}\id\nu \id\Omega
\nonumber
\\
&\!\!\!\!\stackrel{\stackrel{\rm Eq.~\eqref{eq:real_em_abs_simp_a}}{\downarrow}}{=}
-p^{i}_{\rm d}\,h\nu_iB_{\rm 1s\it i}\,N^{\ion{He}{i}}_{\rm 1s}
\left\{N^{i}_{\rm em}-N^{\rm pl}_{\nu_i}\right\}\times P^{i}_{\rm eff}.
\end{align}
Here the effective escape probability is defined as
\bsub
\label{eq:Peff}
\beal
\label{eq:Peff_a}
P^{i}_{\rm eff}(z)&=\int_0^\infty \varphi^i_{\rm V}(\nu, z)\left[1-f_\nu^i(z)\,G^i_{\nu}(z)\right]\id \nu
\\
\label{eq:Peff_b}
G^i_{\nu}(z)&=\frac{\Delta N_\nu(z)}{\Delta N^{i}_{\rm em}(z)},
\end{align}
\esub
where $\Delta N_\nu$ is the overall spectral distortion of the CMB at redshift $z$ introduced by {\it all} the considered line emission and absorption processes. \changeI{Furthermore, we have} $\Delta N^{i}_{\rm em}(z)=N^{i}_{\rm em}(z)-N^{\rm pl}_{\nu_i}(z)$, and $\varphi^i_{\rm V}(\nu, z)=\phi^i_{\rm V}(\nu, z)/\Delta\nu^i_{\rm D}$. 
Note that $P^{i}_{\rm eff}(z)$ can be directly compared to the standard Sobolev escape probability $P^{i}_{\rm S}$ using the relation\footnote{\changeII{This relation is obtained using the quasi-stationary solution for the population of the initial level.}} $P^{i}_{\rm eff,S}(z)=p^{i}_{\rm d}\,P^{i}_{\rm eff}(z)/[1-p^{i}_{\rm sc}\,P^{i}_{\rm eff}(z)]$ with $p^{i}_{\rm sc}=1-p^{i}_{\rm d}$ \citep[e.g. see][]{Chluba2008b}.
%
Below we will give the results for $P^{i}_{\rm eff}(z)$ when including different combinations of the considered \changeI{modifications.}
Note that for most practical purposes one can use $P^{i}_{\rm eff,S}(z)\approx p^{i}_{\rm d}\,P^{i}_{\rm eff}(z)$, since usually $p^{i}_{\rm sc}\,P^{i}_{\rm eff}(z)\ll 1$.

\subsubsection{Numerical results for the $\ion{He}{i}\;\HeIlevel{2}{1}{P}{1}-\HeIlevel{1}{1}{S}{0}$ resonance}
\label{sec:Pesc_corr_HeI_num_res_a}
First we consider the escape of photons from the $\ion{He}{i}\;\HeIlevel{2}{1}{P}{1}-\HeIlevel{1}{1}{S}{0}$ resonance. From the results of Sect.~\ref{sec:sol_F} it is already clear that one does not expect very large corrections due to time-dependent aspects of the problem or the thermodynamic corrections factor. This is because the \ion{H}{i} continuum opacity (even at rather early stages of helium recombination) has such a large impact on the shape of the spectral distortion (cf. Fig.~\ref{fig:Fnu_2400}).
In addition it is clear that the cross-talk with the $\ion{He}{i}\;\HeIlevel{2}{3}{P}{1}-\HeIlevel{1}{1}{S}{0}$ resonance cannot be very important, since in comparison with the line center the \changeI{$\ion{He}{i}\;\HeIlevel{2}{1}{P}{1}-\HeIlevel{1}{1}{S}{0}$} absorption profile drops by several orders of magnitude until $\xD\sim -700$, \changeI{and since there is hardly any photons from the $\ion{He}{i}\;\HeIlevel{2}{3}{P}{1}-\HeIlevel{1}{1}{S}{0}$ line close to the $\ion{He}{i}\;\HeIlevel{2}{1}{P}{1}-\HeIlevel{1}{1}{S}{0}$ resonance}. 
The feedback of photons emitted by higher transitions in helium on the blue side of the $\ion{He}{i}\;\HeIlevel{2}{1}{P}{1}-\HeIlevel{1}{1}{S}{0}$ resonance will be much more important. We will account for this effect in an approximate way in Sect.~\ref{sec:feedback_higher}.

In Fig.~\ref{fig:P.ST} we present the numerical results of our computation. In the case of the $\ion{He}{i}\;\HeIlevel{2}{1}{P}{1}-\HeIlevel{1}{1}{S}{0}$ the relative difference to the 1D-integral approximation for the modified escape probability when including the \ion{H}{i} continuum opacity \citep{Jose2008} is only of the order of a few percent, where part of the correction is due to the 1D-integral approximation itself. 
For the dynamics of helium recombination the impact of such correction is negligible. Here we will not consider them any further, \changeI{however, there is no principle difficulty to take them into account}.

\subsubsection{Numerical results for the $\ion{He}{i}\;\HeIlevel{2}{3}{P}{1}-\HeIlevel{1}{1}{S}{0}$ resonance}
\label{sec:Pesc_corr_HeI_num_res_b}
As mentioned above (Sect.~\ref{sec:sol_F}) for the $\ion{He}{i}\;\HeIlevel{2}{3}{P}{1}-\HeIlevel{1}{1}{S}{0}$ transition corrections due to time-dependence and the thermodynamic correction factor are not important.
However, the feedback of photons from the $\ion{He}{i}\;\HeIlevel{2}{1}{P}{1}-\HeIlevel{1}{1}{S}{0}$ resonance should have a strong impact, in particular at early stages of helium recombination, when the \ion{H}{i} continuum opacity between the two resonances is not yet becoming too large, \changeII{so that most of the $\ion{He}{i}\;\HeIlevel{2}{1}{P}{1}-\HeIlevel{1}{1}{S}{0}$ photons do reach the $\ion{He}{i}\;\HeIlevel{2}{3}{P}{1}-\HeIlevel{1}{1}{S}{0}$ line center}.

In Fig.~\ref{fig:P.ST} we present the numerical results of our computation for the escape probability in the $\ion{He}{i}\;\HeIlevel{2}{3}{P}{1}-\HeIlevel{1}{1}{S}{0}$.
One can see that when including the effect of the hydrogen continuum opacity but neglecting the cross-talk between the lines, the escape probability closely follows the curve computed with the 1D-integral approximation given earlier  \citep{Jose2008}.
There is some 1\% - 4\% difference between the analytic approximation (diamonds) and the numerical result for this case (crosses) at redshift $z\sim 1800-2200$, which could be avoided when using the full 2D-integral expression given by \citet{Jose2008} (Eq.~B.1 in their paper), but again such precision is not required for the recombination dynamics of helium.

When also taking the cross-talk between the lines into account, the effective escape probability is strongly reduced at $z\gtrsim 2200$. As we will see below, this is mainly due to the feedback of $\ion{He}{i}\;\HeIlevel{2}{1}{P}{1}-\HeIlevel{1}{1}{S}{0}$ photons on the $\ion{He}{i}\;\HeIlevel{2}{3}{P}{1}-\HeIlevel{1}{1}{S}{0}$ transition.
Also one can see, that at $z\lesssim 2200$ the effective escape probability again follows the case of no feedback.
As we will explain in Sect.~\ref{sec:feedback_21P1_23P1} this is due to the fact that because of the strong absorption of $\ion{He}{i}\;\HeIlevel{2}{1}{P}{1}-\HeIlevel{1}{1}{S}{0}$ photons in the \ion{H}{i} continuum the feedback is simply stopped.
For comparison we also show the simple analytic approximation, Eq.~\eqref{eq:feedback_approx}, which includes the effect of feedback from the $\ion{He}{i}\;\HeIlevel{2}{1}{P}{1}-\HeIlevel{1}{1}{S}{0}$ line and the absorption of photons between the line (dash-dotted curve).
Again the agreement is sufficient at the desired level of accuracy. For a derivation of this approximation we refer the reader to Sect.~\ref{sec:Pesc_corr_ana}.

\subsection{Escape and feedback in the \ion{H}{i} Lyman continuum}
\label{sec:Pesc_corr_HI}
If we now consider the escape problem in the \ion{H}{i} continuum, then we can define the net rate connecting the 1s-state with the continuum as (see Appendix~\ref{app:cont} for derivation)
\beal
\label{eq:dNi_dt_con}
\left.\Abl{N_{\rm 1s}}{t}\right|_{\rm cont}&\equiv
- \int\frac{1}{c}\left.\Abl{N_{\nu}}{t}\right|^{\rm rec}_{\rm 1s}\id\nu \id\Omega
\nonumber
\\[1mm]
&=\left\{N^{\ion{H}{i}}_{\rm 1s}\,R^{\rm pl}_{\rm 1sc}-N_{\rm e}\,N_{\rm p}\,R^{\rm pl}_{\rm c1s}\right\}
\times P^{\rm c}_{\rm eff},
\end{align}
where $R^{\rm pl}_{\rm c1s}$ and $R^{\rm pl}_{\rm 1sc}$ are the photo-recombination and -ionization rates of the 1s-state in a blackbody ambient radiation field.
The continuum escape probability is also given by the expressions Eq.~\eqref{eq:Peff}, with appropriate replacements (see derivation in Appendix~\ref{app:cont}). In particular one has to replace $\varphi^i_{\rm V}(\nu, z)$ with
\beal
\label{eq:def_phi_c}
\varphi_{\rm c}(\nu, z)=\frac{4\pi N^{\rm pl}_{\nu_{\rm c}}}{R^{\rm pl}_{\rm 1sc}}\frac{\sigma_{{\rm 1s c}}(\nu)}{f_\nu^{\rm c}(z)}
\end{align}
Note that $\varphi^i_{\rm c}(\nu, z)$ is normalized like $\int_0^\infty \varphi^i_{\rm c}(\nu, z)\id\nu=1$.

With equation~\eqref{eq:Peff} it now in principle is possible to compute the escape probability for the \ion{H}{i} Lyman continuum including possible time-dependent corrections. However, it is already clear that the {\it direct escape} in the \ion{H}{i} Lyman continuum is very small, and that it in particular does not lead to any large correction to the recombination dynamics \citep[e.g. see][]{Chluba2007b}. 
For the direct escape of photons it is therefore sufficient to use the analytic expression \citep{Chluba2007b} $P^{\rm c}_{\rm eff}\approx1/[1+\tau^{\rm esc}_{\rm c}]$ with $\tau^{\rm esc}_{\rm c}=\frac{c\,\sigma_{\rm 1sc}(\nu_{\rm c})\,N^{\ion{H}{i}}_{\rm 1s}}{H}\,\frac{k\Tg}{h\nu_{\rm c}}$ for estimates in our computations.
A similar expression can be applied to include the direct escape of photons in the $\ion{He}{i}\;\HeIlevel{1}{1}{S}{1}$ continuum, but again this has a very small impact on the dynamics of helium recombination. In addition, for detailed computations one should simultaneously include the effect of the hydrogen continuum opacity into this problem \citep{Switzer2007I}. 

However, the feedback from photons that were emitted by the main resonances of helium still has to be taken into account and does lead to some pre-recombinational emission by hydrogen, as we will explain in more detail below (Sect.~\ref{sec:cont_feedback}).

\section{Corrections to the net rates and escape probabilities: analytic considerations}
\label{sec:Pesc_corr_ana}
%
In this Section we now discuss the analytic approximations that can be used to include the {\it feedback} \changeII{process} into the multi-level recombination code. As we already argued in Sect.~\ref{sec:Pesc_corr}, for the helium recombination problem time-dependent corrections are not so important. Here we therefore use the  the quasi-stationary approximation. If necessary it is straightforward to compute more refined cases using the analytic expression given above. However, for more detailed computations also other corrections, \changeII{(e.g. related to partial frequency redistribution and electron scattering)} will probably become more important (see discussion below).

We will consider to types of feedback, first the feedback between different resonance of neutral helium. This will lead to a delay of helium recombination, which due to the additional absorption in the \ion{H}{i} Lyman continuum is suppressed very strongly. 
This problem was already discussed earlier in \citet{Switzer2007I}, however our final correction to the ionization history seems to be smaller (see Fig.~\ref{fig:tNe}).

The second type feedback is the one on the \ion{H}{i} Lyman continuum which leads to additional pre-recombinational emission from hydrogen during the epoch of $\ion{He}{ii}\rightarrow \ion{He}{i}$ recombination. Part of this correction was already taken into account earlier by \citet{Jose2008}, but as we have seen  in Sect.~\ref{sec:DI_0} (e.g. Fig.~\ref{fig:Fnu_0}, \changeII{solid lines}) in their computations some photons still escaped until $z=0$. All these photons will still be re-absorbed in the \ion{H}{i} continuum\footnote{This was already pointed out by \citet{Jose2008}.} and lead to additional feedback, but as we show below (Sect.~\ref{sec:cont_feedback}), no net change in the ionization history.

\subsection{Net rates for the multi-level code}
\label{sec:net_rate}
For our multi-level code the net rates in the resonances are important. Here we are interested in those resonances leading to the ground state of hydrogen or helium.
The change in the population $N_i$ of level $i$ due to the transition $i\rightarrow \rm 1s$ is given by \citep[cf.][]{Chluba2008b}
\bsub
\label{eq:net_i}
\beal
\left. \frac{\id N_i}{\id t}\right|_{i\rightarrow \rm 1s}
&=A_{i\rm 1s}\frac{g_i\,N_{\rm 1s} }{g_{\rm 1s}} [ \bar{n}^{i}-n_{\rm L}^i]
\\
&=h\nu_i B_{{\rm 1s}i}\,N_{\rm 1s} [ \bar{N}^{i}-N_{\rm L}^i], 
\end{align}
\esub
where $A_{i\rm 1s}$ denotes the Einstein A-coefficient of the transition $i\rightarrow \rm 1s$, $\bar{n}^{i}=\frac{c^2}{2\nu_i^2}\bar{N}^{i}$ is the average occupation number of the photon field over the line \changeII{absorption} profile. Furthermore, we have introduced the line occupation number $n^{i}_{\rm L}=\frac{c^2}{2\nu_i^2}N^{i}_{\rm L}= 1/[\frac{g_i\,N_{\rm 1s}}{g_{\rm 1s} N_i}-1] \approx \frac{g_{\rm 1s} N_i}{g_i\,N_{\rm 1s}}$. The last approximation is possible when induced effects can be neglected, however, in particular during the pre-recombinational epochs of the considered atomic species one should use the full expression for $n^{i}_{\rm L}$. Note that for Eq.~\eqref{eq:net_i} it is still assumed that the effect of stimulated transitions can be captured using equilibrium values (i.e. $[1+n^{i}_{\rm L}]\approx [1+n^{\rm pl}_{\nu_i}]$).

In the Sobolev approximation for the isolated resonance one will have \citep[e.g. see][]{Chluba2008b}
\beal
\label{eq:Nbar_Sobolev}
\bar{N}^{i}_{\rm cr}=N^{i}_{\rm L}-\Delta N^{i}_{\rm L}\,P^i_{\rm S}, 
\end{align}
where $\Delta N^{i}_{\rm L}=N^{i}_{\rm L}-N^{\rm pl}_{\nu_i}$ and $P^i_{\rm S}=\frac{1-e^{-\tau^i_{\rm S}}}{\tau^i_{\rm S}}$ is the Sobolev escape probability.
On the other hand, in the no line scattering approximation one finds \citep[e.g. see][]{Chluba2008b}
\beal
\label{eq:Nbar_coh}
\bar{N}^{i}_{\rm coh}
=N^{i}_{\rm L}-\Delta N^{i}_{\rm L}\,\frac{p^i_{\rm d}\,P^i_{\rm d}}{1-p^i_{\rm em}\,P^i_{\rm d}}
\end{align}
where $p^i_{\rm em}=1-p^i_{\rm d}$ and $P^i_{\rm d}=\frac{1-e^{-\tau^i_{\rm d}}}{\tau^i_{\rm d}}$ with $\tau^i_{\rm d}=p^i_{\rm d}\,\tau^i_{\rm S}$. 
Note that in the limit $p^i_{\rm d}\rightarrow 1$ one has $\bar{N}^{i}_{\rm cr}\equiv\bar{N}^{i}_{\rm coh}$.
It is valid for the low probability $\ion{He}{i}\;\HeIlevel{n}{3}{P}{1}-\HeIlevel{1}{1}{S}{0}$ intercombination lines and $\ion{He}{i}\;\HeIlevel{n}{1}{D}{2}-\HeIlevel{1}{1}{S}{0}$ series during helium recombination.
Physically this limit is equivalent to the approximation of complete redistribution in the line for each scattering event. However, here the complete redistribution is achieved via transitions to higher levels rather than attributing it to a normal scattering (i.e. ${\rm 1s}\rightarrow{\rm 2p}\rightarrow{\rm 1s}$) event.

Also one should point out that for $\tau^i_{\rm S}\wedge \tau^i_{\rm d}\gg1$ one again finds $\bar{N}^{i}_{\rm cr}\approx \bar{N}^{i}_{\rm coh}$.
However, the optically thin limit is \changeII{reached} (slightly) earlier in the case of the no line scattering approximation. Nevertheless, in both cases one finds that the escape probability scales like $P\approx 1-\frac{1}{2}\,\tau^i_{\rm S}$ for $\tau^i_{\rm d}\wedge \tau^i_{\rm S}\ll 1$, but the scaling for intermediate cases can be rather different.

Below we now will give the solution for $\bar{N}^{i}$ including feedback and \ion{H}{i} continuum absorption in the no scattering approximation. The equations are applicable to both hydrogen and helium recombination, however during hydrogen recombination there is no continuum opacity that can affect the evolution of the photon distribution between the \ion{H}{i} resonances significantly.

\subsection{Feedback between isolated resonances}
\label{sec:Pesc_corr_HeI_QS_line_feed}
In the standard quasi-stationary approximation for an isolated resonance one has to set $\Theta^i_{\rm a}=1$ and $f_\nu^{i}=1$ in Eq.~\eqref{app:kin_abs_em_Sol_phys_asym_F}. 
This then yields $F_\nu^{i, 1}=1-e^{-\tau^i_{\rm abs}(\nu, z_{\rm s}, z)}$.
Furthermore, one should assume that $\tau^i_{\rm abs}(\nu, z', z)=\tau_{\rm d}^i[\chi^i_{\nu'}-\chi^i_\nu]$, with $\chi^i_\nu=\int_0^\nu \varphi^i(\nu')\id\nu'$ and $\tau_{\rm d}^i=p_{\rm d}^i\,\tau_{\rm S}^i$, where $\tau_{\rm S}^i$ is the standard Sobolev optical depth.
Inserting this into Eq.~\eqref{eq:Peff_a} for $z_{\rm s}\rightarrow\infty$ one then has $P^{\rm S, \it i}_{\rm eff}(z)=P^{i}_{\rm d}(z)=[1-e^{-\tau^i_{\rm d}}]/\tau^i_{\rm d}$ \citep[e.g. see also][]{Chluba2008b}.


%

In this approximation the spectral distortion on the red side of the resonance $i$ is given by 
\beal
\label{eq:DN_i_red}
\Delta N^i_{\nu<\nu_i}=\Delta N^i_{-}=\Delta N^{i}_{\rm em}(z)\left[1-e^{-\tau^i_{\rm d}}\right] , 
\end{align}
which is constant with frequency. 
Note that for the quadrupole and intercombination lines the factor $1-e^{-\tau^i_{\rm d}}\lesssim 1$ is important.

For Eq.~\eqref{eq:DN_i_red} it was assume that on the distant blue side ($\nu\gg \nu_i$) of the resonance $i$ the spectrum is given by the CMB blackbody and that the only distortion is created by the line itself. 
If we also allow for some distortion, $\Delta N^i_{+}$, on the very distant blue side of the line then we have
\beal
\label{eq:DN_i_red_prime}
\Delta N^i_{-}=\Delta N^{i}_{\rm em}(z)\left[1-e^{-\tau^i_{\rm d}}\right] + \Delta N^i_{+}\,e^{-\tau^i_{\rm d}}.
\end{align}
%
If one assumes that $\nu_i\gg\nu_{i-1}$ then under quasi-stationary conditions the approximate behavior of the spectral distortion in the vicinity of the next lower-lying resonance $j=i-1$ is given by
\beal
\label{eq:DN_nu_feedback}
\Delta N^{\rm Sf, \it j}_{\nu} = \Delta N^{j}_{\rm em}(z)[1-e^{-\tau^j_{\rm d}[1-\chi^j_\nu]}]
+\Delta N^{i}_{-}(z)\,e^{-\tau^j_{\rm d}[1-\chi^j_\nu]},
\end{align}
with the limiting cases 
\beal
\label{eq:DN_nu_feedback_limits}
\Delta N^{\rm Sf, \it j}_{\nu}=
\begin{cases}
\Delta N^{j}_{-}(z)+\Delta N^{i}_{-}(z)\,e^{-\tau^j_{\rm d}}
&\text{for $\nu<\nu_j$}
\\[1mm]
\Delta N^{i}_{-}(z)
&\text{for $\nu\gg \nu_j$}
\end{cases}
\end{align}
%
For this solution it was assumed that the cross-talk between the resonances is negligible, meaning that in the region around resonance $j$ the contribution of the opacity from $i$ can be omitted. 
Note also that in our formulation $\Delta N^{i}_{-}(z)\equiv \Delta N^{j}_{+}(z)$.

The expression \eqref{eq:DN_nu_feedback} then implies
\beal
\label{eq:F_nu_feedback}
G^{\rm Sf, \it j}_{\nu} = [1-e^{-\tau^j_{\rm d}[1-\chi^j_\nu]}]
+\frac{\Delta N^{i}_{-}(z)}{\Delta N^{j}_{\rm em}(z)}\,e^{-\tau^j_{\rm d}[1-\chi^j_\nu]}
\end{align}
and with Eq.~\eqref{eq:Peff} for $f^j_\nu=1$
\beal
\label{eq:P_feedback}
P^{\rm Sf,\it j}_{\rm eff} = P^j_{\rm d}\left[1-\frac{\Delta N^{i}_{-}(z)}{\Delta N^{j}_{\rm em}(z)}\right],
\end{align}
so that $\Delta \bar{N}^{j}(z)= \bar{N}^{j}(z)-N^{\rm pl}_{\nu_j}=\Delta N^{j}_{\rm em}(z)\times [1-P^{\rm Sf,\it j}_{\rm eff}]$.

To include this into our multi-level code, we want to replace $\Delta N^{j}_{\rm em}(z)$ with the expression related to $\Delta N^{j}_{\rm L}=N^{j}_{\rm L}-N^{\rm pl}_{\nu_j}$.
For this we assume that the solution for the population of the considered initial level $j$ is given by the quasi-stationary solution. This implies that \citep[see][]{Chluba2008b}
\beal
\label{eq:DNem}
\Delta N^{j}_{\rm em}(z)=\frac{\Delta N^{j}_{\rm L}(z)-p^j_{\rm em}\Delta \bar{N}^{j}(z)}{p^j_{\rm d}}.
\end{align}
Inserting this into $\Delta \bar{N}^{j}(z)=\Delta N^{j}_{\rm em}(z)\times [1-P^{\rm Sf,\it j}_{\rm eff}]$ and solving for $\Delta \bar{N}^{j}(z)$ one then finds
\bsub
\beal
\label{eq:DNbar}
\Delta \bar{N}^{j}(z)
&=\frac{1-P^j_{\rm d}}{1-p^j_{\rm em}\,P^j_{\rm d}}\,\Delta N^{j}_{\rm L}(z)
+\frac{p^j_{\rm d}\,P^j_{\rm d}}{1-p^j_{\rm em}\,P^j_{\rm d}}\,\Delta N^{i}_{-}(z)
\nonumber\\
&=\Delta N^{j}_{\rm L}(z)
-\frac{p^j_{\rm d}\,P^j_{\rm d}}{1-p^j_{\rm em}\,P^j_{\rm d}}\,
\left[\,\Delta N^{j}_{\rm L}(z)-\Delta N^{i}_{-}(z)\right]
\\
&=\Delta N^{j}_{\rm L}(z)[1-P^{\rm Sf,\it j}_{\rm eff, S}]
\\
P^{\rm Sf,\it j}_{\rm eff, S}&=\frac{p^j_{\rm d}\,P^j_{\rm d}}{1-p^j_{\rm em}\,P^j_{\rm d}}
\,
\left\{1-\frac{\Delta N^{i}_{-}(z)}{\Delta N^{j}_{\rm L}(z)}\right\}.
\end{align}
\esub
Here $P^{\rm Sf,\it j}_{\rm eff, S}$ is the effective Sobolev escape probability which can be directly used in the multi-level code and includes the effect of feedback.
However, $\Delta N^{i}_{-}(z)$ still depends on $\Delta N^{i}_{\rm em}(z)$, but with Eq.~\eqref{eq:DNbar} it is now also possible to obtain an explicit expression for $\Delta N^{j}_{\rm em}(z)$, which then also can be applied to compute $\Delta N^{i}_{\rm em}(z)$ knowing $\Delta N^{i+1}_{-}(z)$. 
We find
\beal
\label{eq:DNem_explicit}
\Delta N^{k}_{\rm em}(z)&=\Delta N^{k}_{\rm L}(z)
+\frac{p^k_{\rm em}\,P^k_{\rm d}}{1-p^k_{\rm em}\,P^k_{\rm d}}\,
\left[\,\Delta N^{k}_{\rm L}(z)-\Delta N^{k+1}_{-}(z)\right].
\end{align}
For $p^k_{\rm em}\,P^k_{\rm d}\ll 1$ the second term can be neglected. Note that this limit can be reached for $p^k_{\rm em}\ll 1$ and/or $P^k_{\rm d}\ll 1$.
However, when $P^k_{\rm d}\sim 1$ the latter term should be included.
In particular even without feedback ($\Delta N^{k+1}_{-}(z)=0$) one will have
\beal
\label{eq:DNem_ex_nofb}
\Delta N^{k}_{\rm em}(z)&=\frac{\Delta N^{k}_{\rm L}(z)}{1-p^k_{\rm em}\,P^k_{\rm d}}
\approx \begin{cases}
\Delta N^{k}_{\rm L}(z) &\text{for $p^k_{\rm em}\,P^k_{\rm d}\ll 1$}
\\[1mm]
\frac{1}{p^k_{\rm d}}\,\Delta N^{k}_{\rm L}(z) &\text{for $P^k_{\rm d}\rightarrow 1$}.
\end{cases}
\end{align}
This shows that for $p^k_{\rm d}< 1$ one will find $\Delta N^{k}_{\rm L}(z) < \Delta N^{k}_{\rm em}(z)$.
This fact is important in relation with the \ion{H}{i} Lyman series and the $\ion{He}{i}\;\HeIlevel{n}{1}{P}{1}-\HeIlevel{1}{1}{S}{0}$ series during the {\it pre-recombinational} epochs of the considered species.

\subsection{Delay between emission and feedback}
\label{sec:delay}
As a next step we also take into account that the feedback of photons from the resonance $i$ on $j$ occurs at slightly different redshifts. This implies that for $P^{{\rm Sf},j}_{\rm eff, S}$ at redshift $z$ one has to evaluate $\Delta N^{i}_{-}$  at $z_i=\frac{\nu_i}{\nu_j}\,[1+z]-1>z$. In addition one should take the change in the volume element into account so that $\frac{\Delta N^{i}_{-}(z)}{\Delta N^{j}_{\rm L}(z)}\rightarrow \frac{[1+z]^2}{[1+z_i]^2}\,\frac{\Delta N^{i}_{-}(z_i)}{\Delta N^{j}_{\rm L}(z)}=\frac{\nu_j^2}{\nu_i^2}\,\frac{\Delta N^{i}_{-}(z_i)}{\Delta N^{j}_{\rm L}(z)}=\frac{\Delta n^{i}_{-}(z_i)}{\Delta n^{j}_{\rm L}(z)}$, where $\Delta n^k=\frac{c^2 \Delta N^{k}}{2\nu_k^2}$ is the deviation of the photon occupation number from the one of the CMB blackbody.
Then one finally finds 
\bsub
\label{eq:feedback_approx}
\beal
P^{{\rm Sf},j}_{\rm eff,S}(z)
&=P^j_{\rm d,S}(z)\left\{1-\frac{\Delta n^{i}_{-}(z_i)}{\Delta n^{j}_{\rm L}(z)}
\right\}
\\
P^j_{\rm d,S}(z)&=\frac{p^j_{\rm d}\,P^j_{\rm d}}{1-p^j_{\rm em}\,P^j_{\rm d}}.
\end{align}
\esub
With this approximation also the feedback in the \ion{H}{i} Lyman-series was solved earlier \citep{Chluba2007b, Switzer2007I}.
Note that for the \ion{H}{i} Lyman-series one can neglect the factor $1-e^{-\tau^i_{\rm d}}$ since $\tau^i_{\rm d}\gg1$ at all times during hydrogen recombination. Also since $\tau^i_{\rm d}\gg 1$ for all important \ion{H}{i} Lyman-series transitions one will have $\Delta n^{i}_{-}(z)\approx \Delta n^{i}_{\rm em}(z)\approx\Delta n^{i}_{\rm L}(z)$ and $P^j_{\rm d,S}(z)\approx P^j_{\rm S}(z)$, so that $P^{{\rm Sf},j}_{\rm eff,S}(z)\approx P^j_{\rm S}(z)[1-\Delta n^{i}_{\rm L}(z_i)/\Delta n^{j}_{\rm L}(z)]$.

Because during the recombination of hydrogen the whole Lyman series is extremely optically thick, all the photons released in the transition $i$ will be reprocessed in the resonance $j$ with $\nu_j<\nu_i$. In the case of the $\ion{He}{i}\;\HeIlevel{n}{3}{P}{1}-\HeIlevel{1}{1}{S}{0}$ intercombination lines and $\ion{He}{i}\;\HeIlevel{n}{1}{D}{2}-\HeIlevel{1}{1}{S}{0}$ series (see Sect.~\ref{sec:DNe_Ne}) this approximation is not always justified, since the \changeIII{effective} absorption optical depth for the $\ion{He}{i}\;\HeIlevel{n}{3}{P}{1}-\HeIlevel{1}{1}{S}{0}$ transitions with $n>2$ and at high redshift also for the $\ion{He}{i}\;\HeIlevel{n}{1}{D}{2}-\HeIlevel{1}{1}{S}{0}$ series does not exceed unity at any redshift (e.g. see Fig.~\ref{fig:tau_QT}).
Therefore the feedback will not be restricted to $i\rightarrow i-1$, but in some cases (a large) part of the distortion will also feedback like $i\rightarrow i-2$. 
Also during the pre-recombinational epochs of helium and hydrogen one will have to use the full expression Eq.~\eqref{eq:feedback_approx}, \changeII{since even some of the main resonances (e.g. the \ion{H}{i} Lyman series) can become optically thin}.

\subsection{$P^j_{\rm d,S}(z)$ and $\Delta n^{k}_{-}(z)$ for small $\tau_{\rm d}$}
\label{sec:DNminus_cases}
In the pre-recombinational epochs of hydrogen and helium, and during the end of recombination one can have lines with $\tau_{\rm d}\ll 1$ while $\tau_{\rm S}$ can in principle take all values. In particular for the \ion{H}{i} Lyman $\alpha$ line, because of the strong dependence of $p_{\rm d}$ on redshift, one encounters the  situation when $\tau_{\rm d}\ll 1$ and $\tau_{\rm S}\gg 1$.
For $\tau_{\rm d}\ll 1$ one can approximate $P_{\rm d}\approx 1-\frac{1}{2}\,\tau_{\rm d}$. Then one finds
\bsub
\label{eq:PdS_DNm_approx}
\beal
\label{eq:PdS_DNm_approx_a}
P^k_{\rm d,S}(z)&\stackrel{\tau^k_{\rm d}\ll 1}{\approx} 
\frac{1-\frac{\tau^k_{\rm d}}{2}}{1+\frac{p^k_{\rm em}}{2}\tau^k_{\rm S}}
\end{align}
and from Eq.~\eqref{eq:DNem_explicit} with Eq.~\eqref{eq:DN_i_red} including the delay between the emission and feedback redshift one has 
\beal
\label{eq:DNm_approx}
\Delta n^{k}_{-}(z)&\!\stackrel{\tau^k_{\rm d}\ll 1}{\approx} \!
\tau^k_{\rm S}\,\frac{1-\frac{\tau^k_{\rm d}}{2}}{1+\frac{p^k_{\rm em}}{2}\tau^k_{\rm S}}
\left\{
\Delta n^{k}_{\rm L}(z)\,
\!-p^k_{\rm em}
\Delta n^{k+1}_{-}(z_{k+1})
\right\}.
\end{align}
\esub
Here in addition the two limits for $p^k_{\rm em}\tau_{\rm S}\gg 1$ and $p^k_{\rm em}\tau_{\rm S}\ll 1$ exist, resulting in
\bsub
\label{eq:PdS_approx}
\beal
\label{eq:PdS_approx_a}
P^k_{\rm d,S}(z)&
\approx
\begin{cases}
1-\frac{1}{2}\tau^k_{\rm S} &\text{for $\tau^k_{\rm d} \ll 1$, $p^k_{\rm em}\tau^k_{\rm S}\ll 1$}
\\[2mm]
\frac{2}{p^k_{\rm em} \tau^k_{\rm S}}\left[1-\frac{\tau^k_{\rm d}}{2}\right]
&\text{for $\tau^k_{\rm d}\ll 1$, $p^k_{\rm em}\tau^k_{\rm S}\gg 1$}
\end{cases}
\end{align}
and
\beal
\label{eq:DNm_approx_II}
\Delta n^{k}_{-}(z)
&\approx
\left\{
\Delta n^{k}_{\rm L}(z)\,
\!-p^k_{\rm em}
\Delta n^{k+1}_{-}(z_{k+1})
\right\}
\nonumber\\
&\qquad\qquad
\times\begin{cases}
\tau^k_{\rm S}
&\text{for $\tau^k_{\rm d}\ll 1$, $p^k_{\rm em}\tau^k_{\rm S}\ll 1$}
\\[2mm]
\frac{2}{p^k_{\rm em}} \left[1-\frac{\tau^k_{\rm d}}{2}\right]
&\text{for $\tau^k_{\rm d}\ll 1$, $p^k_{\rm em}\tau^k_{\rm S}\gg 1$}.
\end{cases}
\end{align}
\esub
In particular we can see that in the optically thin limit ($\tau^k_{\rm d}\ll 1$ and $p^k_{\rm em}\tau^k_{\rm S}\ll 1$) the effective escape probability and  the occupation number on the red side of the resonance both approach the values expected from the Sobolev approximation.

\subsection{Feedback between isolated resonances in the presence of neutral hydrogen}
\label{sec:Pesc_corr_HeI_QS_line_feed_HI}
While \ion{He}{i} photons redshift from the resonance $i$ to the next lower-lying resonance $j=i-1$ there is also some absorption in the \ion{H}{i} Lyman continuum. In order to take the additional absorption into account we assume that between the resonances the photons only feel the \ion{H}{i} continuum opacity. 
Once the photons, starting at resonance $i$ with $\Delta n^i_{\nu_i}\approx \Delta n^{i}_{-}=\Delta n^{i}_{\rm em}(z_i)\left[1-e^{-\tau^i_{\rm d}(z_i)}\right]$, will have reached the resonance $j$ at redshift $z$, one will have $\Delta n^i_{\nu_j}\approx \Delta n^{i}_{-}(z_i)\,e^{-\tau^{\rm c}_{i \rightarrow j}}$, where we approximate the continuum optical depth by 
\beal
\label{eq:tau_LL_continuum}
\tau^{\rm c}_{i\rightarrow j}(z)
&\approx \frac{c\,\sigma_{\rm 1sc}(\nu_j)\,N^{\ion{H}{i}}_{\rm 1s}}{H}\,\frac{\nu_i-\nu_j}{\nu_j}.
\end{align}
More accurately one can use $\tau^{\rm c}_{i\rightarrow j}(z)=\tau^{\rm c}_{\rm abs}(\nu_j, z_i, z)$, where $\tau^{\rm c}_{\rm abs}(\nu_j, z_i, z)$ has to be computed using Eq.~\eqref{eq:continuum_channel_defs_b}.
This leads to
\beal
\label{eq:PS_feedback}
P^{{\rm Sfc},j}_{\rm eff, S}(z)
\approx P^j_{\rm d, S}(z)
\left[1-\frac{\Delta n^{i}_{\rm -}(z_i)}{\Delta n^{j}_{\rm L}(z)}\,e^{-\tau^{\rm c}_{i \rightarrow j}}\right].
\end{align}
A similar approximation was also given earlier by \citet{Switzer2007I}.  In the case of $\ion{He}{i}\;\HeIlevel{2}{1}{P}{1}-\HeIlevel{1}{1}{S}{0}\rightarrow \ion{He}{i}\;\HeIlevel{2}{3}{P}{1}-\HeIlevel{1}{1}{S}{0}$ feedback it works very well (cf. Fig.~\ref{fig:P.ST} and see Sect.~\ref{sec:feedback_21P1_23P1}).

\subsection{Feedback on the \ion{H}{i} Lyman continuum}
\label{sec:feedback_HILyc}
To include the feedback of \ion{He}{i} photons on the \ion{H}{i} Lyman continuum, we follow a very simple procedure. We assume that every resonance produces a distortion to the photon occupation number of $\Delta n^{i}_{\nu_i}(z')=\Delta n^{i}_{-}(z')$ at $\nu_i$. Due to redshifting this distortion then moves towards lower frequencies, so that at redshift $z$ and frequency $\nu$ one will have a distortion of $\Delta n^{i}_{\nu}(z)\approx \Delta n^{i}_{-}(z_i)$ with $z_i=\frac{\nu_i}{\nu}\,[1+z]-1\geq z$. 
If we now take into account that on their way some of these photons are also absorbed in the \ion{H}{i} Lyman continuum then we will find $\Delta n^{i}_{\nu}(z)\approx \Delta n^{i}_{-}(z_i)\, e^{-\tau^{\rm c}_{\rm abs}(\nu, z_i, z)}$, where $\tau^{\rm c}_{\rm abs}(\nu, z_i, z)$ has to be computed using Eq.~\eqref{eq:continuum_channel_defs_b}.
Then the correction to the ionization rate of the \ion{H}{i} 1s state caused by the distortion from the resonance $i$ is approximately given by
\beal
\label{eq:feedback_HI}
\Delta R^i_{\rm 1sc}&\approx
   4\pi \int _{\nu_{\rm c}}^{\nu_i} \sigma_{{\rm 1s c}}(\nu)\,
   \frac{2\nu^2}{c^2} \Delta n^i_{-} (z_i)\, e^{-\tau^{\rm c}_{\rm abs}(\nu, z_i, z)}\id\nu.
\end{align}
This integral can be computed for every line during the run of the multi-level recombination code. Note that whenever the photons from resonance $i$ are passing through a resonance $j<i$ then the distortion at $\nu\leq \nu_j$ is suppressed by $e^{-\tau_{\rm d}^j}$. For the $\ion{He}{i}\;\HeIlevel{n}{3}{P}{1}-\HeIlevel{1}{1}{S}{0}$ series and at high redshift also for the $\ion{He}{i}\;\HeIlevel{n}{1}{D}{2}-\HeIlevel{1}{1}{S}{0}$ series this strongly increases the distortion at frequencies below the line, a fact that has to be taken into account in the computations.

\begin{figure}
\centering 
\includegraphics[width=0.99\columnwidth]
{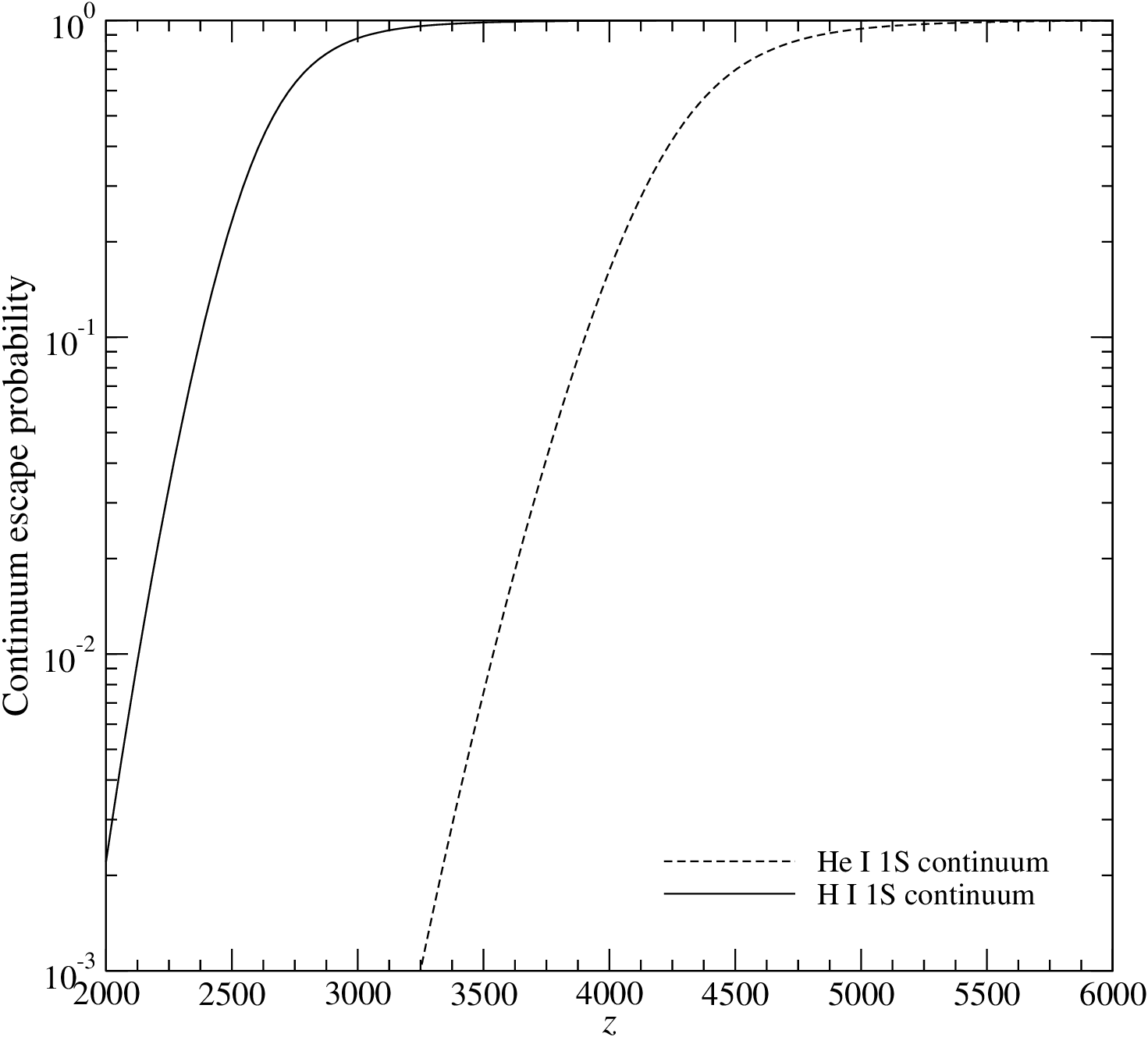}
\caption{Escape probability in the $\ion{He}{i}\;\HeIlevel{1}{1}{S}{0}$ and \ion{H}{i} Lyman continuum. \changeI{The curves were computed using the approximation for the escape probability given in Appendix~\ref{app:approxII}.}}
\label{fig:Pc.HI.HeI}
\end{figure}
\subsection{Feedback by photons escaping the \ion{H}{i} Lyman continuum}
\label{sec:escape_HIcont}
Even though the direct escape of photons in the \ion{H}{i} Lyman continuum is not important during the epoch of hydrogen recombination, in the pre-recombinational epoch of hydrogen ($z\gtrsim 1800$) the escape probability can be close to unity (cf. Fig.~\ref{fig:Pc.HI.HeI}). 
%
%
Due to the feedback of helium photons on the  \ion{H}{i} Lyman continuum one can therefore produce some amount of photons that directly escape below the \ion{H}{i} Lyman continuum frequency.
These photons then feedback in the forrest of \ion{H}{i} Lyman series lines, where for the uppermost transitions one could in principle model this process as a continuation of the \ion{H}{i} continuum cross section \citep[e.g. see Appendix II in][]{Beigman1968}. 
Here we approximately accounted for this process by simply adding the small distortion from the \ion{H}{i} Lyman continuum on the blue side of the uppermost feedback level that we included.
The distortion from the \ion{H}{i} Lyman continuum was computed using the net rate as given by Eq.~\eqref{eq:dNi_dt_con}, and assuming that the photons were emitted at the  \ion{H}{i} Lyman continuum threshold frequency.

In addition, some of the high frequency photons emitted in the pre-recombinational epoch of \ion{He}{i} induced by the feedback of \ion{He}{ii} photons on neutral helium can reach below the \ion{H}{i} Lyman continuum frequency.
We also included these photons in the computations by feeding them back on the blue side of the uppermost feedback level in hydrogen.

For the processes in connection with $\ion{He}{ii}\rightarrow \ion{He}{i}$ feedback we followed a similar approach. However, in this case the distribution of photons from $\ion{He}{iii}\rightarrow \ion{He}{ii}$ recombination can be computed before solving the problem (see Sect.~\ref{sec:HeIII_feedback} for details).


\begin{table}
\caption{First few main transitions to the ground-state of helium. 
The resonances have been ordered according to their transition frequency. In this way one can understand the sequence of feedback between the different lines. 
For this purpose we also give the $\Delta z/ z\sim \Delta \nu/\nu$ that is necessary to reach the next lower-lying resonance. For comparison due to the thermal motion of the atoms one has $\Delta\nu_{\rm D}/\nu\sim \pot{1.7}{-5}\left[\frac{1+z}{2200}\right]^{1/2}$.}
\label{tab:feedback}
\centering
\begin{tabular}{@{}ccccc}
\hline
\hline
Initial Level & Type & $A_{i \rm 1s}$ [$s^{-1}$] & $\nu_{i \rm 1s}$ [Hz] & $|\Delta z/z|$\\
\hline
$\HeIlevel{2}{3}{P}{1}$ & E1 (TS) & 177.58 &  $\pot{5.0691}{15}$ & -- \\
$\HeIlevel{2}{1}{P}{1}$ & E1 (SS) & $\pot{1.7989}{9}$ &  $\pot{5.1305}{15}$ & $\pot{1.2}{-2}$ \\
\hline
$\HeIlevel{3}{3}{P}{1}$ & E1 (TS) & $56.1$ &  $\pot{5.5631}{15}$ & $\pot{7.8}{-2}$ \\
$\HeIlevel{3}{1}{D}{2}$ & E2 (SS) & $\pot{1.298}{3}$ &  $\pot{5.5793}{15}$ & $\pot{2.9}{-3}$ \\
$\HeIlevel{3}{1}{P}{1}$ & E1 (SS) & $\pot{5.6634}{8}$ &  $\pot{5.5824}{15}$ & $\pot{5.6}{-4}$ \\
\hline
$\HeIlevel{4}{3}{P}{1}$ & E1 (TS) & $23.7$ &  $\pot{5.7325}{15}$ & $\pot{2.6}{-2}$ \\
$\HeIlevel{4}{1}{D}{2}$ & E2 (SS) & $748.5$ &  $\pot{5.7394}{15}$ & $\pot{1.2}{-3}$ \\
$\HeIlevel{4}{1}{P}{1}$ & E1 (SS) & $\pot{2.4356}{8}$ &  $\pot{5.7408}{15}$ & $\pot{2.4}{-4}$ \\
\hline
$\HeIlevel{5}{3}{P}{1}$ & E1 (TS) & $12.1$ &  $\pot{5.8100}{15}$ & $\pot{1.2}{-2}$ \\
$\HeIlevel{5}{1}{D}{2}$ & E2 (SS) & $431.4$ &  $\pot{5.8135}{15}$ & $\pot{6.0}{-4}$ \\
$\HeIlevel{5}{1}{P}{1}$ & E1 (SS) & $\pot{1.2582}{8}$ &  $\pot{5.8143}{15}$ & $\pot{1.4}{-4}$  \\
\hline
\hline
\end{tabular}
\end{table}

\begin{figure}
\centering 
\includegraphics[width=0.88\columnwidth]
{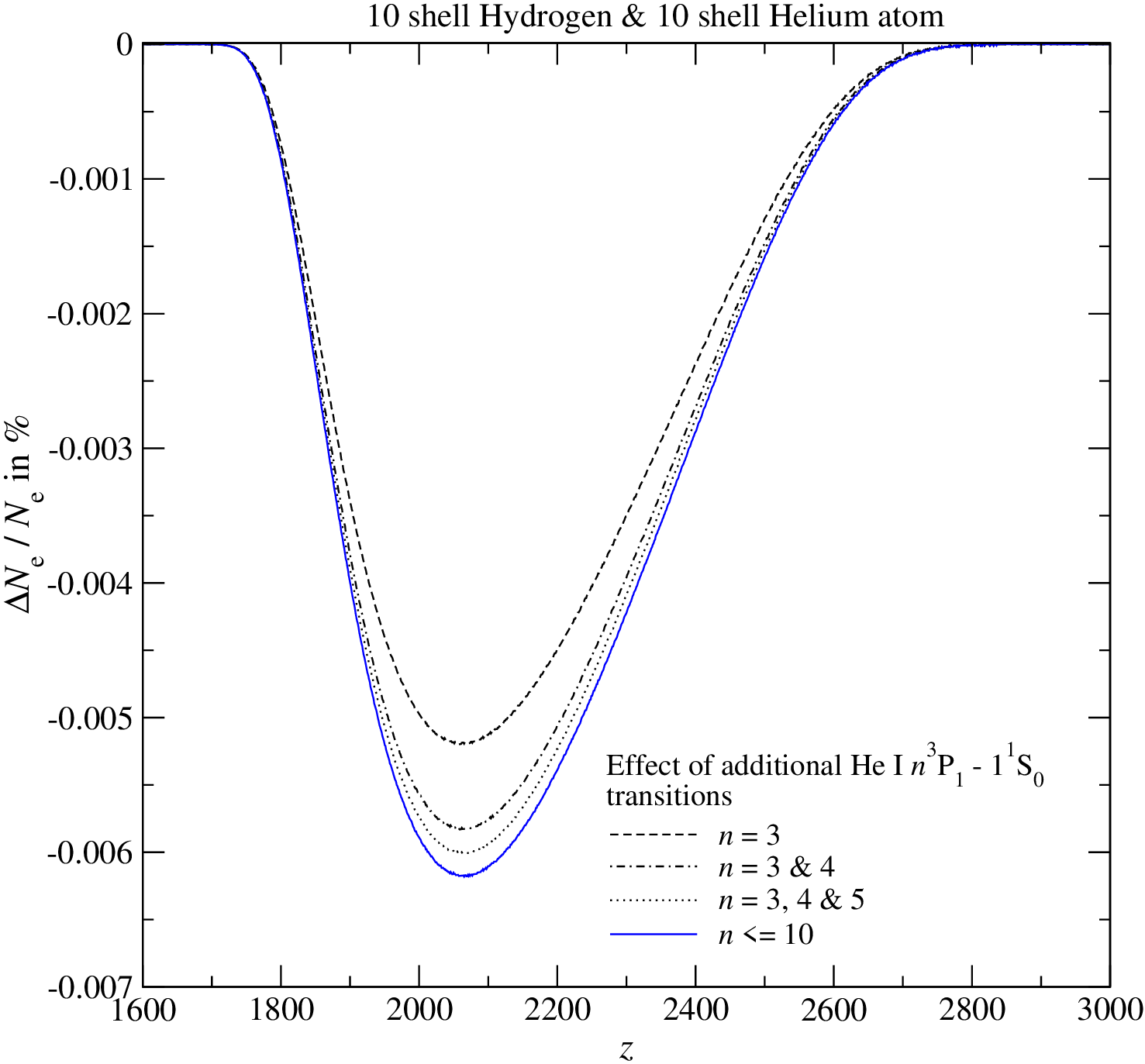}
\\
\includegraphics[width=0.88\columnwidth]
{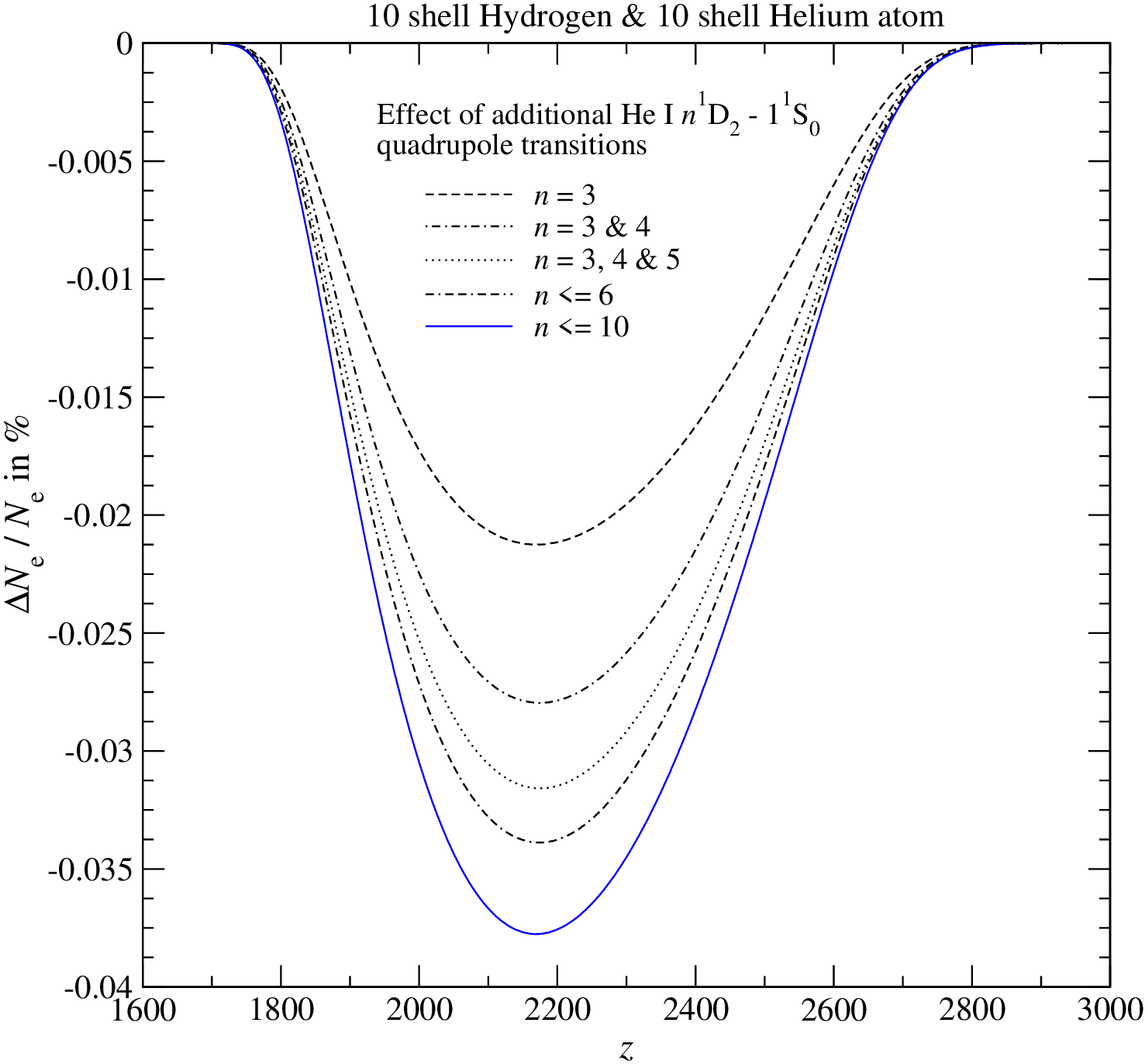}
\\
\includegraphics[width=0.88\columnwidth]
{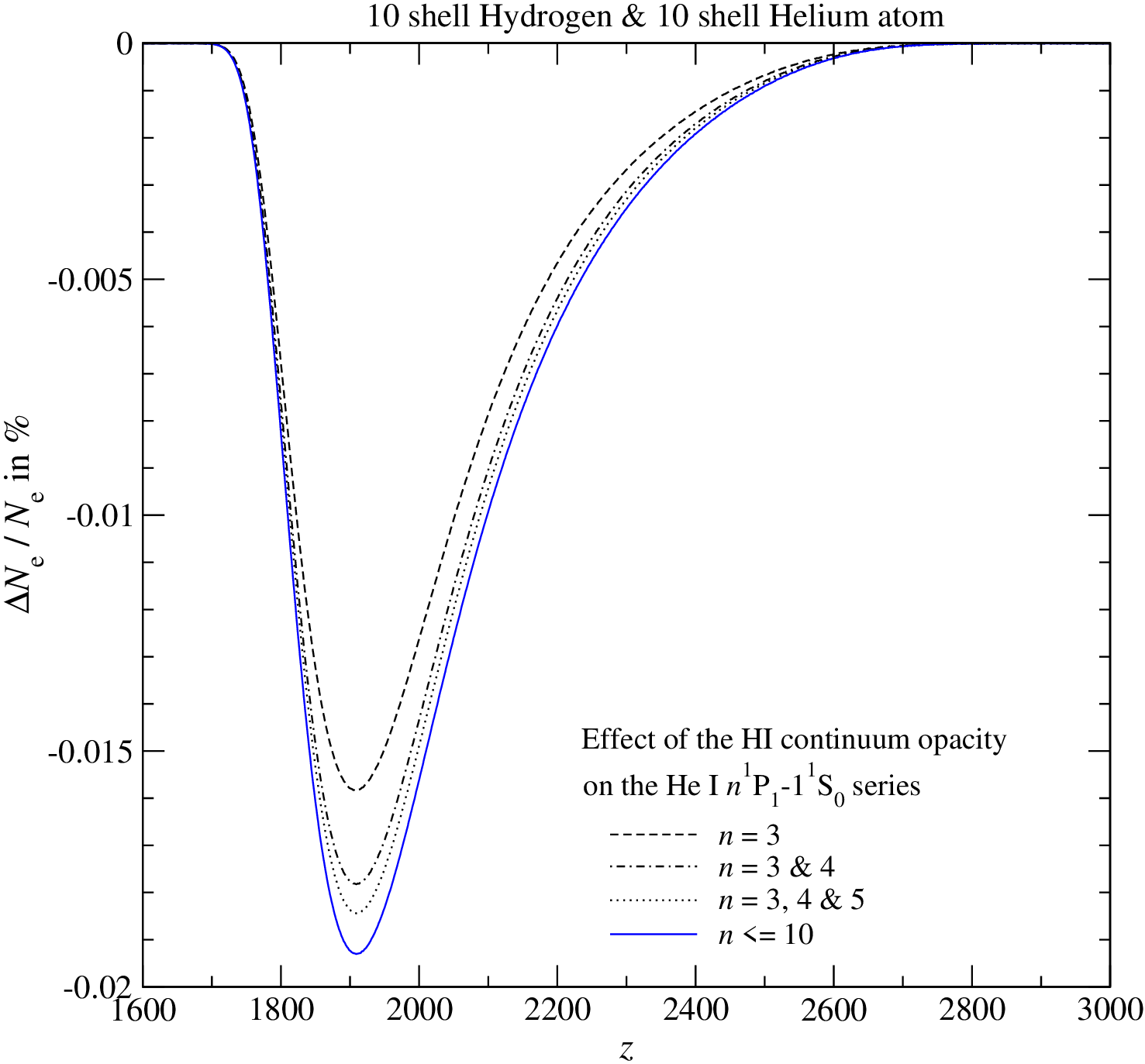}
\caption{Corrections to the free electron fraction during $\ion{He}{ii} \rightarrow \ion{He}{i}$ recombination: effect of additional $\ion{He}{i}\;\HeIlevel{n}{3}{P}{1}-\HeIlevel{1}{1}{S}{0}$ transitions (upper panel), $\ion{He}{i}\;\HeIlevel{n}{1}{D}{1}-\HeIlevel{1}{1}{S}{0}$ electric quadrupole transitions (middle panel), and the effect of the hydrogen continuum opacity on the $\ion{He}{i}\;\HeIlevel{n}{1}{P}{1}-\HeIlevel{1}{1}{S}{0}$ series.
We included 10 shells for hydrogen and 10 shells for helium in our computations, and compared the resulting electron fraction with our reference model \citep{Jose2008}.}
\label{fig:DN.TS}
\end{figure}

\section{Corrections to the ionization history during helium recombination}
\label{sec:DNe_Ne}
Below we will now look at the feedback of photons between the different resonances of helium and their effect on the ionization history. Our main results are given in Fig.~\ref{fig:DN_feed} and \ref{fig:tNe}. At low redshift our final correction seems to be smaller than the one presented in \citet{Switzer2007I}.
However, it is clear that the difference will not be very important for the analysis of future CMB data.

\subsection{Initial refinements of the recombination code}
In this Section we mainly want to discuss the effect of feedback of helium photons on the cosmological recombination history. However, for this we also need to take into account additional processes, which we neglected until now. 
First we also include $\ion{He}{i}\;\HeIlevel{n}{1}{D}{2}-\HeIlevel{1}{1}{S}{0}$ electric quadrupole transitions and $\ion{He}{i}\;\HeIlevel{n}{3}{P}{1}-\HeIlevel{1}{1}{S}{0}$ intercombination lines with $n>2$, which all have transition rates and escape probabilities that should lead to net rate which are comparable to those of the main transitions controlling the dynamics of helium recombination \citep{Switzer2007I}.
In Appendix~\ref{app:atom} we give some more details on the rates that we used. 
In Fig.~\ref{fig:DN.TS} we show the effect of these transitions on the recombination history.
The corrections that we find seem to be in good agreement with the results of \citet{Switzer2007I} for these processes, however the final corrections are negligible.

Then we also refine our modeling of the escape of photons from the $\ion{He}{i}\;\HeIlevel{n}{1}{P}{1}-\HeIlevel{1}{1}{S}{0}$ series with $n>2$. Like in the case of the $\ion{He}{i}\;\HeIlevel{2}{1}{P}{1}-\HeIlevel{1}{1}{S}{0}$ series and the $\ion{He}{i}\;\HeIlevel{2}{3}{P}{1}-\HeIlevel{1}{1}{S}{0}$ the presence of neutral hydrogen increases the escape probabilities of these lines towards the end of helium recombination. We take this process into account using the 1D-integral approximation given by \citet{Jose2008} (see Eq.~(B.3) in their paper) .
We show the effect of this process on the ionization history in Fig.~\ref{fig:DN.TS}. Again in good agreement with \citet{Switzer2007I} we find a negligible correction.
We do not include this speed-up for the escape of photons in the $\ion{He}{i}\;\HeIlevel{n}{1}{D}{2}-\HeIlevel{1}{1}{S}{0}$ and $\ion{He}{i}\;\HeIlevel{n}{3}{P}{1}-\HeIlevel{1}{1}{S}{0}$ series for $n>2$, since the effect of these lines on the dynamics of helium recombination is already very small. Additional refinements of the escape probabilities will not change this very much, since the speed-up in the main resonances will dominate.
The results of \citet{Switzer2007I} also support the correctness of this statement.

\subsection{Total amount of primary \ion{He}{i} photons which are available for feedback}
\label{sec:feed_num_HeI}
As shown by\footnote{It turns out that also with the refinements to our computations introduced in this paper the quoted numbers do not change very much.} \citet{Jose2008}, about 90\% of all helium atoms became neutral via the $\ion{He}{i} \;\HeIlevel{2}{1}{P}{1}-\HeIlevel{1}{1}{S}{0}$ and $\ion{He}{i} \;\HeIlevel{2}{3}{P}{1}-\HeIlevel{1}{1}{S}{0}$ dipole transitions to the ground state.
Direct transition to the ground state from initial levels with principle quantum number $n>2$ allowed about 2\% of the helium atoms to recombine, while the remaining $\sim 8\%$ of helium atoms became neutral via the $\ion{He}{i} \;\HeIlevel{2}{1}{S}{0}-\HeIlevel{1}{1}{S}{0}$ two-photon decay channel.
This implies that in total about $1.08\,\gamma$ per helium atom were emitted in these transition, where $\sim 1\,\gamma$ per helium atom was released above the \ion{H}{i} Lyman $\alpha$ resonance.

It is clear that the photons emitted in $\ion{He}{i} \;N^{2S+1}L_J-\HeIlevel{1}{1}{S}{0}$ transitions all have the possibility to feedback on the \ion{H}{i} Lyman continuum, while a large part of the $\ion{He}{i} \;\HeIlevel{2}{1}{S}{0}-\HeIlevel{1}{1}{S}{0}$ two-photon continuum will feed back only on the \ion{H}{i} Lyman series (see Sect.~\ref{sec:add_corr_HeI}).

\begin{figure}
\centering 
\includegraphics[width=0.99\columnwidth]
{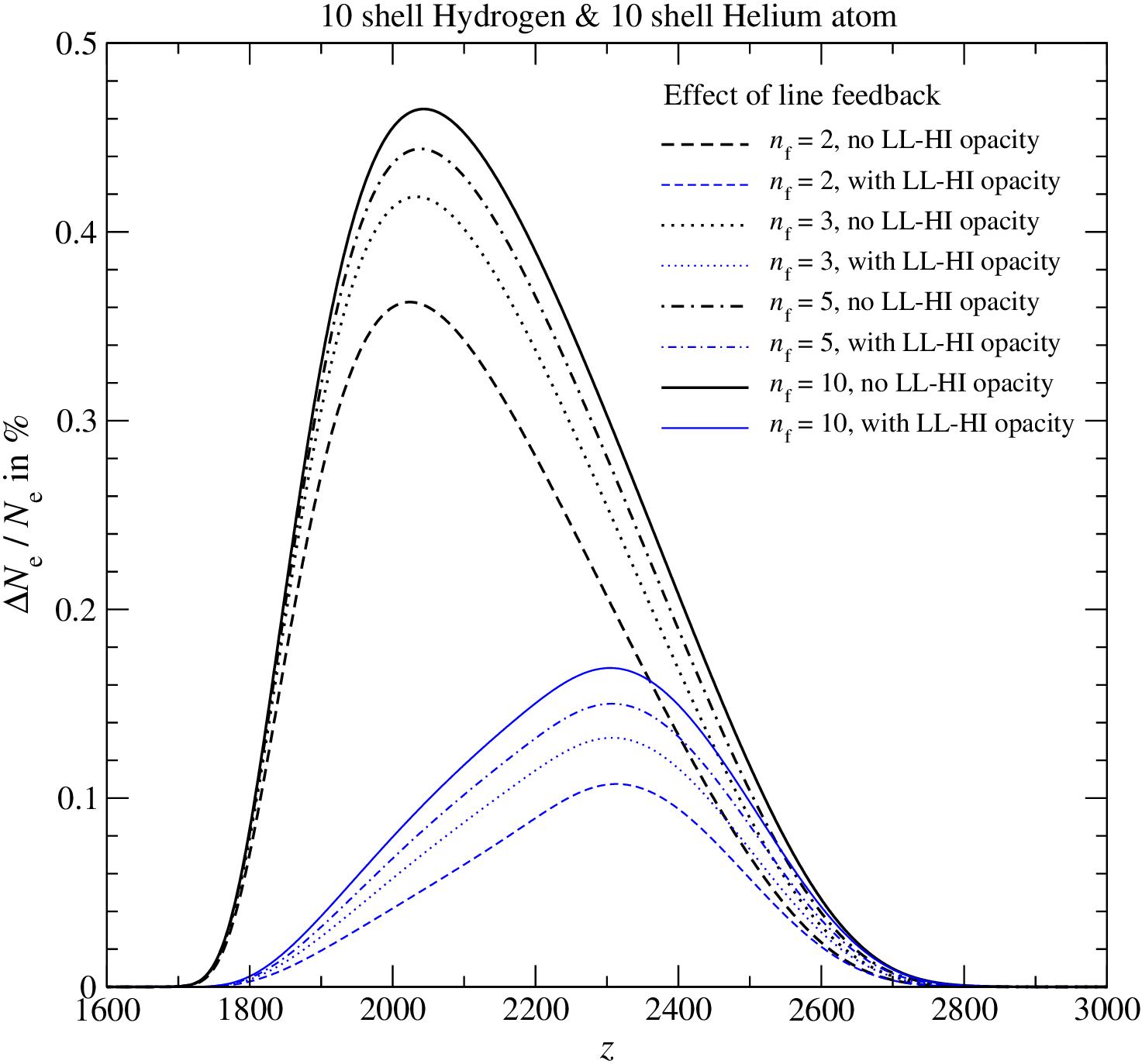}
\caption{Corrections to the free electron fraction during $\ion{He}{ii} \rightarrow \ion{He}{i}$ recombination: the effect of feedback among the resonances of neutral helium.
We show the results for two sets of computations, one that does not include the effect of photon absorption between the lines (labeled 'no LL-HI opacity') and those that do ('with LL-HI opacity').
We included 10 shells for hydrogen and 10 shells for helium in our computations, and compared the resulting electron fraction with our reference model \citep{Jose2008}.}
\label{fig:DN_feed}
\end{figure}

\subsection{$\ion{He}{i}\;\HeIlevel{2}{1}{P}{1}-\HeIlevel{1}{1}{S}{0}\rightarrow \ion{He}{i}\;\HeIlevel{2}{3}{P}{1}-\HeIlevel{1}{1}{S}{0}$ feedback}
\label{sec:feedback_21P1_23P1}
In Table~\ref{tab:feedback} we give the main resonances, that are important for the feedback problem.
%
Here we first consider the feedback of photons from the $\ion{He}{i}\;\HeIlevel{2}{1}{P}{1}-\HeIlevel{1}{1}{S}{0}$ line on the $\ion{He}{i}\;\HeIlevel{2}{3}{P}{1}-\HeIlevel{1}{1}{S}{0}$ resonance. These resonances are separated by $\Delta \nu/\nu\sim 1\%$ or $\sim 700$ Doppler width, so that one can certainly treat them as two isolated resonances.
For this rather large distance one should in principle compute the \ion{H}{i} continuum opacity between the lines with the full integral Eq.~\eqref{eq:continuum_channel_defs_b} rather than the approximation Eq.~
\eqref{eq:tau_LL_continuum}. However, for the accuracy required here we still use the simple approximation, since otherwise for the current version of our code the computations are slowed down significantly.

In Fig.~\ref{fig:DN_feed} we show the results of our numerical computations for this problem. If we neglect the absorption of photons in the \ion{H}{i} continuum while they travel from the $\ion{He}{i}\;\HeIlevel{2}{1}{P}{1}-\HeIlevel{1}{1}{S}{0}$ line towards the $\ion{He}{i}\;\HeIlevel{2}{3}{P}{1}-\HeIlevel{1}{1}{S}{0}$ resonance we obtain the thick dashed curve. This leads to a significant correction during helium recombination, which is largest at $z\sim 2020$ with $\Delta N_{\rm e}/N_{\rm e}\sim +0.36\%$.
However, when we include the effect of the continuum opacity (thin dashed line), a large part of the correction disappears, leaving only $\Delta N_{\rm e}/N_{\rm e}\sim +0.1\%$ at $z\sim 2300$. Looking at Fig.~\ref{fig:tauc} (solid line) this behavior is not surprising, since at $z\sim 2370$ the \ion{H}{i} Lyman continuum becomes optically thick between the two resonances. This stops $\ion{He}{i}\;\HeIlevel{2}{1}{P}{1}-\HeIlevel{1}{1}{S}{0}$ photons from reaching the $\ion{He}{i}\;\HeIlevel{2}{3}{P}{1}-\HeIlevel{1}{1}{S}{0}$ resonance. This effect also can be seen in Fig.~\ref{fig:P.ST}, where at $z\gtrsim 2600$ the reduction in the escape probability from the $\ion{He}{i}\;\HeIlevel{2}{3}{P}{1}-\HeIlevel{1}{1}{S}{0}$ line is only caused by the feedback of photons. At $z\sim 2600$ the effect of the hydrogen continuum opacity starts to set in and until $z\sim 2200$ none of the photons from the $\ion{He}{i}\;\HeIlevel{2}{1}{P}{1}-\HeIlevel{1}{1}{S}{0}$
ever reach the intercombination line.

\subsection{Feedback among higher level transitions}
\label{sec:feedback_higher}
In this Section we now include the effect of feedback from the higher level transitions. Here the interesting aspect is that for each shell $n>2$ one has a feedback sequence $\ion{He}{i}\;\HeIlevel{n}{1}{P}{1}-\HeIlevel{1}{1}{S}{0}\rightarrow\ion{He}{i}\;\HeIlevel{n}{1}{D}{2}-\HeIlevel{1}{1}{S}{0}\rightarrow\ion{He}{i}\;\HeIlevel{n}{3}{P}{1}-\HeIlevel{1}{1}{S}{0}$ where the separation between the lines is rather small.
For example, the $\ion{He}{i}\;\HeIlevel{3}{1}{P}{1}-\HeIlevel{1}{1}{S}{0}$ photons will feedback on the 
$\ion{He}{i}\;\HeIlevel{3}{1}{D}{2}-\HeIlevel{1}{1}{S}{0}$ transition after $\Delta z/z\sim \pot{5.6}{-4}$, since the quadrupole line at $z\sim 2200$ is only about $30$ Doppler width below the $\ion{He}{i}\;\HeIlevel{3}{1}{P}{1}-\HeIlevel{1}{1}{S}{0}$ resonance. Similarly the $\ion{He}{i}\;\HeIlevel{3}{3}{P}{1}-\HeIlevel{1}{1}{S}{0}$ resonance is only $170$ Doppler width away from the $\ion{He}{i}\;\HeIlevel{3}{1}{D}{2}-\HeIlevel{1}{1}{S}{0}$ quadrupole transition (see Table~\ref{tab:feedback} for more examples).
In particular for the $\ion{He}{i}\;\HeIlevel{n}{1}{P}{1}-\HeIlevel{1}{1}{S}{0}\rightarrow\ion{He}{i}\;\HeIlevel{n}{1}{D}{2}-\HeIlevel{1}{1}{S}{0}$ feedback it is questionable if it is really possible to treat the resonances completely independently, since the line broadening due to resonance and (to a smaller extent) electron scattering will likely be larger than the separation of these lines. For example in the case of hydrogen our computations show \citep{Chluba2009b} that the line broadening typically is of the order of a $\text{few}\times 10^2$ Doppler width in the case of Lyman $\alpha$.
For the $\ion{He}{i}\;\HeIlevel{n}{1}{P}{1}-\HeIlevel{1}{1}{S}{0}$ sequence it will be a bit smaller due to the fact that there is less helium than hydrogen, but still it should exceed $\text{few}\times 10$ Doppler width.
However, we will still use our simple approximation, as the correction again turns out to be rather small.

As mentioned above, since the $\ion{He}{i}\;\HeIlevel{3}{3}{P}{1}-\HeIlevel{1}{1}{S}{0}$ line is never really optically thick (cf. Fig.~\ref{fig:tau_QT}), most of the photons from the $\ion{He}{i}\;\HeIlevel{3}{1}{D}{2}-\HeIlevel{1}{1}{S}{0}$ quadrupole line will pass through this resonance and therefore could also feedback on the  $\ion{He}{i}\;\HeIlevel{2}{1}{P}{1}-\HeIlevel{1}{1}{S}{0}$ line.
However it takes about $\Delta z/z \sim 8\%$ to travel from the $\ion{He}{i}\;\HeIlevel{3}{3}{P}{1}-\HeIlevel{1}{1}{S}{0}$ to the $\ion{He}{i}\;\HeIlevel{2}{1}{P}{1}-\HeIlevel{1}{1}{S}{0}$ line. 
Looking at Fig.~\ref{fig:tauc} we can see that at $z\lesssim 2600-2700$ photons emitted by transitions with $n=3$ will never reach the $\ion{He}{i}\;\HeIlevel{2}{1}{P}{1}-\HeIlevel{1}{1}{S}{0}$ resonance, because they will be re-absorbed in the \ion{H}{i} continuum before. 
Similar comments also apply for the higher $\ion{He}{i}\;\HeIlevel{n}{3}{P}{1}-\HeIlevel{1}{1}{S}{0}$ series and at early times even for the photons from the $\ion{He}{i}\;\HeIlevel{n}{1}{D}{2}-\HeIlevel{1}{1}{S}{0}$ quadrupole transitions.
For example, because of the small fraction of neutral hydrogen atoms present during \ion{He}{i} recombination photons emitted in transitions $n=4$ to $n=1$ will no longer reach the $\ion{He}{i}\;\HeIlevel{3}{1}{P}{1}-\HeIlevel{1}{1}{S}{0}$ resonance at $z\lesssim 2300$, and at $z\lesssim 1900-2000$ $\ion{He}{i}\;\HeIlevel{3}{1}{P}{1}-\HeIlevel{1}{1}{S}{0}$ photons will not be able to feedback on the $\ion{He}{i}\;\HeIlevel{3}{1}{D}{2}-\HeIlevel{1}{1}{S}{0}$ resonance (cf. Fig.~\ref{fig:tauc}).
For our computations we also took these aspects of the problem into account.

In Figure~\ref{fig:DN_feed} we also show the results of our computations including the feedback for the higher transition. $n_{\rm f}=3$ means that we took the feedback between the main resonances with $n\leq 3$ into account, starting the sequence with the $\ion{He}{i}\;\HeIlevel{n_{\rm f}}{1}{P}{1}-\HeIlevel{1}{1}{S}{0}$ line.
Again we considered the cases with and without including the hydrogen continuum opacity between the subsequent resonances.
The reabsorption of photons between the lines leads to a very large suppression of the feedback correction, which for $n_{\rm f}=10$ reaches $\Delta N_{\rm e}/N_{\rm e}\sim 0.17\%$ at $z\sim 2300$ instead of  $\Delta N_{\rm e}/N_{\rm e}\sim 0.46\%$ at $z\sim 2045$.
We also ran cases including more than 10 shells in the computation, and found that the correction already converges for $n_{\rm f}\sim 10$. However, given that the final result is rather small we did not investigate this in more detail.
One can also see from Fig.~\ref{fig:DN_feed} that the largest single contribution to the feedback is coming from the second shell, i.e. the feedback sequence $\ion{He}{i}\;\HeIlevel{2}{1}{P}{1}-\HeIlevel{1}{1}{S}{0}\rightarrow \ion{He}{i}\;\HeIlevel{2}{3}{P}{1}-\HeIlevel{1}{1}{S}{0}$.

\begin{figure}
\centering 
\includegraphics[width=0.99\columnwidth]
{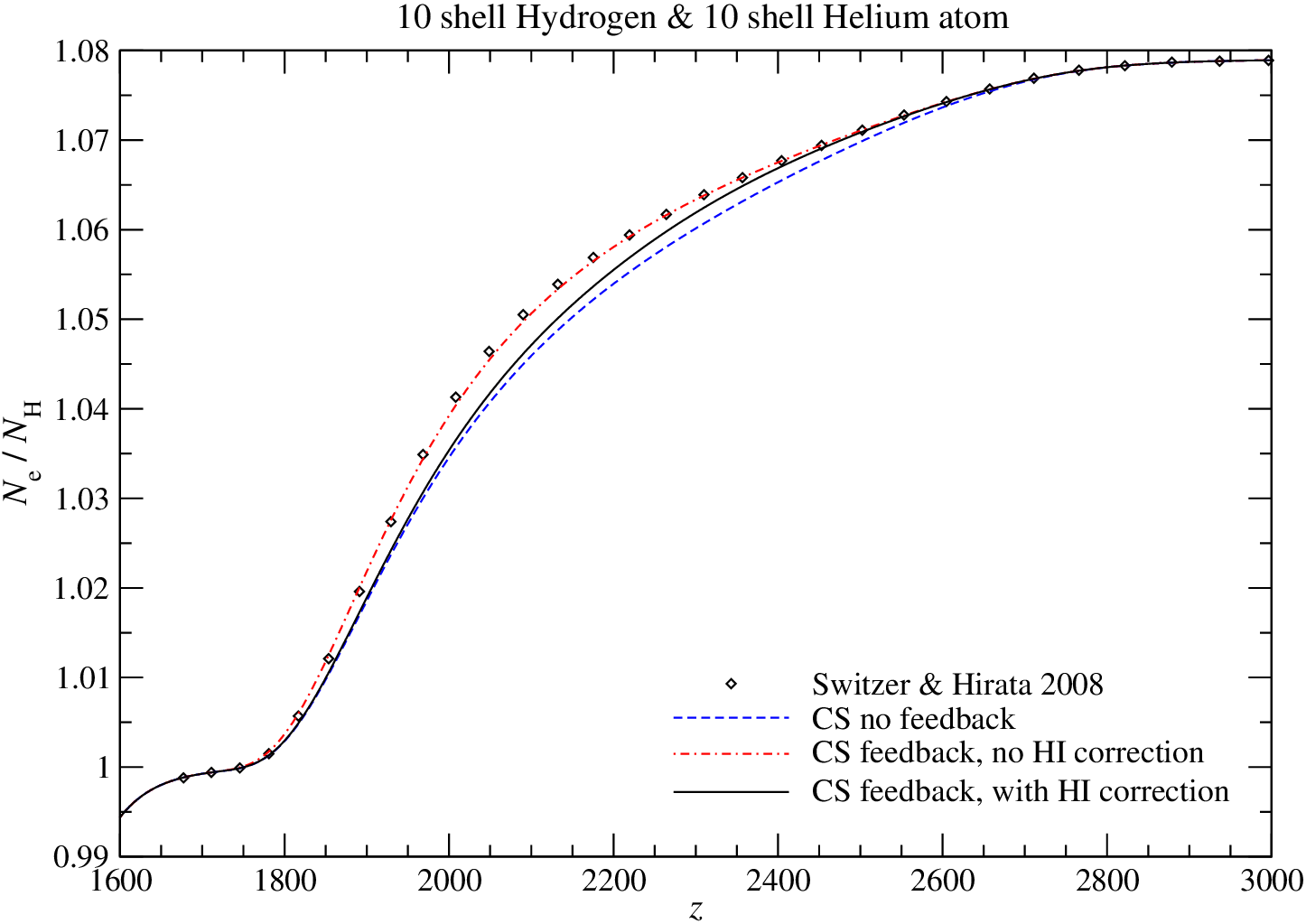}
\caption{Ionization history during $\ion{He}{ii}\rightarrow \ion{He}{i}$ recombination. The results were obtained for 10 shells for hydrogen and 10 shells for helium including the feedback between all 10 shells of helium. We also changed our cosmology to match the one of \citet{Switzer2007II}. }
\label{fig:tNe}
\end{figure}

In Fig.~\ref{fig:tNe} we also give the direct comparison with the results presented in \citet{Switzer2007II}. Our feedback correction seems to be slightly smaller at $z\lesssim 2400$, when including the effect of photon absorption in the \ion{H}{i} Lyman continuum between subsequent resonances. Surprisingly, our result seems to be very comparable to the one of \citet{Switzer2007II}, when we do not include this additional continuum absorption.
However, when including the \ion{H}{i} continuum opacity between sub-sequent resonances, our result is slightly smaller.
Nevertheless one should mention that these kind of changes in the helium recombination history will not be very important for the analysis of future CMB data.

\subsection{Feedback of \ion{He}{i} photons in the \ion{H}{i} Lyman continuum}
\label{sec:cont_feedback}
We also computed the recombination history including the feedback of \ion{He}{i} photons on the  \ion{H}{i} Lyman continuum, using  the approximation discussed in Sect.~\ref{sec:feedback_HILyc}.
As we already mentioned above all \ion{He}{i} photons are feeding back on hydrogen during its {\it pre-recombinational epoch}. Therefore one does not expect any large effect in the ionization history. We found that the correction does not exceed $\Delta N_{\rm e}/N_{\rm e} \sim 10^{-6}-10^{-5}$, so that for computations of the CMB power spectra this process is negligible.
However, as we show in Sect.~\ref{sec:pre-rec}, this process does lead to some traces in the cosmological recombination radiation, which should still be present as distortion of the CMB today. 
%

\begin{figure}
\centering 
\includegraphics[width=0.99\columnwidth]
{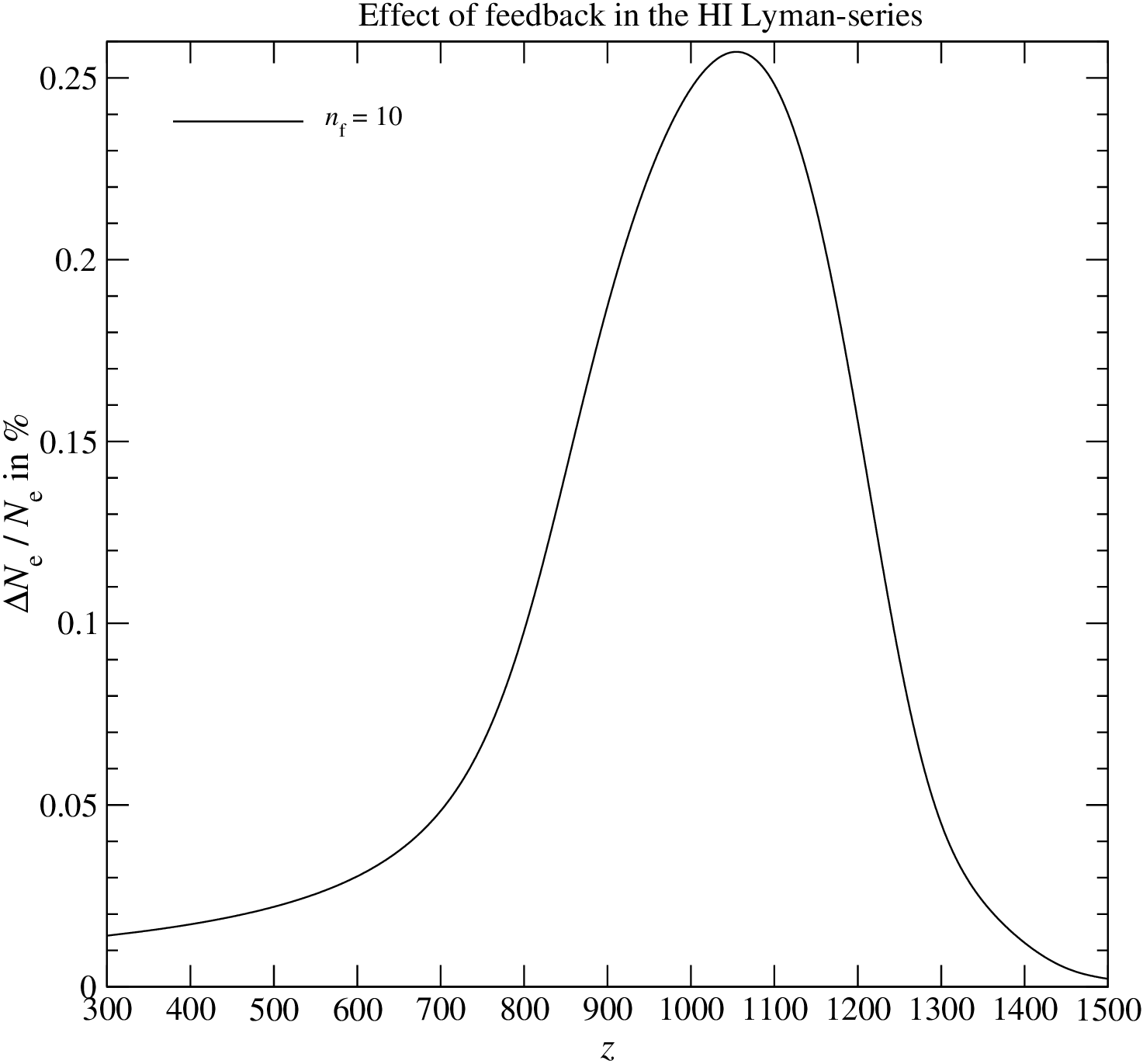}
\caption{Effect of Lyman series feedback on the ionization history during $\ion{H}{i}$ recombination. The results were obtained including 20 shells for hydrogen and 20 shells for helium and accounting for the feedback between the first 10 shells of hydrogen.}
\label{fig:NeHI}
\end{figure}
We also checked whether the feedback of the helium induced pre-recombinational features from the higher \ion{H}{i} Lyman series will have an effect on the dynamics of hydrogen recombination. Like in the normal recombinational epoch \citep{Chluba2007b} these additional photons will feed back on the next lower-lying Lyman series transition, however, due to the fact that at high redshift the Lyman series is not completely optically thick, the feedback is no longer restricted to $\text{Ly}n\rightarrow \text{Ly}(n-1)$, but can go beyond that. 
We refined our computation of the \ion{H}{i} Lyman series feedback problem including this process, but found no significant correction during the pre-recombinational epoch. 

Furthermore, we also included the small feedback correction due to photons that are emitted in the \ion{H}{i} Lyman continuum during the pre-recombinational epoch (see Sect.~\ref{sec:escape_HIcont} for more details). %
Again the correction to the ionization history was negligible. However, as we explain below (Sect.~\ref{sec:pre-rec}) this process leaves some interesting traces in the recombinational radiation spectrum.
Our final result for the effect of feedback during hydrogen recombination is shown in Fig.~\ref{fig:NeHI}. 
Note that the result is a bit bigger (by about $\Delta N_{\rm e}/\Delta N_{\rm e}\sim 0.04\%$ at $z\sim 1050$) than presented in \citet{Chluba2007b} because we slightly improved the accuracy of our numerical treatment for the normal \ion{H}{i} Lyman series feedback.

\begin{figure}
\centering 
\includegraphics[width=0.99\columnwidth]{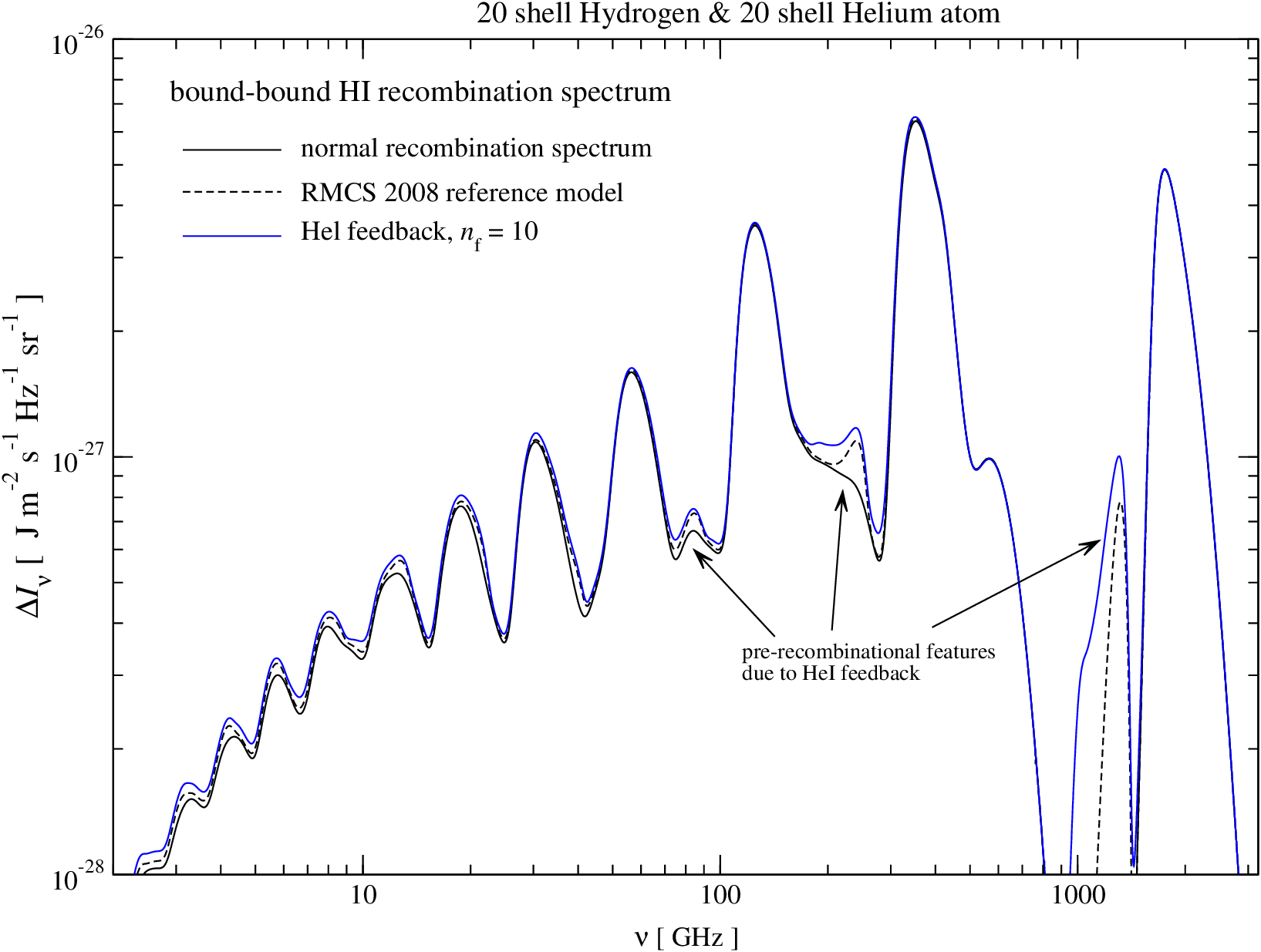}
\\[1mm]
\includegraphics[width=0.99\columnwidth]{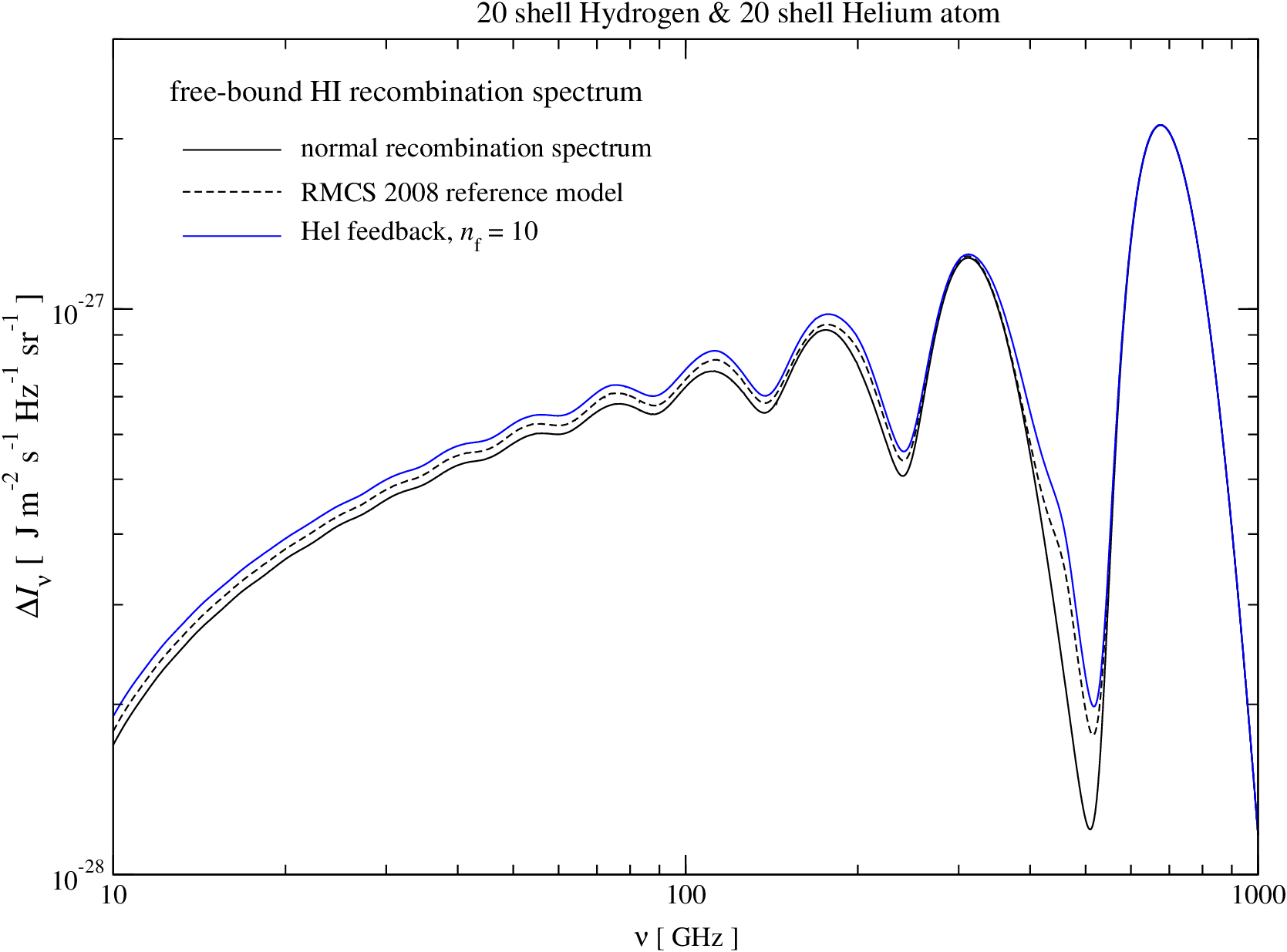}
\\[1mm]
\includegraphics[width=0.99\columnwidth]{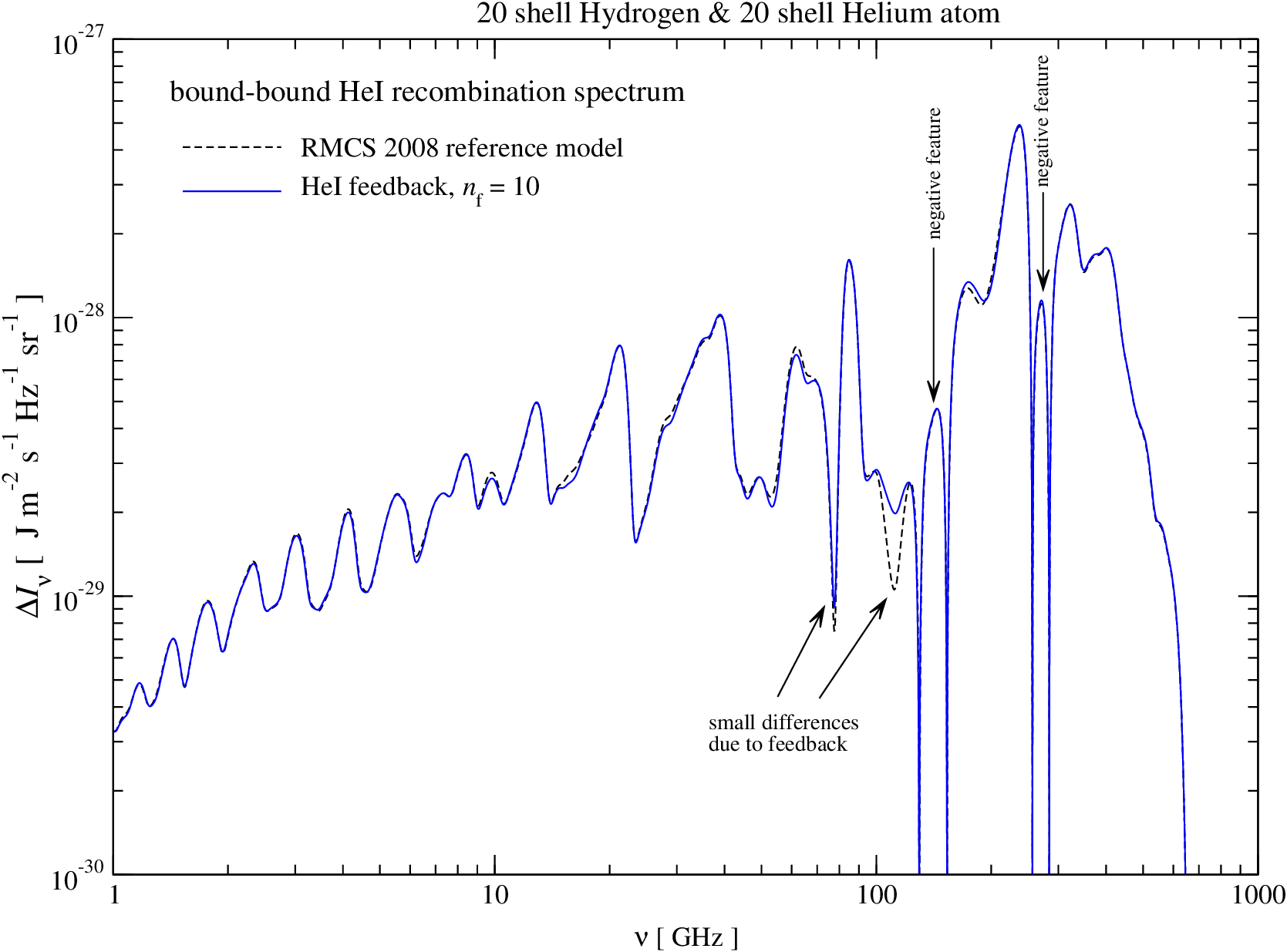}
\caption{Contributions to the cosmological recombination spectrum. We show the \ion{H}{i} bound-bound spectrum (upper panel), the  \ion{H}{i} free-bound spectrum (middle panel), and the  \ion{He}{i} bound-bound spectrum (lower panel). We included 20 shells of hydrogen and 20 shells of helium in our computations and allowed for helium feedback from the {\it first ten} shells. The curves marked with 'RMCS 2008' where computed using our reference model \citep{Jose2008}. Those marked as 'normal recombination spectrum' do not include the effects of the hydrogen continuum opacity on the dynamics of helium recombination. \changeIII{Note that for the \ion{He}{i} bound-bound spectrum we plotted the modulus of the distortion. There are two negative features at $\nu\sim 140\,$GHz and $\nu\sim 270\,$GHz.}}
\label{fig:DI}
\end{figure}
\section{Pre-recombinational emission from hydrogen due to the feedback of primary \ion{He}{i} photons}
\label{sec:pre-rec}
In this Section we present the results for the cosmological recombination spectrum from hydrogen and \ion{He}{i}, when including the full feedback of {\it primary} \ion{He}{i} photons from the $\HeIlevel{n}{1}{P}{1}-\HeIlevel{1}{1}{S}{0}$, $\HeIlevel{n}{3}{P}{1}-\HeIlevel{1}{1}{S}{0}$ and $\HeIlevel{n}{1}{D}{2}-\HeIlevel{1}{1}{S}{0}$ series with $n\leq 10$.
Figure~\ref{fig:DI} shows the results of our computations. Starting with the helium bound-bound contribution one can see that not very much has changed with respect to our reference spectrum \citep{Jose2008} when including the \ion{He}{i} feedback process. The most important difference is that there is {\it no} high frequency spectral distortion due to direct transitions from levels with $n\geq2$ to the ground state. These photons have all died in the \ion{H}{i} Lyman continuum during the pre-recombinational epoch of \ion{H}{i}, and also for our reference spectrum we neglected them \changeII{here}.

If we look at the bound-bound contribution from hydrogen (Fig.~\ref{fig:DI}, upper panel), then we can see that the feedback of helium photons leads to a pre-recombinational \ion{H}{i} feature at $\nu\sim 1300\,$GHz. Part of this feature was already given by \citet{Jose2008}, however, due to the additional reabsorption of high frequency photons from neutral helium, this signature of helium feedback increased.
As we will see below the total amount of photons appearing in that feature practically doubled.
\changeII{Note, that also at lower frequencies the helium feedback leads to some notable changes in the bound-bound emission from \ion{H}{i}.}

In Fig.~\ref{fig:DI} we also present the changes in the \ion{H}{i} free-bound contribution, which were not presented elsewhere so far.
One can see that here no strong additional helium induced pre-recombinational features are visible. This is because the \changeII{\ion{H}{i}} free-bound continua are rather broad at all times (typical $\Delta\nu/\nu_i \sim k\Tg/h\nu_i \sim 0.17\times\frac{1+z}{2500}\,\frac{n^2}{4}$) and therefore overlap strongly, so that \ion{He}{i} feedback only leads to an increase in the overall amplitude of the free-bound continuum.
We also checked the emission in the 2s-1s two-photon continuum, but the changes were so small that we simply neglected this contribution here.

\begin{table}
\caption{Total number (density) of photons, $N_{\gamma}$, produced by different transitions in hydrogen. Here $N_{\rm H}$ and $N_{\rm He}$ denote the total number (density) of hydrogen and helium nuclei. Contributions marked with 'rec' are from the normal recombinational epoch. Those marked with 'pre-rec I' are the additional photons computed for our reference model \citep{Jose2008}, while those marked with 'pre-rec II' give the additional photons when accounting for the full \ion{He}{i} feedback.
We also give the numbers for the whole bound-bound and free-bound spectrum, as well as the total difference (bound-bound $+$ free-bound) in the number of photons.}
\label{tab:Ng}
\centering
\begin{tabular}{@{}lrrrrr}
\hline
\hline
Line & $N_{\gamma}/N_{\rm H}$ & $N_{\gamma}/N_{\rm He}$ \\
\hline
\ion{H}{i} Ly$\alpha$ (rec) & 42.5\% &  -- \\
\ion{H}{i} Ly$\alpha$ (pre-rec I) & 3.3\% & 42\% \\
\ion{H}{i} Ly$\alpha$ (pre-rec II) & 3.1\% &  39\% \\
\ion{H}{i} Ly$\alpha$ (all) & 48.9\% & -- \\
%
\hline
\ion{H}{i} Ly series (rec) & 42.7\% &  -- \\
\ion{H}{i} Ly series (pre-rec I) & 3.6\% & 46\% \\
\ion{H}{i} Ly series (pre-rec II) & 3.4\% &  43\% \\
\ion{H}{i} Ly series (all) & 49.7\% & -- \\
\hline
\ion{H}{i} 2s-1s (rec)& $2\times 57.2\%$ & -- \\
\ion{H}{i} 2s-1s (pre-rec I)& $\sim 0$ & -- \\
\ion{H}{i} 2s-1s (pre-rec II)& $2\times 0.1\%$ & $1\%$\\
\ion{H}{i} 2s-1s (all)& $2\times 57.3\%$ & -- \\
\hline
\ion{H}{i} free-bound (rec)& $\sim 100\%$ & -- \\
\ion{H}{i} free-bound (pre-rec I)& 3.6\% & 46\% \\
\ion{H}{i} free-bound (pre-rec II)& 3.5\% & 44\% \\
\ion{H}{i} free-bound (all)& 107.1\% & -- \\
\hline
\ion{H}{i} bound-bound (dipole, low-$\nu$, rec)& $1.94$ & -- \\
\ion{H}{i} bound-bound (dipole, low-$\nu$, pre-rec I) & 5.3\% & 67\% \\
\ion{H}{i} bound-bound (dipole, low-$\nu$, pre-rec II) & 5.1\% & 65\% \\
\ion{H}{i} bound-bound (dipole, low-$\nu$, all)& 2.04 & -- \\
\hline
\ion{H}{i} bound-bound (dipole, rec)& $2.37$ & -- \\
\ion{H}{i} bound-bound (dipole, pre-rec I) & 8.9\% & 113\% \\
\ion{H}{i} bound-bound (dipole, pre-rec II) & 8.5\% & 108\% \\
\ion{H}{i} bound-bound (dipole, all)& 2.54 & -- \\
\hline
\hline
total difference (pre-rec I)& 12.7\% &  1.6 \\
total difference (pre-rec II)& 12.2\% &  1.6 \\
total difference (pre-rec I+II)& 24.9\% &  3.2 \\
\hline
\hline
\end{tabular}
\end{table}
\subsection{Counting the number of additional photons}
\label{sec:count_HeI}
We can now look at the number of additional photons that are created due to the feedback on hydrogen.
It is clear that every helium photon that is absorbed in the \ion{H}{i} Lyman continuum will at least be replaced by {\it one} photon in the free-bound spectrum of hydrogen.
Since so far we only took the feedback of photons emitted in the $\ion{He}{i}\;\HeIlevel{n}{1}{P}{1}-\HeIlevel{1}{1}{S}{0}$, $\ion{He}{i}\;\HeIlevel{n}{3}{P}{1}-\HeIlevel{1}{1}{S}{0}$ and $\ion{He}{i}\;\HeIlevel{n}{1}{D}{2}-\HeIlevel{1}{1}{S}{0}$ series with $n\leq 10$ into account, and because these amount to about
%
0.9 photons per helium atom available for the feedback process, one expects that the free-bound emission from hydrogen increases by about $7\%$. 
Simply computing the total amount of photons emitted in the free-bound continuum of hydrogen, we indeed find that in total $\sim~7.1\%$ more photons are created in the \ion{H}{i} free-bound continuum (Table~\ref{tab:Ng}). 
About half of this number is already appearing for our reference model, in which only part of the feedback correction is included.

We then also looked at the number of photons emitted in the \ion{H}{i} 2s-1s two-photon continuum, finding that this number in not affected much due to helium feedback (cf. Table~\ref{tab:Ng}). This is because at early times the Lyman $\alpha$ line and the other Lyman series transitions are more optically thin and therefore much more efficient than the 2s-1s two-photon channel. 
As shown in \citet{Jose2006} the 2s-1s two-photon channel only starts to become active at redshift $z\lesssim 1600-1700$ where practically no additional helium photons are produced.

We also computed the total number of additional photon emitted in the Lyman series and found about $7\%$ per hydrogen atom. Together with the contribution from the 2s-1s two-photon channel this yields $\sim 7.1\%$
per hydrogen atom as it should be, since all electrons that entered the hydrogen atom via the continuum, should reach the ground state via the Lyman series or alternatively the 2s-1s two-photon channel. 
%
%
From Table~\ref{tab:Ng} one can also see that most (i.e. $\sim 6.4\%$) of these additional high frequency photons are produced in the Lyman $\alpha$ line, but also several additional photons are created in the higher \ion{H}{i} Lyman-series.  

%

\subsubsection{Total number of photons related to \ion{He}{i}}
\label{sec:HeI_total_num}
We finally also computed the total number of additional photons which are related to the presence of neutral helium in our Universe (see Table~\ref{tab:Ng}).
Per \ion{He}{i} atom we found about 3.2 additional photons in the \ion{H}{i} recombination radiation. 
Since in total about $\sim 0.9$ ionizing quanta per helium atom were absorbed in the feedback process, and then eventually replaced by a \ion{H}{i} Lyman $\alpha$ resonance or 2s-1s two-photon continuum photon (see Sect.~\ref{sec:count_HeI}), about 2.3 {\it extra} photons per helium atom appeared in the \ion{H}{i} recombination spectrum. These photons would be absent without the inclusion of the \ion{He}{i} feedback process.
Furthermore, of these 2.3 {\it extra} photon per helium atom about $\sim 0.9\,\gamma$ were produced in the free-bound continuum and the remaining $\sim 1.4\,\gamma$ were emitted in \changeII{transitions between excited states with principle quantum number $n\geq2$}.
Furthermore, one can say that per \ion{He}{i} feedback photon about $2.3/0.9\sim 2.6$ extra photons were released by \ion{H}{i}.

In our reference model per \ion{He}{i} atom we find at total of $\sim 1.85$ photons\footnote{In this number we neglected the high frequency resonance photons that will eventually be reprocessed by \ion{H}{i} in the feedback problem.} in the normal \ion{He}{i} bound-bound spectrum, $\sim 0.16$ photons in the $\ion{He}{i}\;\HeIlevel{2}{1}{S}{0}-\HeIlevel{1}{1}{S}{0}$ two-photon decay spectrum, and $\sim 1$ photon in the \ion{He}{i} free-bound continuum.
This implies a total of $\sim 3\,\gamma$ per \ion{He}{i} atom.
Including the full feedback caused by the photons from the $\ion{He}{i}\;\HeIlevel{n}{1}{P}{1}-\HeIlevel{1}{1}{S}{0}$, $\ion{He}{i}\;\HeIlevel{n}{3}{P}{1}-\HeIlevel{1}{1}{S}{0}$ and $\ion{He}{i}\;\HeIlevel{n}{1}{D}{2}-\HeIlevel{1}{1}{S}{0}$ series with $n\leq 10$, we removed a total of $\sim 0.9$ high frequency photons per helium atom, but add $\sim 3.2$ new photons to the \ion{H}{i} recombination spectrum. Therefore in total the number of photons related to the presence of \ion{He}{i} increased to a total of $6.2\,\gamma$ per helium atom, or by a factor of $\sim 1.7$ relative to the case with no interaction between helium and hydrogen for which a total of $3.7\,\gamma$ per helium nucleus are emitted (see Table~\ref{tab:Ng_HeII}). 
This is a large increase in the total number of photons that are related to the presence of \ion{He}{i} in the Universe. One should also mention, that the absolute number should increase when including more shells for both the hydrogen and neutral helium atom into the computation.
Also, the feedback from higher level transitions and the photons emitted in the $\ion{He}{i}\;\HeIlevel{2}{1}{S}{0}-\HeIlevel{1}{1}{S}{0}$ two-photon continuum will slightly increase this number (see Sections below). We leave a detailed computation for a future paper.

\subsection{Additional corrections to the \ion{He}{i} feedback problem}
\label{sec:add_corr_HeI}

\subsubsection{Feedback from higher level transitions}
Here we restricted our computation to the feedback of photons from the $\ion{He}{i}\;\HeIlevel{n}{1}{P}{1}-\HeIlevel{1}{1}{S}{0}$, $\ion{He}{i}\;\HeIlevel{n}{3}{P}{1}-\HeIlevel{1}{1}{S}{0}$ and $\ion{He}{i}\;\HeIlevel{n}{1}{D}{2}-\HeIlevel{1}{1}{S}{0}$ series with $n\leq 10$.
We computed the additional number of photons emitted in these series from $n>10$ and found about $0.02$ photons per helium atom.
Taking the feedback of these into account one therefore expects $8\%\times 2\% \times 2.6 \sim 0.4\%$ extra photons per hydrogen atom (or $\sim 0.05\,\gamma$ per helium atom).
As we discuss below also the feedback from photons in the $\ion{He}{i}\;\HeIlevel{2}{1}{S}{0}-\HeIlevel{1}{1}{S}{0}$ two-photon continuum should lead to another small increase in the total number of photons that are related to the presence of helium in the Universe.

\subsubsection{Feedback of photon from the $\ion{He}{i}\;\HeIlevel{2}{1}{S}{0}-\HeIlevel{1}{1}{S}{0}$ two-photon continuum}
In our computation so far we did not include the correction due to the feedback of photons emitted in the $\ion{He}{i}\;\HeIlevel{2}{1}{S}{0}-\HeIlevel{1}{1}{S}{0}$ two-photon continuum.
It is clear that $\sim 1/2$ of these photons are never able to feedback on hydrogen, since they are directly emitted below the \ion{H}{i} Lyman $\alpha$ line with transition energy $10.2\,$eV.
Therefore, as mentioned in the Sect.~\ref{sec:feed_num_HeI}, in principle per helium atom there are in addition $\sim 0.08$ photons available for the feedback on hydrogen.
Of this number about $0.044$ photons\footnote{We estimated this number by integrating the $\ion{He}{i}\;\HeIlevel{2}{1}{S}{0}-\HeIlevel{1}{1}{S}{0}$ two-photon profile over the appropriate range of frequencies.} per helium atom will feedback in the hydrogen Lyman continuum, while the remaining photons could directly affect the \ion{H}{i} Lyman series.
Again this feedback will not introduce any important change in the ionization history, but will only lead to the production of additional \ion{H}{i} photons.
Assuming that {\it all} these photons are absorbed one should therefore find another $8\%\times 8\%\times 2.6 \sim 1.7\%$ more photons per hydrogen atom, or $\sim 0.21$ additional photons per helium atom.
The final number is expected to be a bit smaller, \changeII{because not all photons will be absorbed and since the feedback in the \ion{H}{i} Lyman series will not produced as many secondary photons}, but we leave a detailed calculation for some future work.


\section{Feedback of photon from $\ion{He}{iii}\rightarrow\ion{He}{ii}$ recombination}
\label{sec:HeIII_feedback}
It is clear that also the feedback of high frequency photons from $\ion{He}{iii}\rightarrow\ion{He}{ii}$ recombination will lead to some additional changes to the CMB distortions from cosmological recombination.
\changeII{Like in} the case of \ion{He}{i} feedback on hydrogen, there is no significant additional corrections to the ionization history caused by this process, since \ion{He}{ii} feedback is again occurring in the {\it pre-recombinational} epoch, but this time of \ion{He}{i}. \changeII{However, several secondary photons are produced, increasing the total contribution of helium related photons to the cosmological recombination spectrum.}
A rigorous computation of this problem is beyond the scope of this paper, however, there are several interesting aspects, that  we would like to mention, before demonstrating the principle importance of this feedback process for the total contribution of helium-related photons to the cosmological recombination spectrum (see Sect.~\ref{sec:spec_HeII}).

\subsection{Total amount of high frequency $\ion{He}{ii}$  related photons that are initially available for feedback}
\label{sec:feed_num_HeII}
As mentioned in the introduction, all $\ion{He}{ii}$ photons emitted at energies above\footnote{In principle \ion{He}{ii} photons could also feedback on the $\ion{He}{i}\;\HeIlevel{2}{1}{S}{0}-\HeIlevel{1}{1}{S}{0}$ two-photon continuum, but this transition is never optically thick during recombination so that we neglected it here.} $\sim 20.96\,$eV (corresponding to the transition energy of the $\ion{He}{i}\;\HeIlevel{2}{3}{P}{1}-\HeIlevel{1}{1}{S}{0}$ resonance) in principle can feedback on \ion{He}{i}, while those photons emitted at energies above $\sim 10.2\,$eV (i.e. the \ion{H}{i} Lyman $\alpha$ transition energy) may affect hydrogen. These statements are rather simple, but the details turn out to be more involved. Below we provide some estimates for the total number of \ion{He}{ii} photons that are available in the feedback problem. 

\subsubsection{Initial amount of high frequency $\ion{He}{ii}$ photons that are available for the feedback on \ion{He}{i}}
According to our computations, for $\ion{He}{ii}$ recombination about 45\% of $\ion{He}{ii}$ electrons went through the \ion{He}{ii} 2s-1s two-photon continuum, while the remaining $\sim 55\%$ recombined via the $\ion{He}{ii}$ Lyman $\alpha$ channel (with a very small addition due to the higher $\ion{He}{ii}$ Lyman series).
Furthermore, about $0.2\,\gamma$ photons\footnote{This number was estimated from the result for the 100 shell hydrogen atom \citep{Chluba2006b}.} per helium nucleus were emitted in the $\ion{He}{ii}$ Balmer continuum, and in total $\sim 4.69\,\gamma$ were produced per $\ion{He}{ii}$ atom. These numbers were obtained including 100 shells for the \ion{He}{ii} atom.

All the \ion{He}{ii} Lyman $\alpha$ photons (with energy $\sim 40.8\,$eV) are able to directly ionize $\ion{He}{i}$ atoms. After significant redshifting, these photons in principle could also directly ionize hydrogen atoms, but it turns out that all \ion{He}{ii} Lyman $\alpha$ quanta will already disappear in the $\ion{He}{i}\;\HeIlevel{1}{1}{S}{0}$ continuum before that (see Sect.~\ref{sec:time-HeII-feed}).
To simplify matters, for our more detailed estimates of the effect on the cosmological recombination spectrum (Sects.~\ref{sec:number_HeII}) we will {\it only} take these photons into account. 

It is clear that also the \ion{He}{ii} 2s-1s two-photon continuum will lead to some interesting feedback. About\footnote{We estimated the numbers given below by integrating the $\ion{He}{ii}$ 2s-1s two-photon profile over the appropriate range of frequencies.} $37\%$ of the photons emitted in the  \ion{He}{ii} 2s-1s two-photon continuum have energies above the $\ion{He}{i}\;\HeIlevel{1}{1}{S}{0}$ continuum.
Therefore, in total one should have $\sim 0.33\, \gamma$ per helium nucleus with energy larger than the \ion{He}{i} ionization potential $E\sim 24.59\,$eV. 

In addition, those photons emitted in the \ion{He}{ii} 2s-1s two-photon continuum with energies $20.96\,\text{eV}\lesssim h\nu\lesssim 24.59\,$eV (i.e. about $0.1\, \gamma$ per helium nucleus) may be able to directly feedback to the $\ion{He}{i}\;\HeIlevel{n}{1}{P}{1}-\HeIlevel{1}{1}{S}{0}$, $\ion{He}{i}\;\HeIlevel{n}{3}{P}{1}-\HeIlevel{1}{1}{S}{0}$ and $\ion{He}{i}\;\HeIlevel{n}{1}{D}{2}-\HeIlevel{1}{1}{S}{0}$ series, depending on the optical depth in these resonances at different stages during recombination. However, since the $\ion{He}{i}\;\HeIlevel{2}{1}{P}{1}-\HeIlevel{1}{1}{S}{0}$, $\ion{He}{i}\;\HeIlevel{2}{3}{P}{1}-\HeIlevel{1}{1}{S}{0}$ and $\ion{He}{i}\;\HeIlevel{2}{1}{D}{2}-\HeIlevel{1}{1}{S}{0}$ resonances are not optically thick at $z\gtrsim 3000$ (see Fig.~\ref{fig:tau_QT}), most of these \ion{He}{ii} photons will survive and instead be available for feedback onto hydrogen at lower redshift.
Also it is clear that the $\ion{He}{ii}$ Balmer continuum photons will not feedback on $\ion{He}{i}$, but could affect the recombination of hydrogen (see next Section).

\begin{figure}
\centering 
\includegraphics[width=0.99\columnwidth]
{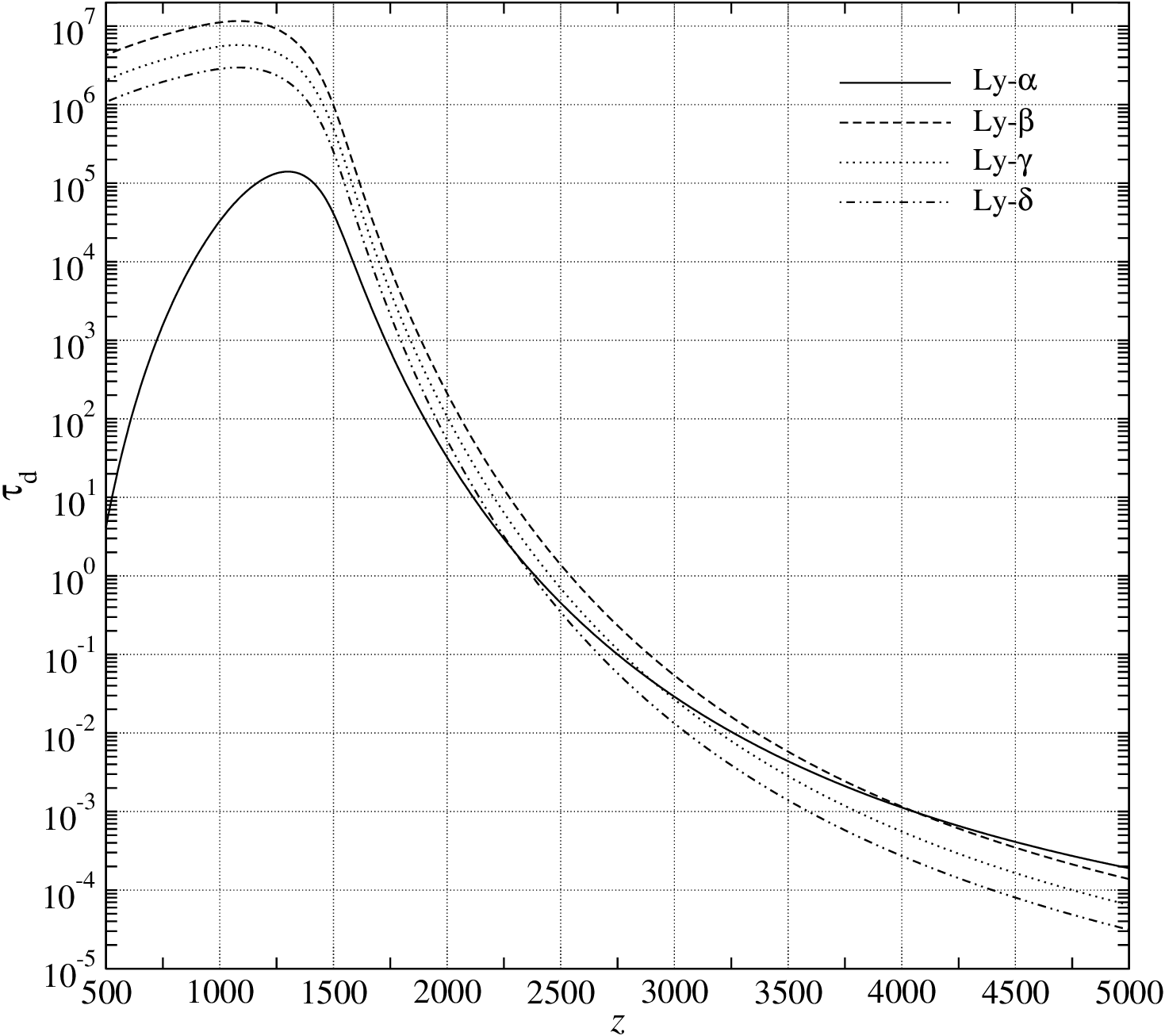}
\caption{\changeIII{Effective} absorption optical depth, \changeI{$\tau^i_{\rm d}=p^i_{\rm d}\tau^i_{\rm S}$}, in the $\ion{H}{i}$ Lyman series. \changeI{An interpretation and definition of $\tau^i_{\rm d}$ is given in Sect.~\ref{sec:sol_F_2400}.}}
\label{fig:taud_Ly}
\end{figure}
\subsubsection{Initial amount of high frequency $\ion{He}{ii}$ photons that are available for the feedback on \ion{H}{i}}
Regarding the possibility of direct ionization of hydrogen atoms by high frequency $\ion{He}{ii}$ photons one will have those photons from the \ion{He}{ii} 2s-1s two-photon continuum emitted in the frequency range $13.6\,\text{eV}\lesssim h\nu\lesssim 20.96\,$eV, amounting to about $0.21\,\gamma$ per helium nucleus. 
Furthermore, those \ion{He}{ii} 2s-1s photons emitted at energies $10.2\,\text{eV}\lesssim h\nu\lesssim 13.6\,$eV ($\sim 0.09\,\gamma$ per helium nucleus) will be able to feedback on the \ion{H}{i} Lyman series. And finally, all those photons emitted in the \ion{He}{ii} Balmer continuum in principle will also be able to feedback on the \ion{H}{i} Lyman series. This implies a total of $\sim 0.5$ potential feedback photons per helium nucleus (or $0.04$ photons per hydrogen nucleus) that may {\it directly} feedback on hydrogen.
Even though these photons should lead to secondary \ion{H}{i} Lyman series photons in the pre-recombinational epoch which then again feedback on hydrogen itself, here we will not consider this part of the feedback process any further.

Also the numbers given above should be considered as estimates since at $z\gtrsim 2500$ none of the \ion{H}{i} Lyman series transitions is optically thick (cf. Fig.~\ref{fig:taud_Ly}), implying that the feedback process will not re-process all these photons.
Similarly, some photons may escape the feedback problem in neutral helium and therefore add to the total amount of {\it primary} feedback photons available for hydrogen.
In addition, practically every ionizing \ion{He}{ii} photon that is re-processed by \ion{He}{ii} will be replaced by some {\it secondary} high frequency \ion{He}{i} photon (in addition to some secondary low-frequency photons that do not affect the recombination problem anymore).
This leads to a {\it double-reprocessing} of $\ion{He}{ii}$ photons (see Sect.~\ref{sec:double}).
A detailed analysis is beyond the scope of this paper.

\subsection{Double-reprocessing of $\ion{He}{ii}$ photons}
\label{sec:double}
The probably most interesting aspect of the \ion{He}{ii} feedback problem is the {\it double-reprocessing} of $\ion{He}{ii}$ photons, first by neutral helium and then later by hydrogen.
This is because in addition to the {\it primary} photons initially released by \ion{He}{ii} there will be {\it secondary} photons produced by both \ion{He}{i} and \ion{H}{i}. Those from hydrogen will only feedback on hydrogen itself but likely will not produce many more (low frequency) photons.
However, the feedback of secondary \ion{He}{i} photons on hydrogen is expected to be more important. 
The total number of these secondary \ion{He}{i} photons will be very close to the total number of \ion{He}{ii} photons that were re-processed. 
This can be understood because in the pre-recombinational epoch of \ion{He}{i} the $\ion{He}{i}\;\HeIlevel{2}{1}{S}{0}-\HeIlevel{1}{1}{S}{0}$ two-photon channel will not be able to compete with the main resonances connecting to the ground state. Therefore it is clear that practically every  \ion{He}{ii} feedback photon, independent of the way it was re-processed, will be replaced by a high frequency \ion{He}{i} photon, which then again could be re-processed by hydrogen. 
Similarly, every high frequency photon that is re-processed by hydrogen will eventually be re-appearing as a \ion{H}{i} Lyman $\alpha$ photon.

Using the numbers given in the previous section one therefore expects a maximum of $\sim 0.98$ primary feedback photons per helium nucleus, and $\sim 0.12\,$ primary plus secondary feedback photons per hydrogen atom that are related to \ion{He}{ii}.

In every feedback there is the potential possibility to produce {\it several additional} photons at low frequencies because electrons that reach highly excited states (via excitations from the ground state or direct recombination after feedback) can cascade towards lower states via {\it many} intermediate levels, emitting low frequency photons on their way. Therefore this process can lead to an important increase of the helium-related photons in the cosmological recombination spectrum.

\begin{figure}
\centering 
\includegraphics[width=0.99\columnwidth]
{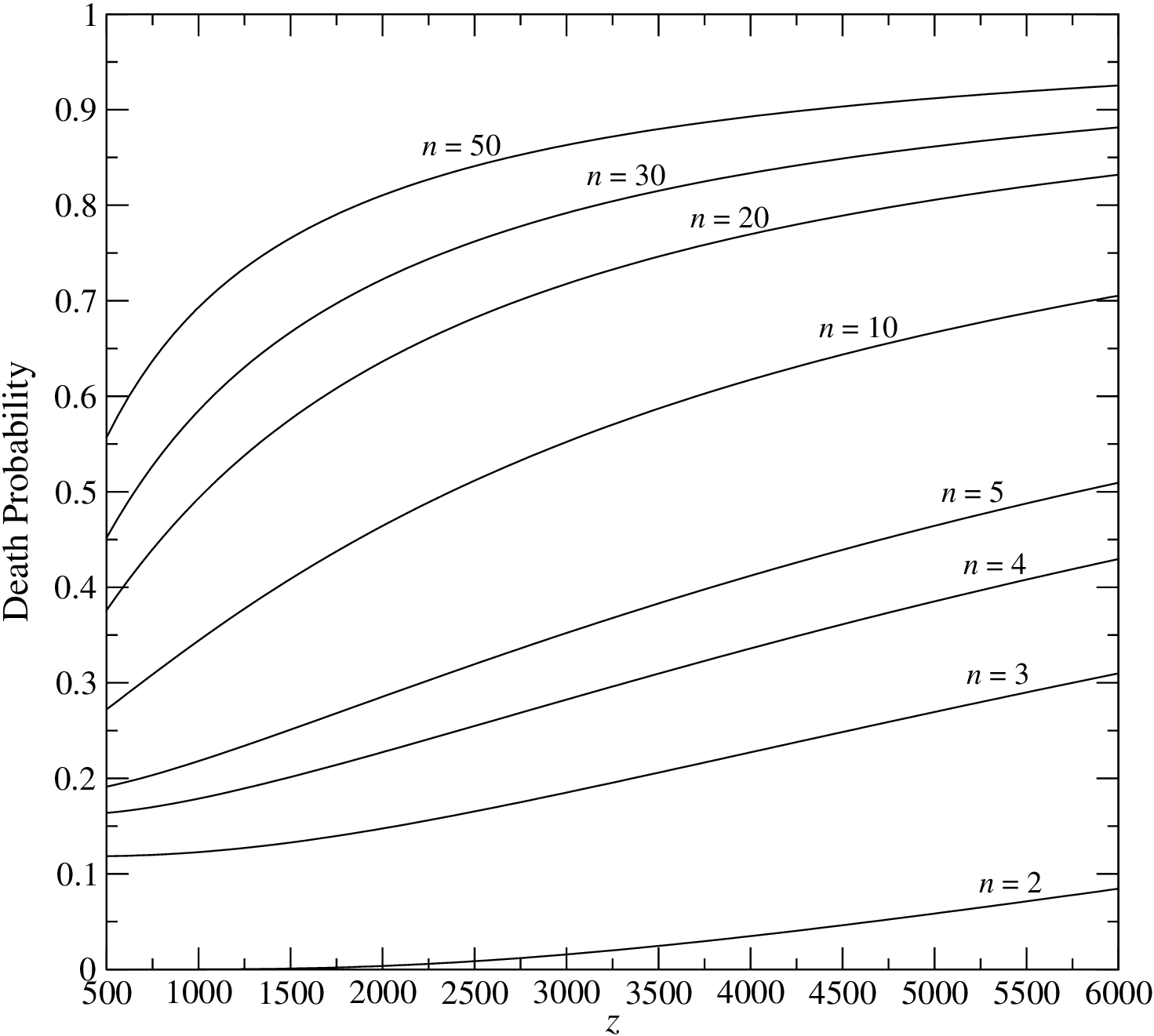}
\caption{Death probabilities in the \ion{H}{i} Lyman series. We included 200 shells into the sums and assumed that the ambient radiation field is given by the CMB blackbody spectrum.}
\label{fig:pd.Lyn}
\end{figure}

\subsection{Direct recombinations to the ground state of \ion{He}{i} and \ion{H}{i} and the efficiency of photon production}
\label{sec:direct}
If we consider the case of hydrogen, then there is a principle difference in the efficiency of (low frequency) photon production if the feedback process occurs in the \ion{H}{i} Lyman continuum or the Lyman series. 
The efficiency of photon production will be larger if the feedback photon disappeared in the Lyman continuum, while for feedback in the Lyman series the total number of \changeII{(low frequency)} photons is not changing very much.
Therefore it is also clear that the total efficiency of secondary photon production will depend on whether the \ion{H}{i} Lyman continuum is {\it optically thick} or {\it thin}.
A similar statement also applies to the photon production efficiency of \ion{He}{i}. \changeII{Furthermore, the same physical argument explains why the total number of photons is not affected very much in the case of intra-species feedback.}

To understand this point, let us consider a feedback photon that dies in the \ion{H}{i} Lyman continuum. 
It liberates an electron which then can be captured to some highly excited state, sub-sequently emitting {\it several} low frequency photons on it way back to the ground state. 
Here the important point is that it in principle can reach states with large angular momentum quantum number. These states can only depopulate via {\it many} transitions with {\it small} $\Delta n$, i.e. preferably emitting {\it low} frequency photons.

If on the other hand, a photon is feeding back on one of the \ion{H}{i} Lyman series transitions, then even for Ly-$n$ with $n>2$ the electron will very likely remain bound and only reach a few intermediate states preferably with small angular momentum quantum number, before again ending in one of the Lyman resonances. The low $l$-states will nearly always depopulate via {\it a few} transitions with {\it large} $\Delta n$, i.e. only emitting some photon at {\it high} frequencies.
%

\begin{figure}
\centering 
\includegraphics[width=0.99\columnwidth]
{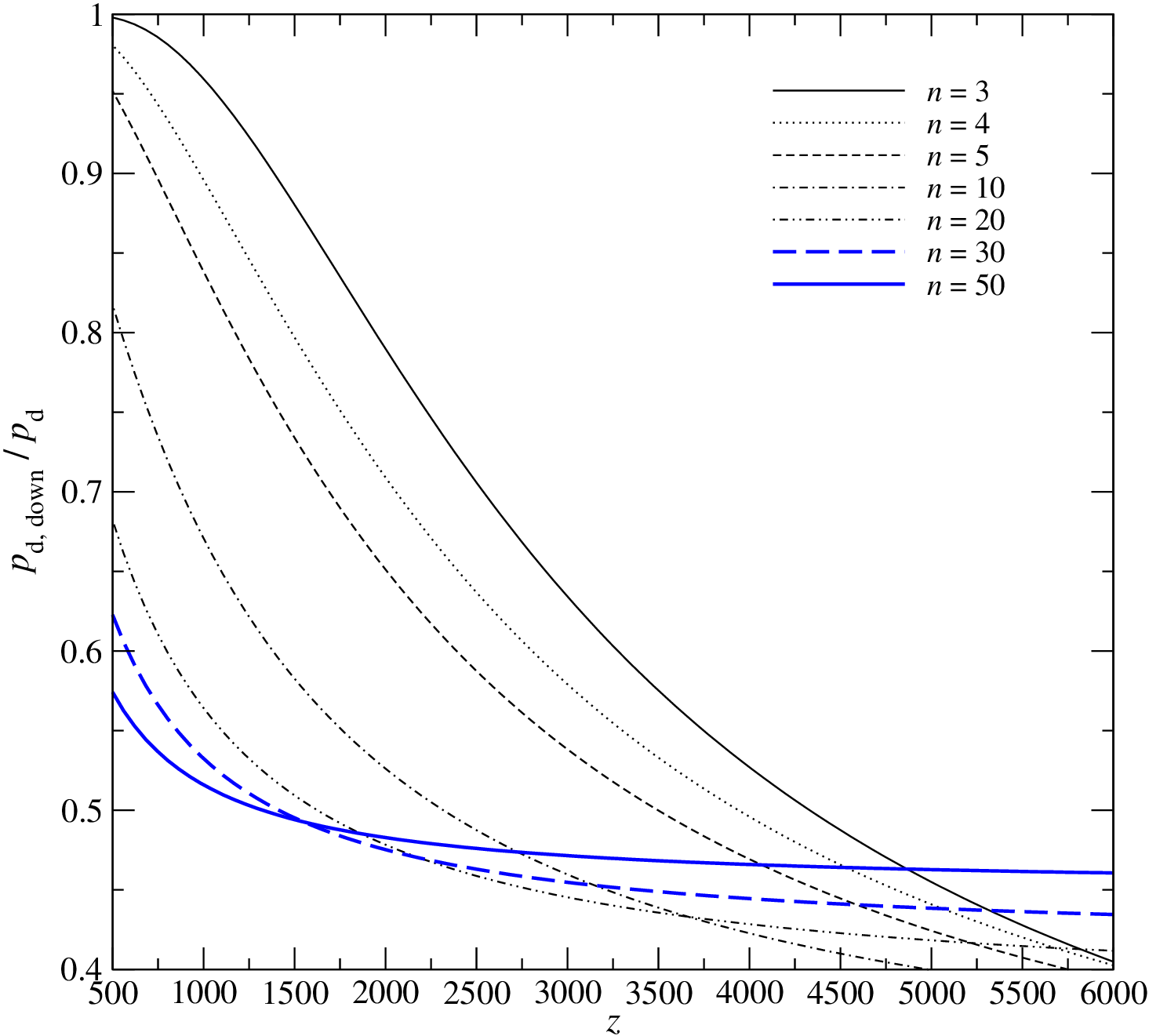}
\caption{Relative contribution of transitions towards lower levels to the total death probabilities in the \ion{H}{i} Lyman series. We included 200 shells into the sums and assumed that the ambient radiation field is given by the CMB blackbody spectrum.}
\label{fig:pd.Lyn.down}
\end{figure}

\begin{figure}
\centering 
\includegraphics[width=0.99\columnwidth]
{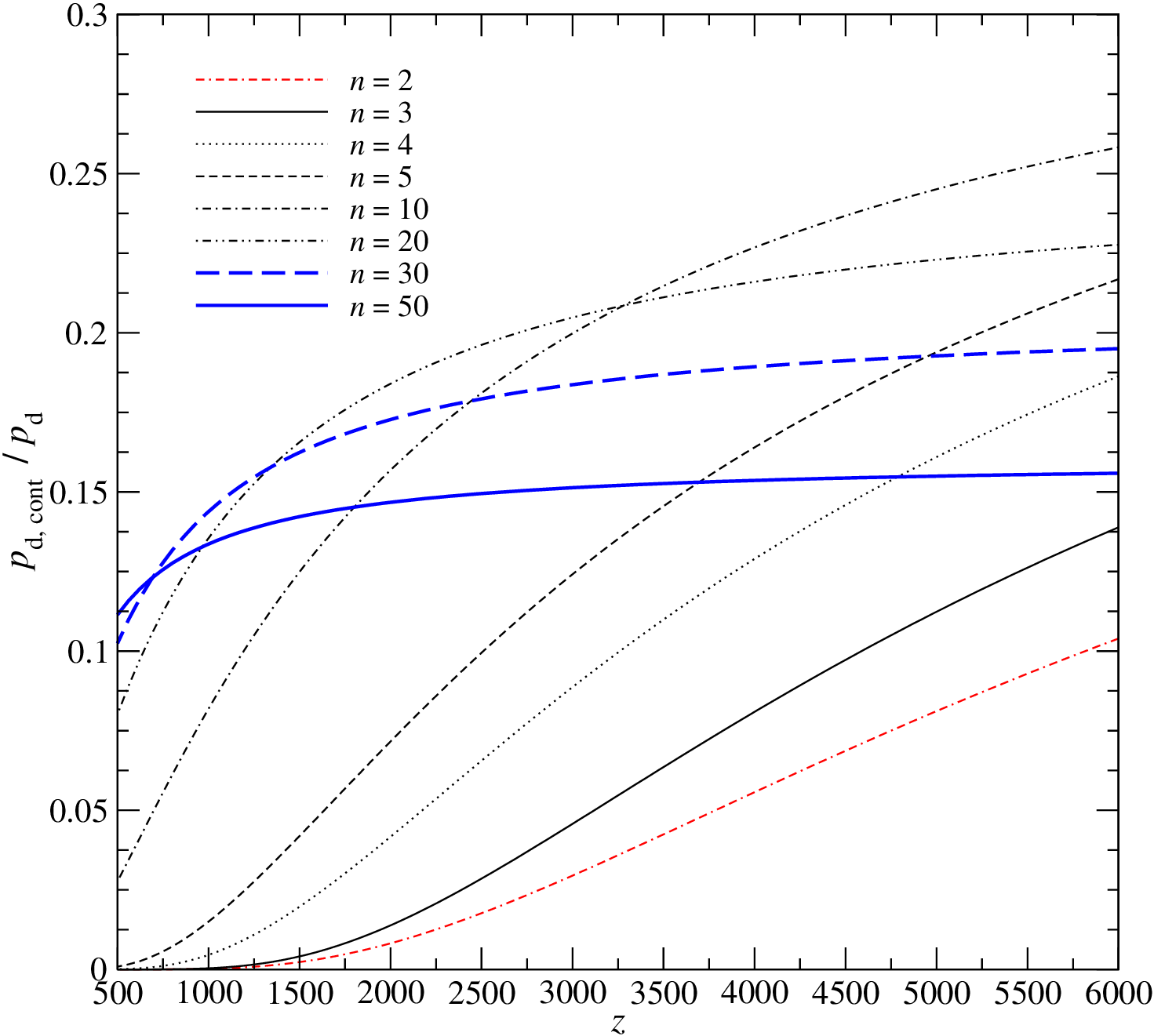}
\caption{Relative contribution of continuum transitions to the total death probabilities in the \ion{H}{i} Lyman series. We included 200 shells into the sums and assumed that the ambient radiation field is given by the CMB blackbody spectrum.}
\label{fig:pd.Lyn.cont}
\end{figure}
To support these statement, in Fig.~\ref{fig:pd.Lyn} we give the death probabilities for several \ion{H}{i} Lyman series transitions. In Fig.~\ref{fig:pd.Lyn.down} and \ref{fig:pd.Lyn.cont} we also show the partial contribution to the death probably coming only from transitions towards lower levels and the continuum, respectively.
Even though for levels with $n>2$ the death probability is very large in comparison to the \ion{H}{i} Lyman $\alpha$ line, less than $15\%-25\%$ of the transitions end with a free electron. In all cases about $40\%-50\%$ of transitions go towards lower levels, and the remaining $\sim 25\%-35\%$ lead to some higher levels.

As we will see below (Sects.~\ref{sec:number_HeII}), this aspect of the problem changes the total number of secondary \changeII{low frequency} photons that are produced due to feedback of \ion{He}{ii} photons.
%

\begin{figure}
\centering 
\includegraphics[width=0.99\columnwidth]
{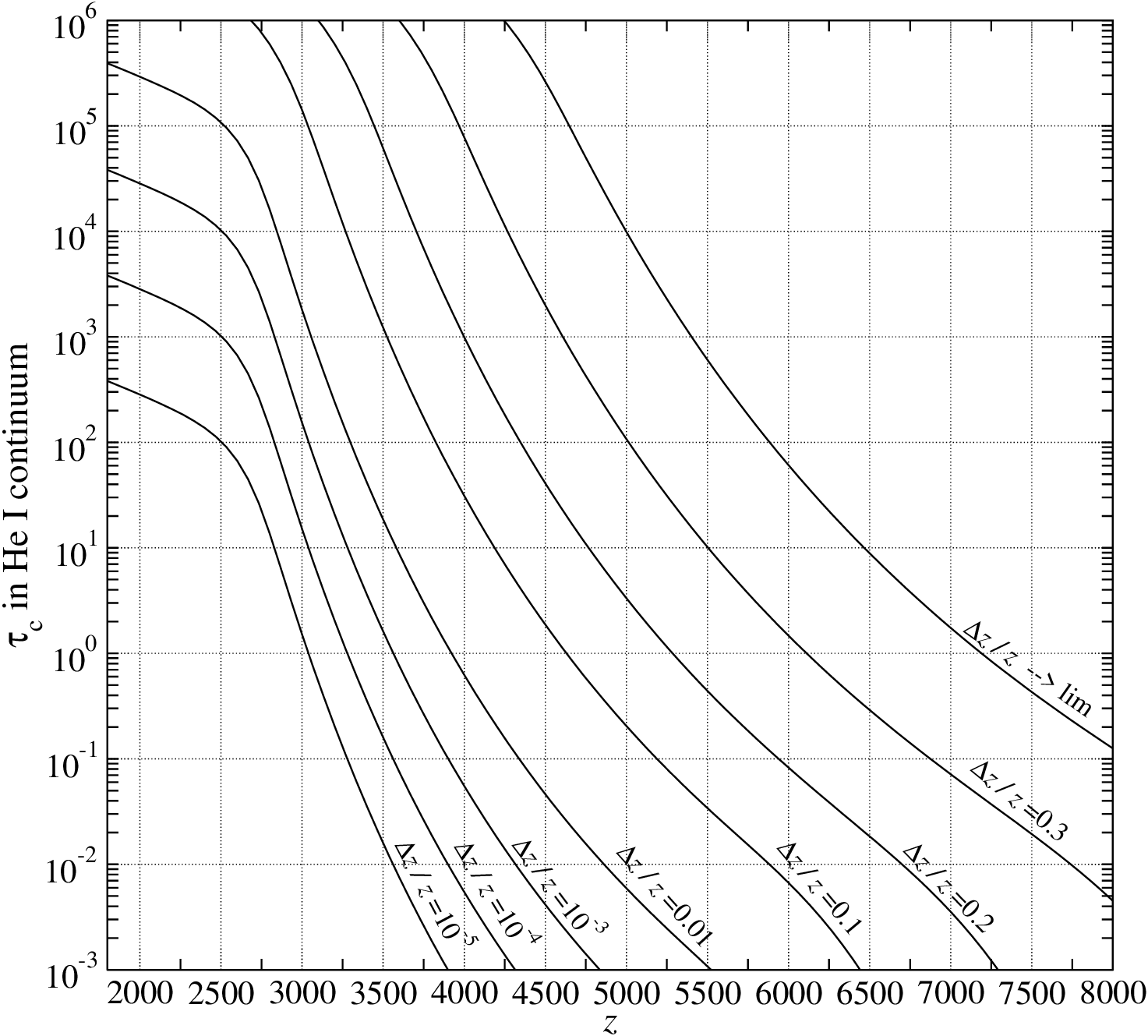}
\caption{Absorption optical depth in the $\ion{He}{i}\;\HeIlevel{1}{1}{S}{0}$ continuum for a \ion{He}{ii} Lyman $\alpha$ photon that was emitted at redshift $z$ and traveled until $z'=z-\Delta z$. At high redshifts ($z\gtrsim 4500$) we used the Saha solution for the $\ion{He}{i}\;\HeIlevel{1}{1}{S}{0}$ population and at low redshifts ($z\lesssim 4500$) the {\sc Recfast} solution \citep{SeagerRecfast1999}. The curve labeled '$\Delta z/z\rightarrow \rm lim$' denotes the cases when the photons reach below the continuum threshold.}
\label{fig:tauc_HeI}
\end{figure}

\begin{figure}
\centering 
\includegraphics[width=0.99\columnwidth]
{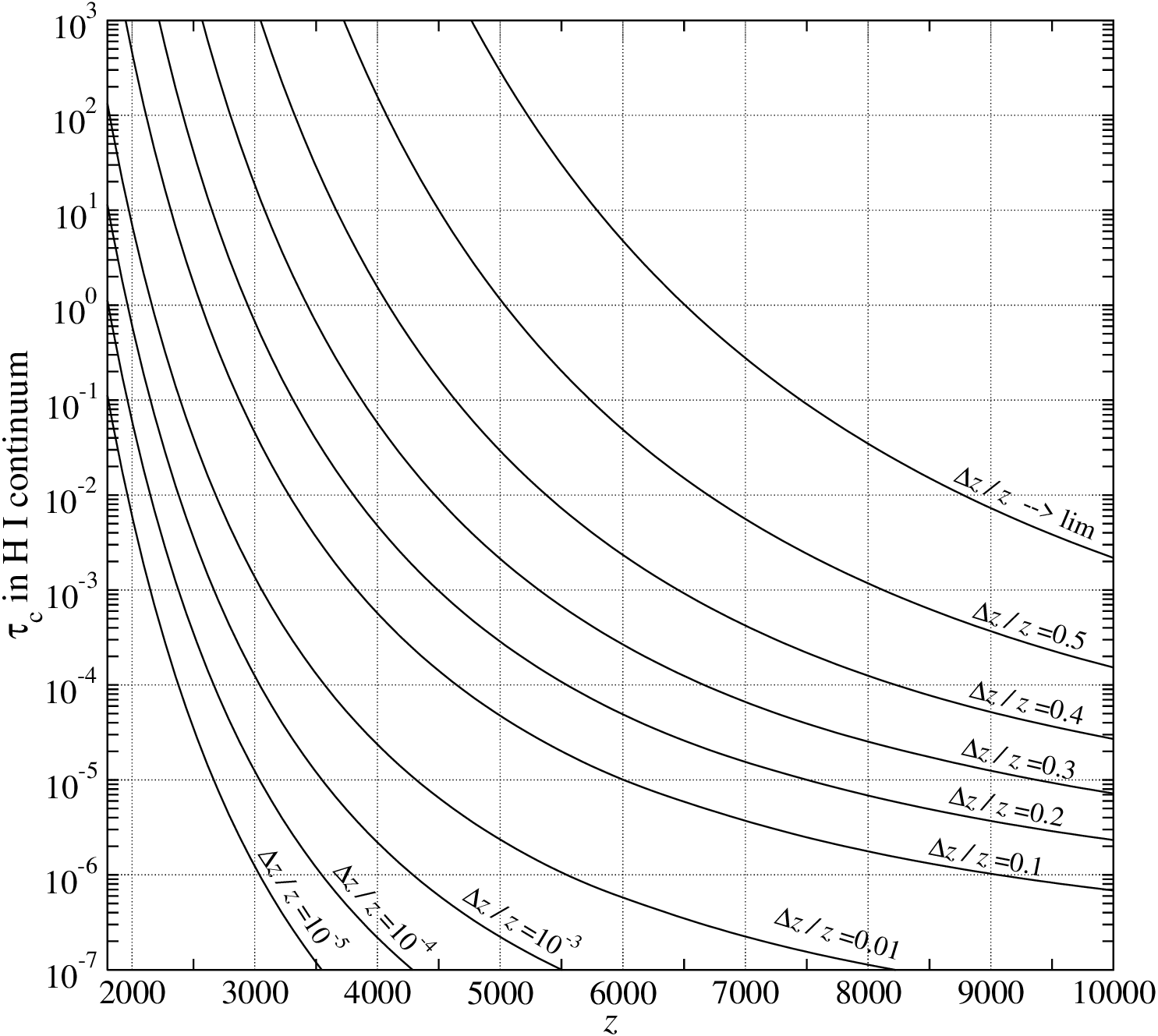}
\caption{Absorption optical depth in the $\ion{H}{i}$ 1s continuum for a \ion{He}{ii} Lyman $\alpha$ photon that was emitted at redshift $z$ and traveled until $z'=z-\Delta z$. At high redshifts ($z\gtrsim 4500$) we used the Saha solution for the $\ion{H}{i}$ 1s population and at low redshifts ($z\lesssim 4500$) the solution obtained with our implementation of {\sc Recfast}. The curve labeled '$\Delta z/z\rightarrow \rm lim$' denotes the cases when the photons reach below the continuum threshold.}
\label{fig:tauc_HI}
\end{figure}

%

\subsection{Possible speed-up of $\ion{He}{iii}\rightarrow\ion{He}{ii}$ recombination by \ion{He}{i}}
Like in the case of $\ion{He}{ii}\rightarrow\ion{He}{i}$ recombination, the absorption of \ion{He}{ii} photons by the small amount of \ion{He}{i} already present at $z\sim 6000$ could speed up $\ion{He}{iii}\rightarrow\ion{He}{ii}$ recombination ($5000\lesssim z \lesssim 7000$). 
However, there are several differences that make this process much less important than in the case of hydrogen.
Although the dependence of the  photon-ionization cross section for neutral helium on frequency is not as strong as in the case of hydrogen\footnote{In the energy range $\nu_{\rm c}^{\ion{He}{i}}\approx 24.587\,{\rm eV}\leq h\nu \lesssim 55\,{\rm eV}$ we found the fit $\sigma_{\rm 1s}^{\ion{He}{i}}\approx \pot{7.4}{-18}\,\text{cm}^{2}\,[\nu/\nu_{\rm c}^{\ion{He}{i}}]^{-1.77}$ to the data of \citet{Samson1994} which is accurate to $\pm 5\%$.} \citep{Samson1994}, the amount of helium is about $\sim 13$ times smaller than the amount of hydrogen.
Also during the $\ion{He}{iii}\rightarrow\ion{He}{ii}$ recombination \ion{He}{i} is in addition destroyed, because \ion{He}{ii} becomes ionized. This effect is also increased because  the ratio of the ionization potentials of neutral helium and singly ionized helium is a bit larger than for neutral helium and hydrogen.
This leads to an additional decrease in the amount of neutral helium close to the time of $\ion{He}{iii}\rightarrow\ion{He}{ii}$ recombination.

To understand the importance of the $\ion{He}{i}\;\HeIlevel{1}{1}{S}{0}$ continuum opacity we computed the the absorption optical depth for the \ion{He}{ii} Lyman $\alpha$ line for different $\Delta z/z$ between the emission redshift $z$ and the absorption redshift $z'=z-\Delta z<z$ (see results in Fig.~\ref{fig:tauc_HeI}).
For comparison, the Doppler width of the the \ion{He}{ii} Lyman $\alpha$ line is of the order $\Delta \nu_{\rm D}/\nu\sim \pot{2.8}{-5}\left[\frac{1+z}{6000}\right]^{1/2}$. 
From Fig.~\ref{fig:tauc_HeI} we can see that inside the \ion{He}{ii} Lyman $\alpha$ Doppler core the $\ion{He}{i}\;\HeIlevel{1}{1}{S}{0}$ continuum absorption should become important only at $z\sim 3000-3250$.
At that time practically all $\ion{He}{ii}$ recombined, so that there is no important speed-up of $\ion{He}{iii}\rightarrow\ion{He}{ii}$ recombination expected.

Comparing more carefully with the speed-up seen for \ion{He}{i}, from Figs.~\ref{fig:P.ST} and \ref{fig:tauc} it is clear that even $\tau^{\ion{H}{i}}_{\rm c}\sim 10^{-3}$ already leads to some notable effect.
However, in for the $\ion{He}{i}\;\HeIlevel{1}{1}{S}{0}$ continuum absorption inside the \ion{He}{ii} Lyman $\alpha$ Doppler core this is reached only at $z\sim 4000-4300$, again well after $\ion{He}{ii}$ recombined.
In addition, $\ion{He}{iii}\rightarrow\ion{He}{ii}$ recombination already occurs very close to the Saha case, so that the shape and width of the spectral lines cannot be altered much in addition. 
We therefore believe that this process is negligible in this context.

Finally, also the small amount of neutral hydrogen present already during  $\ion{He}{iii}\rightarrow\ion{He}{ii}$ recombination cannot change anything about this conclusion, since inside the \ion{He}{ii} Lyman $\alpha$ Doppler core the \ion{H}{i} continuum optical depth it completely negligible (cf. Fig.~\ref{fig:tauc_HI}, curve for $\Delta z/z\sim 10^{-5}$).

\subsection{Electron scattering during $\ion{He}{iii}\rightarrow\ion{He}{ii}$ recombination}
\label{sec:escat}
Although \changeII{in terms of the ionization history} during hydrogen and neutral helium recombination electron scattering is not very important, for the  shape, position and width of the \ion{He}{ii} distortions its effect is already notable \citep{Jose2008}.
This should affect the time-dependence of the feedback caused by high frequency \ion{He}{ii} photons on \ion{He}{i}, since, \changeII{for example}, electron recoil will lead to some important drift of the lines towards lower frequencies, increasing the effective rate of redshifting. \changeII{Also the Doppler broadening of the lines will affect the duration of the feedback.}
Furthermore, one also expects that electron scattering could lead to an significant increase in the $\ion{He}{iii}\rightarrow\ion{He}{ii}$ recombination rate.
For the ionization history and computations of the CMB power spectra this again will not be important, but in computations of the CMB spectral distortions introduced during recombination this process will matter.
Also one should include the CMB induced stimulated electron scattering process \citep{Chluba2008d} for accurate computations of the low frequency spectral distortions.

\begin{figure}
\centering 
\includegraphics[width=0.99\columnwidth]
{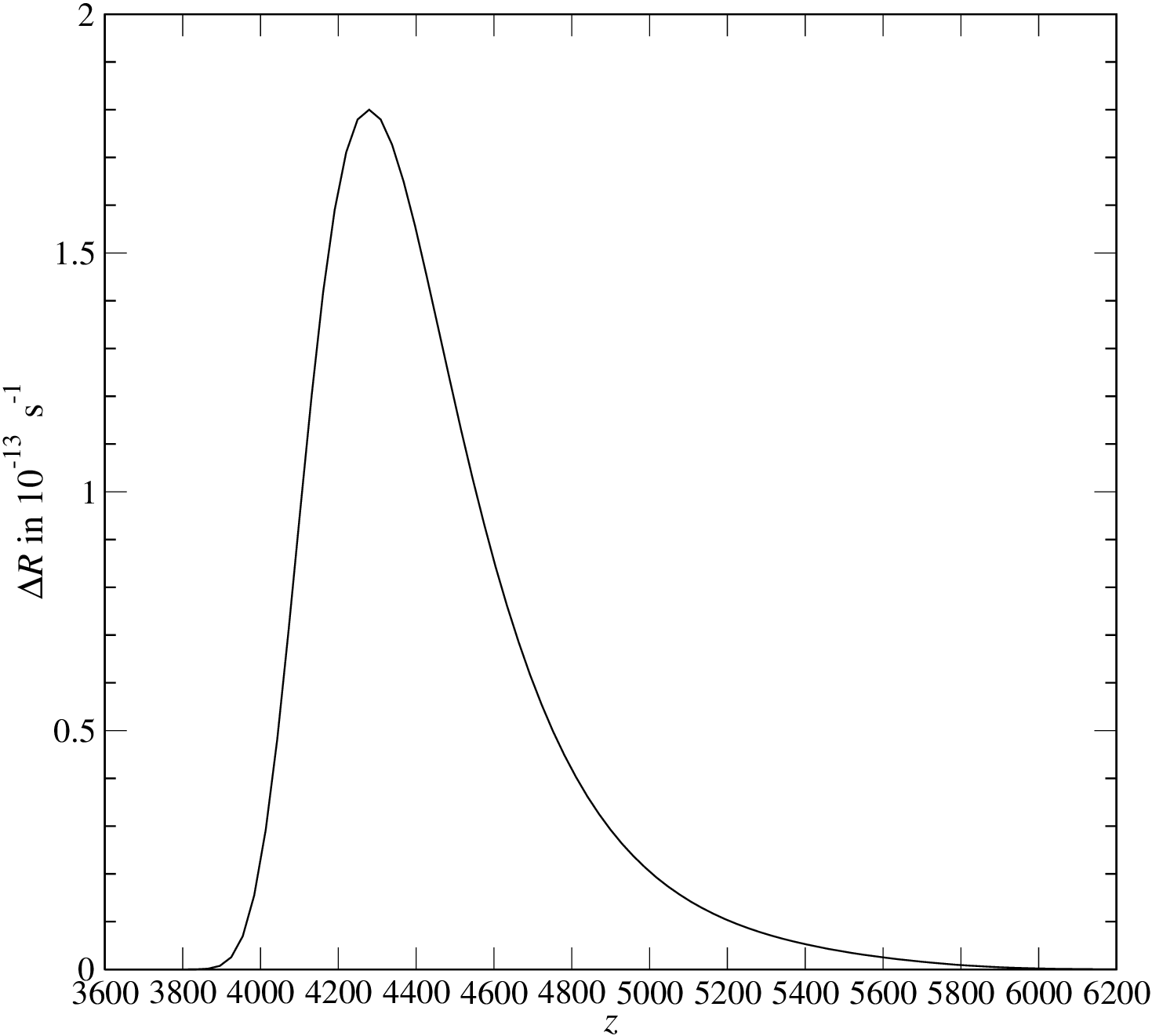}
\caption{Feedback induced correction to the $\ion{He}{i}\;\HeIlevel{1}{1}{S}{0}$ photoionization rate. We only included the effect of the  \ion{He}{ii} Lyman $\alpha$ line in our computations using the pre-computed results from our 100 shell calculations.}
\label{fig:DR_HeI}
\end{figure}
\subsection{Time of $\ion{He}{ii}$ Lyman $\alpha$ feedback on \ion{He}{i}}
\label{sec:time-HeII-feed}
From Fig.~\ref{fig:tauc_HeI} one can understand when most of the \ion{He}{ii} Lyman $\alpha$ feedback is expected to occur. 
Since most of these photons were released at $z\sim 6000$, we can see from Fig.~\ref{fig:tauc_HeI} that they will disappear in the $\ion{He}{i}\;\HeIlevel{1}{1}{S}{0}$ continuum until $\Delta z/z \sim 0.3$, or $z'\sim 4200$.
This is well after the recombination epoch of  \ion{He}{ii}  and still well before the normal recombination epoch of  \ion{He}{i}. Therefore the feedback is most strong in the {\it pre-recombinational} epoch of neutral helium.

We computed the strength of this feedback in more detail using the result for the \ion{He}{ii} Lyman $\alpha$ line from our 100 shell computation.
The resulting correction to the $\ion{He}{i}\;\HeIlevel{1}{1}{S}{0}$ photoionization rate is shown in Fig.~\ref{fig:DR_HeI}. As one can see the feedback is largest at $z\sim 4280$. The width of the feedback curve is about $10\%-15\%$, so that one can expect some rather narrow pre-recombinational, feedback induced lines in the \ion{He}{i} recombination spectrum.

Looking again at Fig.~\ref{fig:tauc_HeI} one can also see that practically all \ion{He}{ii} Lyman $\alpha$ photons will never reach frequencies below the $\ion{He}{i}\;\HeIlevel{1}{1}{S}{0}$ continuum threshold.
This is because practically no photons are released at $z\gtrsim 7000$, and there the maximal optical depth (corresponding to the case $\Delta z/z \rightarrow \rm lim$) exceeds unity already.

One should also mention that the a small fraction of \ion{He}{ii} Lyman $\alpha$ photons is expected to directly disappear in the \ion{H}{i} Lyman continuum (see Fig.~\ref{fig:tauc_HI}), but this should have a negligible effect on the recombination spectrum.

\begin{figure}
\centering 
\includegraphics[width=0.99\columnwidth]
{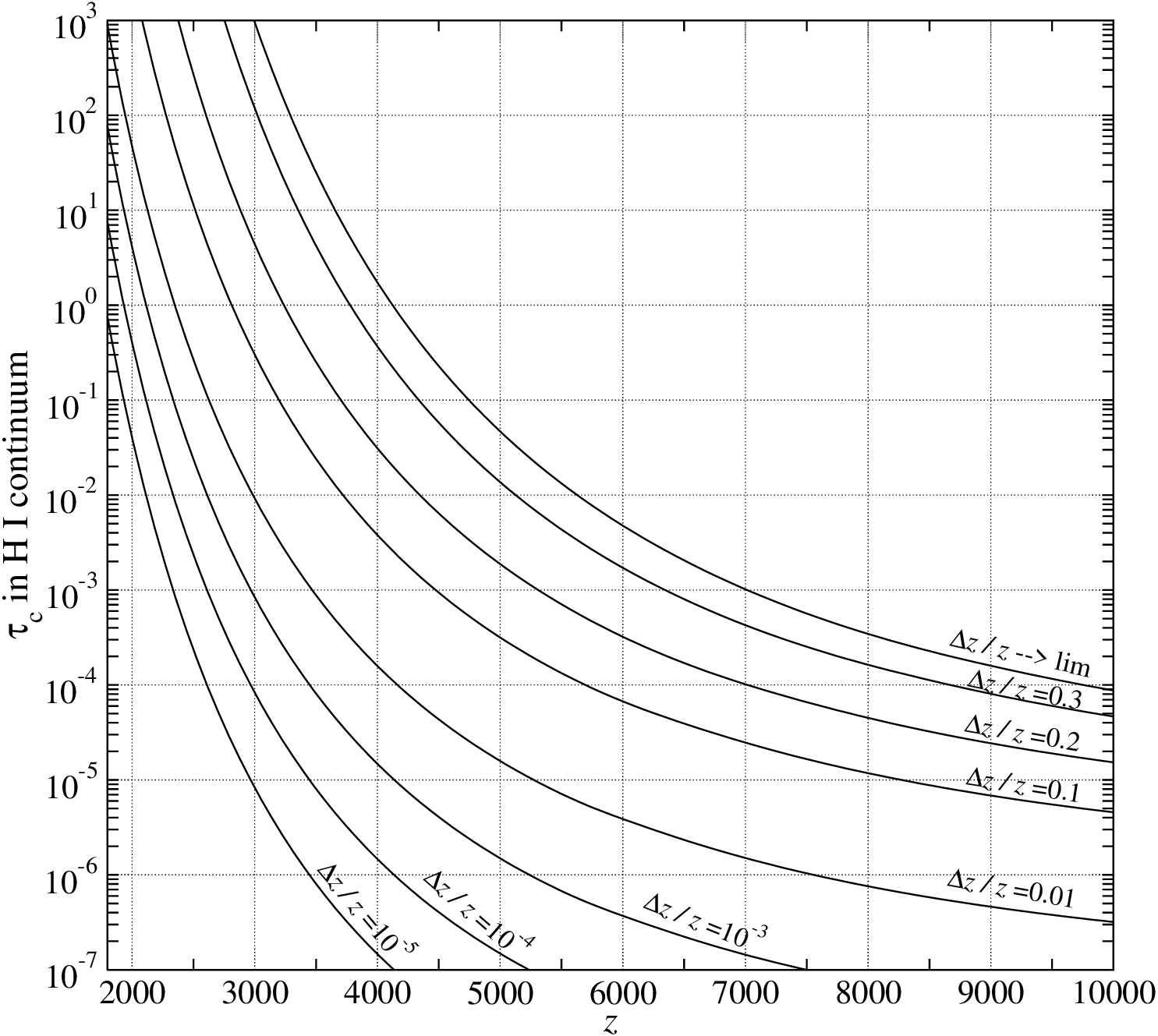}
\caption{Absorption optical depth in the $\ion{H}{i}$ 1s continuum for a $\ion{He}{i}\;\HeIlevel{2}{3}{P}{1}-\HeIlevel{1}{1}{S}{0}$ photon that was emitted at redshift $z$ and traveled until $z'=z-\Delta z$. At high redshifts ($z\gtrsim 4500$) we used the Saha solution for the $\ion{H}{i}$ 1s population and at low redshifts ($z\lesssim 4500$) the solution obtained with our implementation of {\sc Recfast}. The curve labeled '$\Delta z/z\rightarrow \rm lim$' denotes the cases when the photons reach below the continuum threshold.}
\label{fig:tauc_HI_T}
\end{figure}

\subsection{Reprocessing of pre-recombinational $\ion{He}{i}$ photons}
\label{sec:repro_HeI}
As explained above (Sect.~\ref{sec:double}), every \ion{He}{ii} photon that was reprocessed by \ion{He}{i} will eventually be replaced by some photon in resonances connecting to the $\ion{He}{i}\;\HeIlevel{1}{1}{S}{0}$ state. Since at $z\gtrsim 2500$ the \ion{H}{i} continuum opacity between these \ion{He}{i} resonances is not important (cf. Fig.\ref{fig:tauc}), it is clear that for these secondary \ion{He}{i} photons one only has to look at the possible feedback process from the $\ion{He}{i}\;\HeIlevel{2}{3}{P}{1}-\HeIlevel{1}{1}{S}{0}$ intercombination line.

In Fig.~\ref{fig:tauc_HI_T} we present the optical depth in the \ion{H}{i} Lyman continuum for photon emitted at the different redshift in the $\ion{He}{i}\;\HeIlevel{2}{3}{P}{1}-\HeIlevel{1}{1}{S}{0}$ line. Again we consider several values for the considered redshift interval that the photon traveled. It is clear that photons emitted at $z\lesssim 3700-4000$ will all be absorbed in the \ion{H}{i} Lyman continuum before they can reach below the threshold frequency (cf. curve labeled '$\Delta z/z\rightarrow \rm lim$').
However, most of the secondary \ion{He}{i} photons emitted at $z\gtrsim 4000$ will in fact survive the absorption in the  \ion{H}{i} Lyman continuum and instead die in the \ion{H}{i} Lyman series at lower redshift. This therefore leads to a very extended feedback processes of secondary \ion{He}{i} photons on hydrogen. 
As we will see below (Sect.~\ref{sec:spec_HeII}) the associated spectral distortion at high frequencies indeed is rather broad, reflecting this fact.

\begin{figure}
\centering 
\includegraphics[width=0.99\columnwidth]{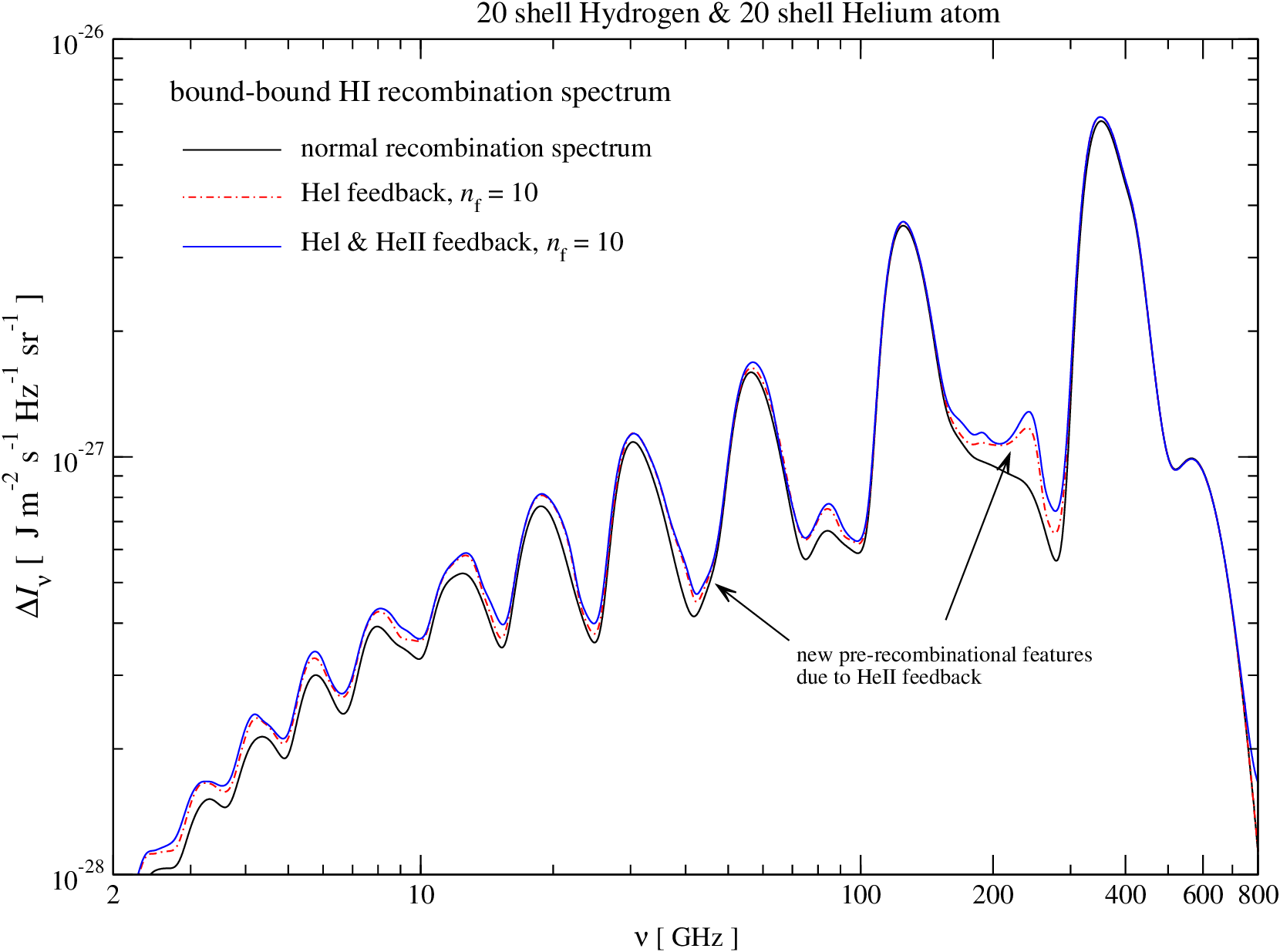}
\\[1mm]
\includegraphics[width=0.99\columnwidth]{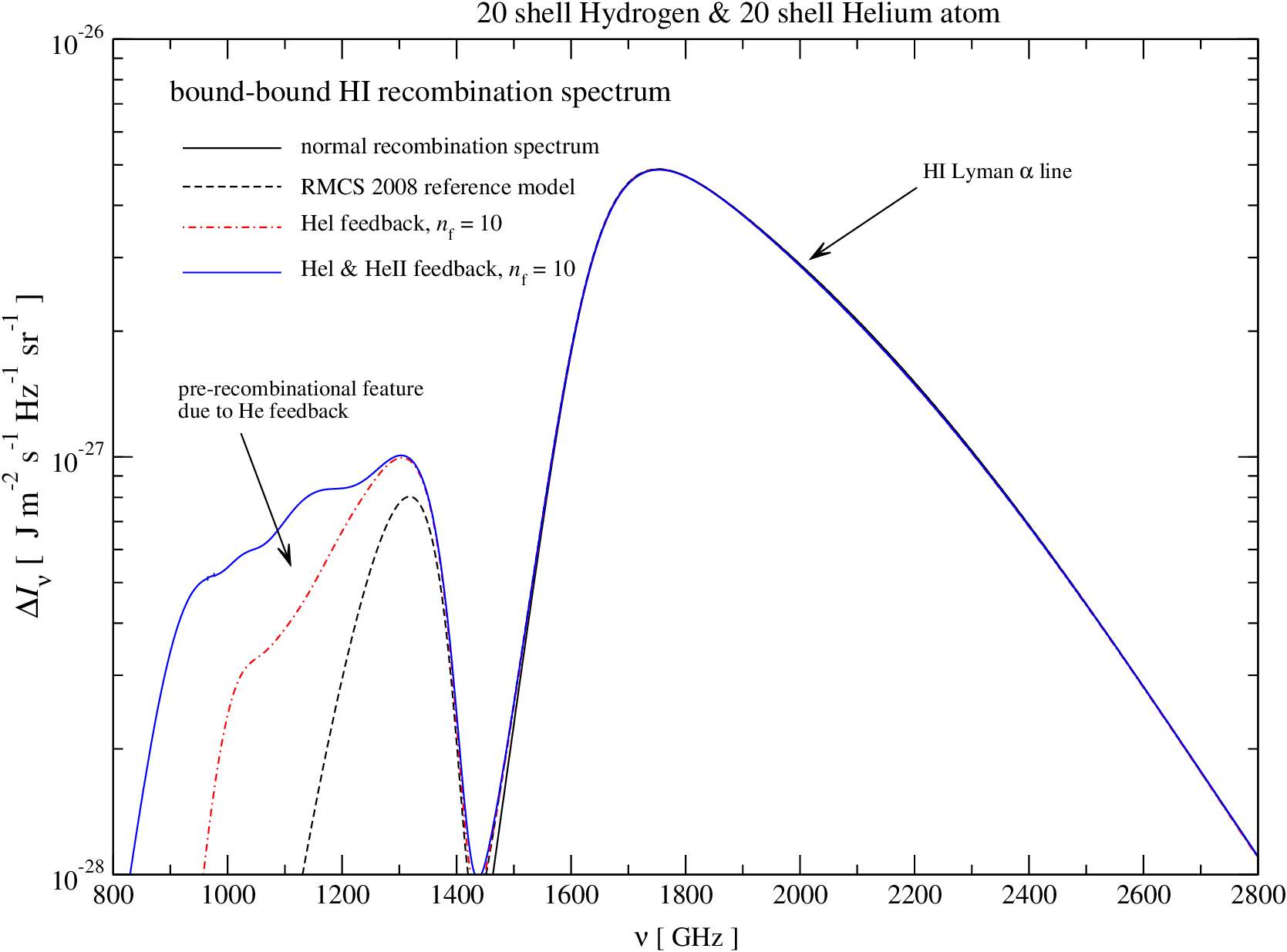}
\\[1mm]
\includegraphics[width=0.99\columnwidth]{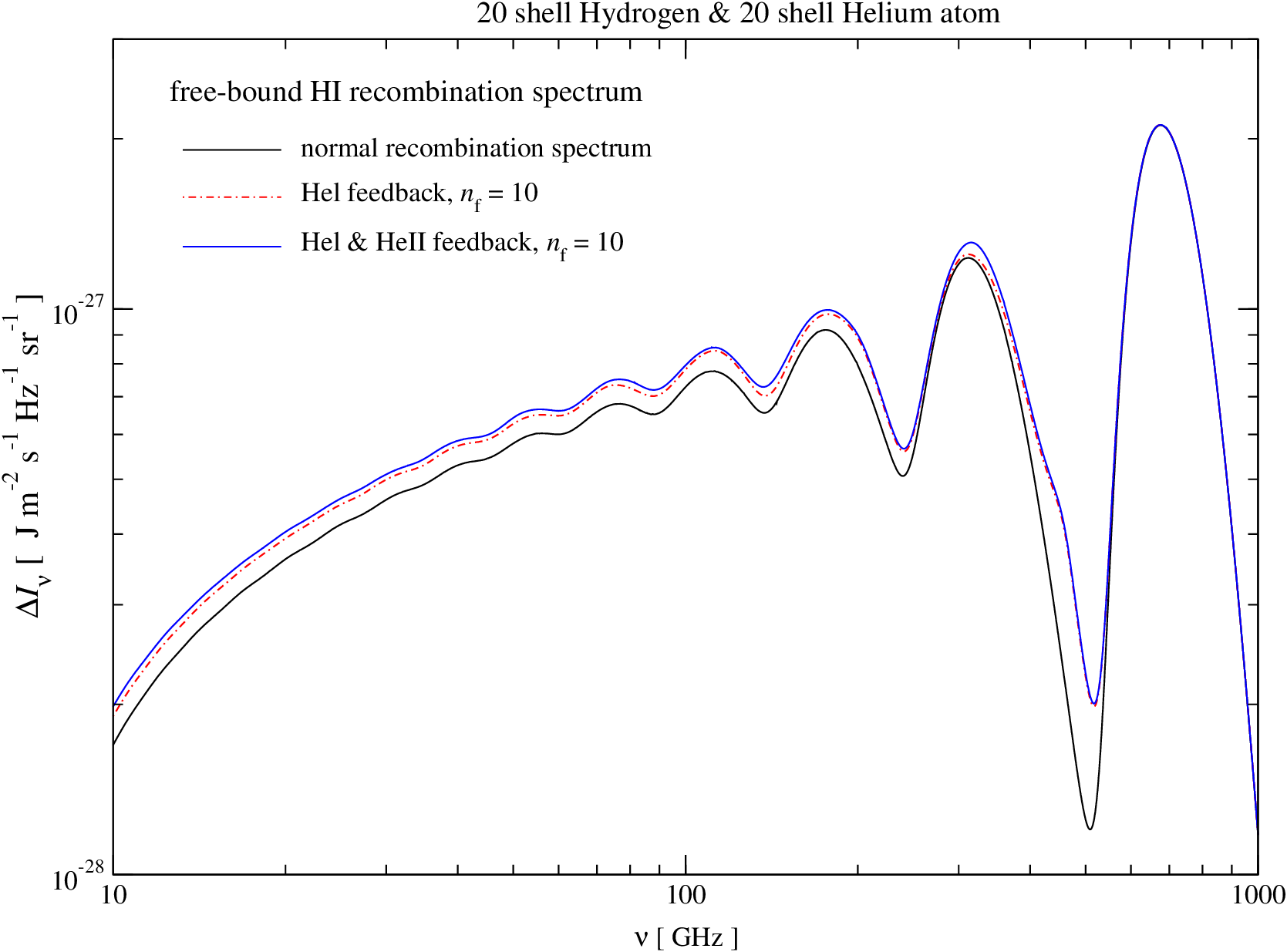}
\caption{Contributions to the cosmological recombination spectrum from hydrogen. We show the low frequency \ion{H}{i} bound-bound spectrum (upper panel), the  high frequency \ion{H}{i} bound-bound spectrum (middle panel), and the  \ion{H}{i} free-bound spectrum (lower panel). We included 20 shells of hydrogen and 20 shells of helium in our computations and allowed for helium feedback from the {\it first ten} shells. The curves marked with 'RMCS 2008' where computed using our reference model \citep{Jose2008}. Those marked as 'normal recombination spectrum' do not include the effects of the hydrogen continuum opacity on the dynamics of helium recombination. For the cases with \ion{He}{ii} feedback we only included the effect of the \ion{He}{ii} Lyman $\alpha$ line and allowed for escape in the \ion{H}{i} and \ion{He}{i} continua.}
\label{fig:DI.HI.20}
\end{figure}

\subsection{Changes in the recombination spectrum due to \ion{He}{ii} Lyman $\alpha$ feedback}
\label{sec:spec_HeII}
\begin{figure}
\centering 
\includegraphics[width=0.99\columnwidth]
{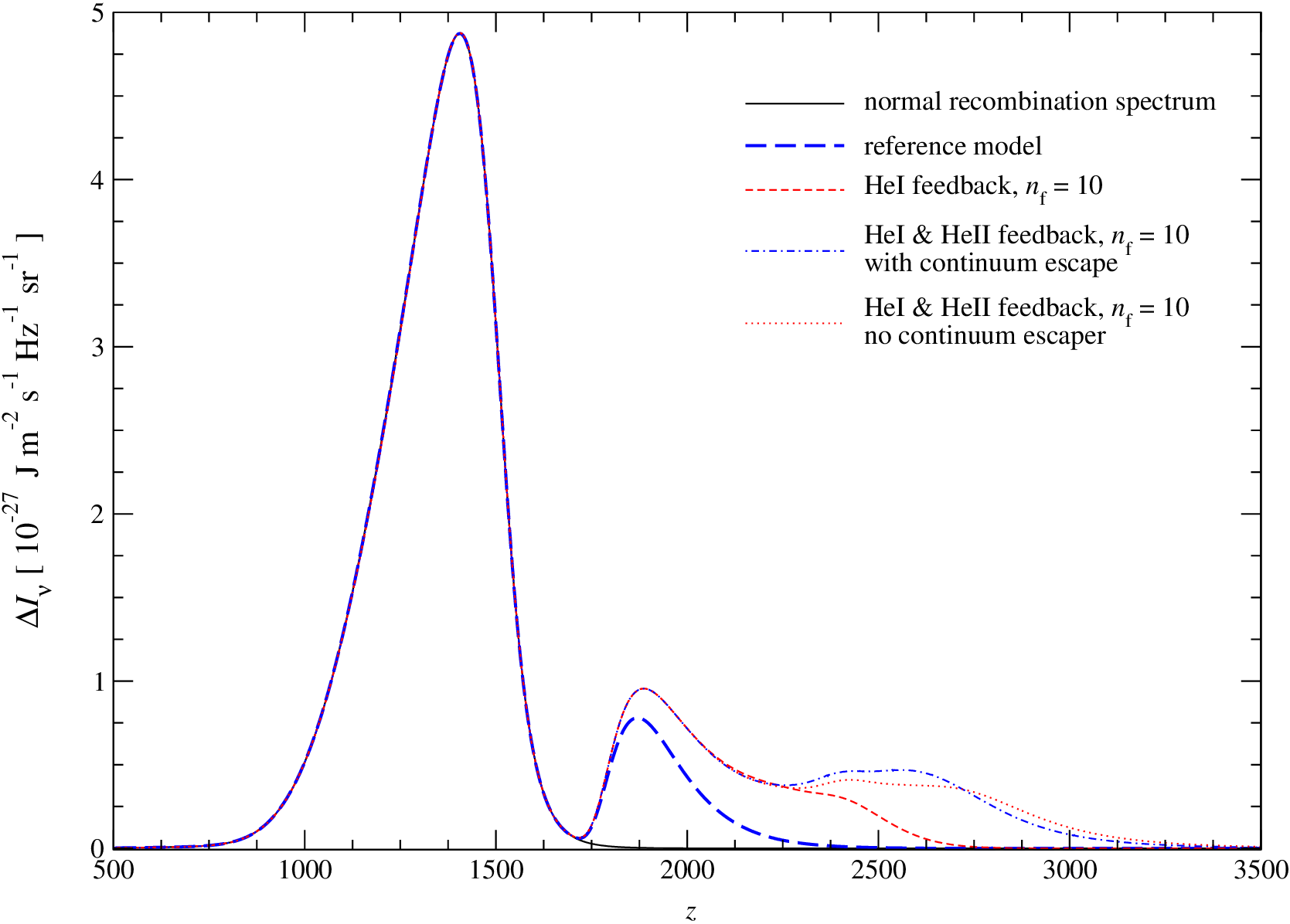}
\\[2mm]
\includegraphics[width=0.99\columnwidth]
{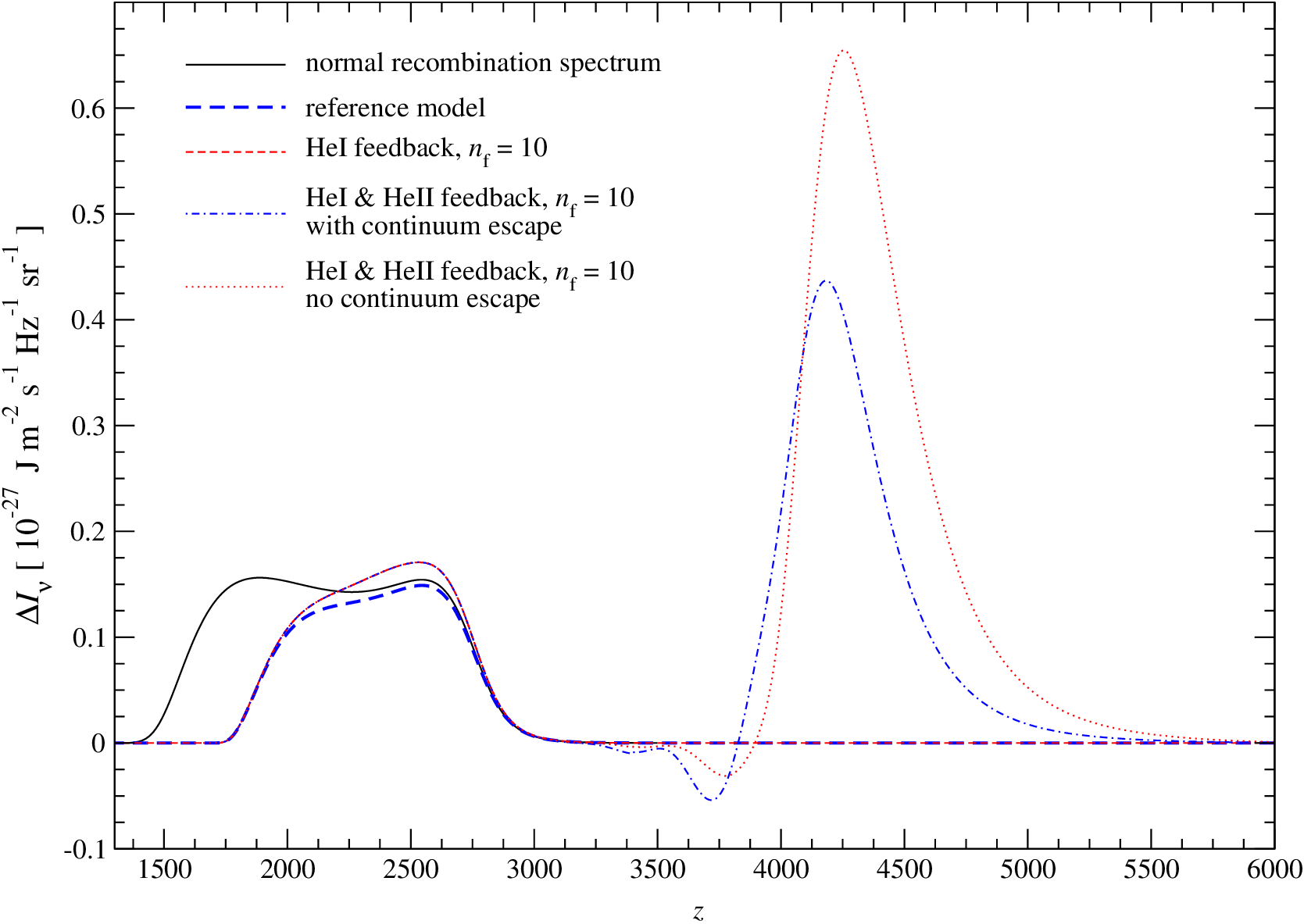}
\caption{Feedback induced correction to the \ion{H}{i} Lyman $\alpha$ line (upper panel) and $\ion{He}{i}\;\HeIlevel{2}{1}{P}{1}-\HeIlevel{1}{1}{S}{0}$ line (lower panel). We only included the effect of the  \ion{He}{ii} Lyman $\alpha$ line in our computations. We show the spectral distortion $\Delta I_\nu$ as a function of the emission redshift, which can be computed with the relation $\nu(z=0)=\nu_i/[1+z_{\rm em}]$. In both cases we subtracted the phase space density on the far blue side of the resonances. Furthermore, we show the cases with and without allowing photon escape in the \ion{H}{i} and \ion{He}{i} continua.}
\label{fig:DI_HI_Lya_HeI}
\end{figure}

\begin{figure}
\centering 
\includegraphics[width=0.99\columnwidth]{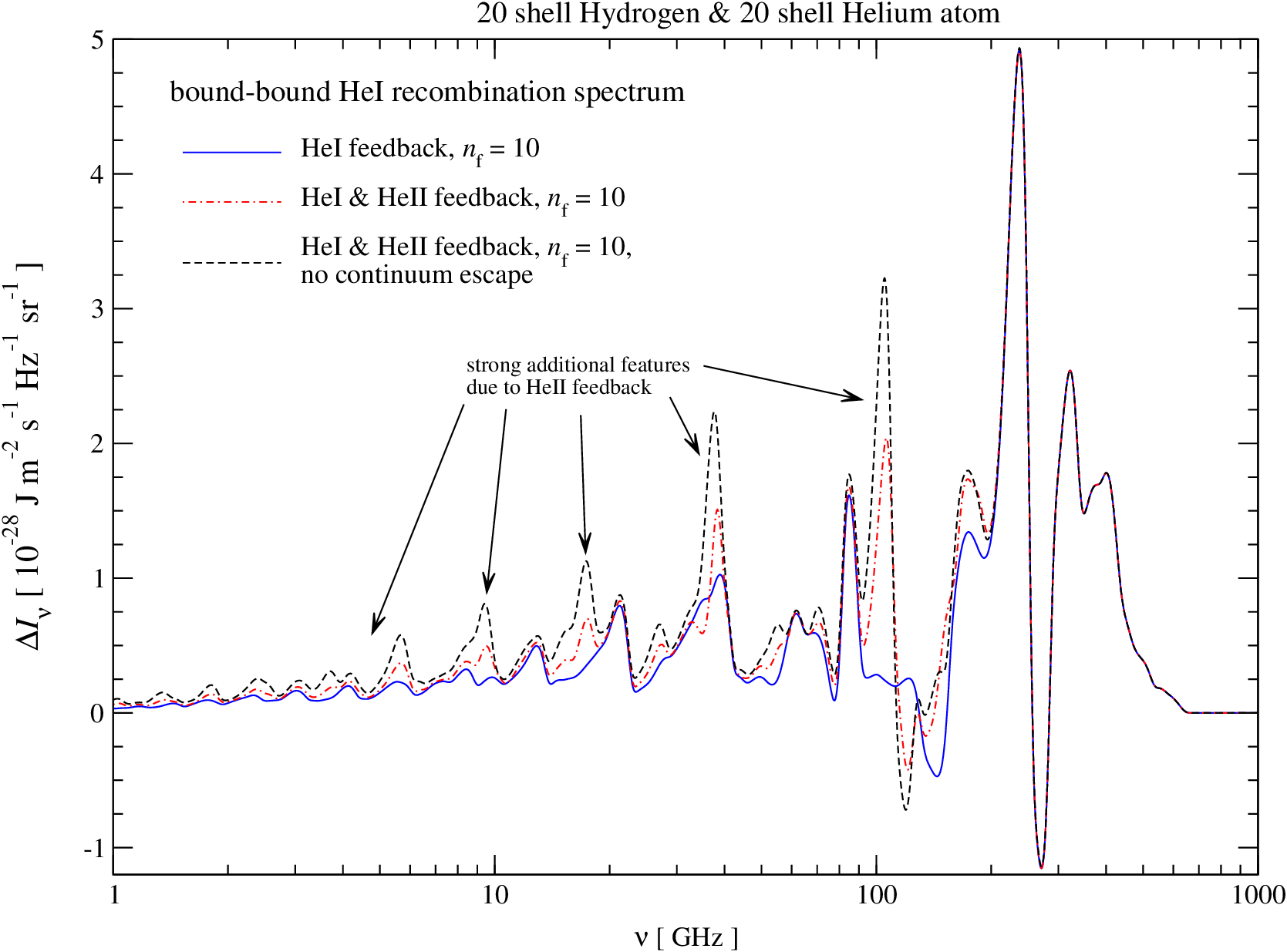}
\caption{Contributions to the cosmological recombination spectrum from neutral helium. We show \ion{He}{i} bound-bound spectrum. We included 20 shells of hydrogen and 20 shells of helium in our computations and allowed for helium feedback from the {\it first ten} shells. 
%
%
For the cases with \ion{He}{ii} feedback we only included the effect of the \ion{He}{ii} Lyman $\alpha$ line.}
\label{fig:DI.HeI.20}
\end{figure}

In this Section we want to {\it demonstrate} the effect of \ion{He}{ii} feedback on the cosmological recombination spectrum. 
Taking the whole list of processes mentioned above into account is far beyond the scope of this paper, however, it is rather straightforward to account for the feedback effect from the \ion{He}{ii} Lyman $\alpha$ line, using pre-computed results for the changes in the $\ion{He}{i}\;\HeIlevel{1}{1}{S}{0}$ photoionization rate (see Fig.~\ref{fig:DR_HeI} and Sect.~\ref{sec:time-HeII-feed}).
The main results of these computation are shown in Fig.~\ref{fig:DI.HI.20}-\ref{fig:DI.total}.
The most important aspect is that there are several additional, rather {\it narrow spectral features} in the \ion{He}{i} bound-bound spectrum (cf. Fig.~\ref{fig:DI.HeI.20}), which locally increase the contribution coming from helium significantly (see Fig.~\ref{fig:DI.total}). This may render a determination of the {\it primordial} helium abundance using the cosmological recombination spectrum \citep[e.g. see][]{Chluba2008T0, Sunyaev2009} a bit easier.

\subsubsection{Changes in the \ion{H}{i} recombination spectrum due to \ion{He}{ii} Lyman $\alpha$ feedback}
\label{sec:spec_HI_HeII}
As in the case of \ion{He}{i} feedback on hydrogen, one can see \changeII{(see Fig.~\ref{fig:DI.HI.20})} that the feedback of quanta related to \ion{He}{ii} leads to the production of several additional photons in the recombination spectrum of hydrogen. 
In particular, the pre-recombinational feature at high frequencies (see Fig.~\ref{fig:DI.HI.20}, middle panel) changes significantly, while a low frequencies, the additions are generally smaller (see Fig.~\ref{fig:DI.HI.20}, upper panel). Only at frequencies around $\nu\sim 200\,$GHz, one can see significant changes.
As explained above (Sect.~\ref{sec:time-HeII-feed}), \ion{He}{ii} Lyman $\alpha$ photons do not directly feed back on hydrogen, since these quanta are all absorbed in the $\ion{He}{i}\,\HeIlevel{1}{1}{S}{0}$ continuum before. 
Furthermore, most of the secondary \ion{He}{i} high frequency photons feed back on the \ion{H}{i} Lyman series (Sect.~\ref{sec:repro_HeI}), so that, in agreement with our computations, the low frequency photon production efficiency is smaller (see Sect.~\ref{sec:direct}).
The free-bound spectral distortion (see Fig.~\ref{fig:DI.HI.20}, lower panel) like in the case of \ion{He}{i} feedback (e.g. see Fig.~\ref{fig:DI}) again shows practically no additional features, but only a small additional increase in the total amplitude. 
%

In Fig.~\ref{fig:DI_HI_Lya_HeI}, as an example, we also demonstrate the changes in the \ion{H}{i} Lyman $\alpha$ line as a function of redshift. In addition to the cases shown in the previous figures, we also ran a case for which we did not include the escape of photons in the \ion{H}{i} Lyman and $\ion{He}{i}\,\HeIlevel{1}{1}{S}{0}$ continuum.
Although for the distortions from hydrogen this does not make a large difference, for the distortions from helium it is very important (see next Section).

\subsubsection{Changes in the \ion{He}{i} recombination spectrum due to \ion{He}{ii} Lyman $\alpha$ feedback}
\label{sec:spec_HeI_HeII}
In Fig.~\ref{fig:DI.HeI.20} we show the \ion{He}{i} bound-bound recombination spectrum for different cases of the feedback process.
One can clearly see that the feedback of \ion{He}{ii} Lyman $\alpha$ photons affects the \ion{He}{i} emission very strongly, leading to several new features at practically all frequencies.
It is also clear that the escape in the $\ion{He}{i}\,\HeIlevel{1}{1}{S}{0}$ continuum is very important for the correct shape and strength of the pre-recombinational features.
Without the inclusion of $\ion{He}{i}\,\HeIlevel{1}{1}{S}{0}$ continuum escape the pre-recombinational features are overestimated by about a factor of 2.
This again is due to the fact that every feedback photon that can escape the $\ion{He}{i}\,\HeIlevel{1}{1}{S}{0}$ continuum will subsequently feed back on bound-bound transitions connecting to the ground state of helium, for which the low frequency photon production rate is expected to be smaller (see explanation in Sect.~\ref{sec:direct}).

As an example, in Fig.~\ref{fig:DI_HI_Lya_HeI} we give the distortion due to the $\ion{He}{i}\,\HeIlevel{2}{1}{P}{1}-\HeIlevel{1}{1}{S}{0}$ resonance only.  
One can see that the maximum of the pre-recombinational feedback is reached at $z\sim 4200$, corresponding to the maximum in the correction to the $\ion{He}{i}\;\HeIlevel{1}{1}{S}{0}$ photoionization rate (see Fig.~\ref{fig:DR_HeI}).
Figure~\ref{fig:DI_HI_Lya_HeI} also shows that the pre-recombinational feature from a given resonance is expected \changeII{to have a larger} amplitude than the recombinational feature itself, even though the total emission remains comparable. This is because the feedback process occurs over a rather short time interval, owing to the narrowness of the \ion{He}{ii} Lyman $\alpha$ line. As explained in Sect.~\ref{sec:escat}, the inclusion of electron scattering into the problem may alter this aspect of the problem.
In addition, the number of ionizing \ion{He}{ii} photons per \ion{He}{i} atom is comparable to unity, so that for full feedback in the $\ion{He}{i}\,\HeIlevel{1}{1}{S}{0}$ continuum the number of \ion{He}{i} quanta would practically be doubled. However, as mentioned above the efficiency of low frequency photon production is decreased due to the partial escape in the $\ion{He}{i}\,\HeIlevel{1}{1}{S}{0}$ continuum.

\begin{figure}
\centering 
\includegraphics[width=0.99\columnwidth]{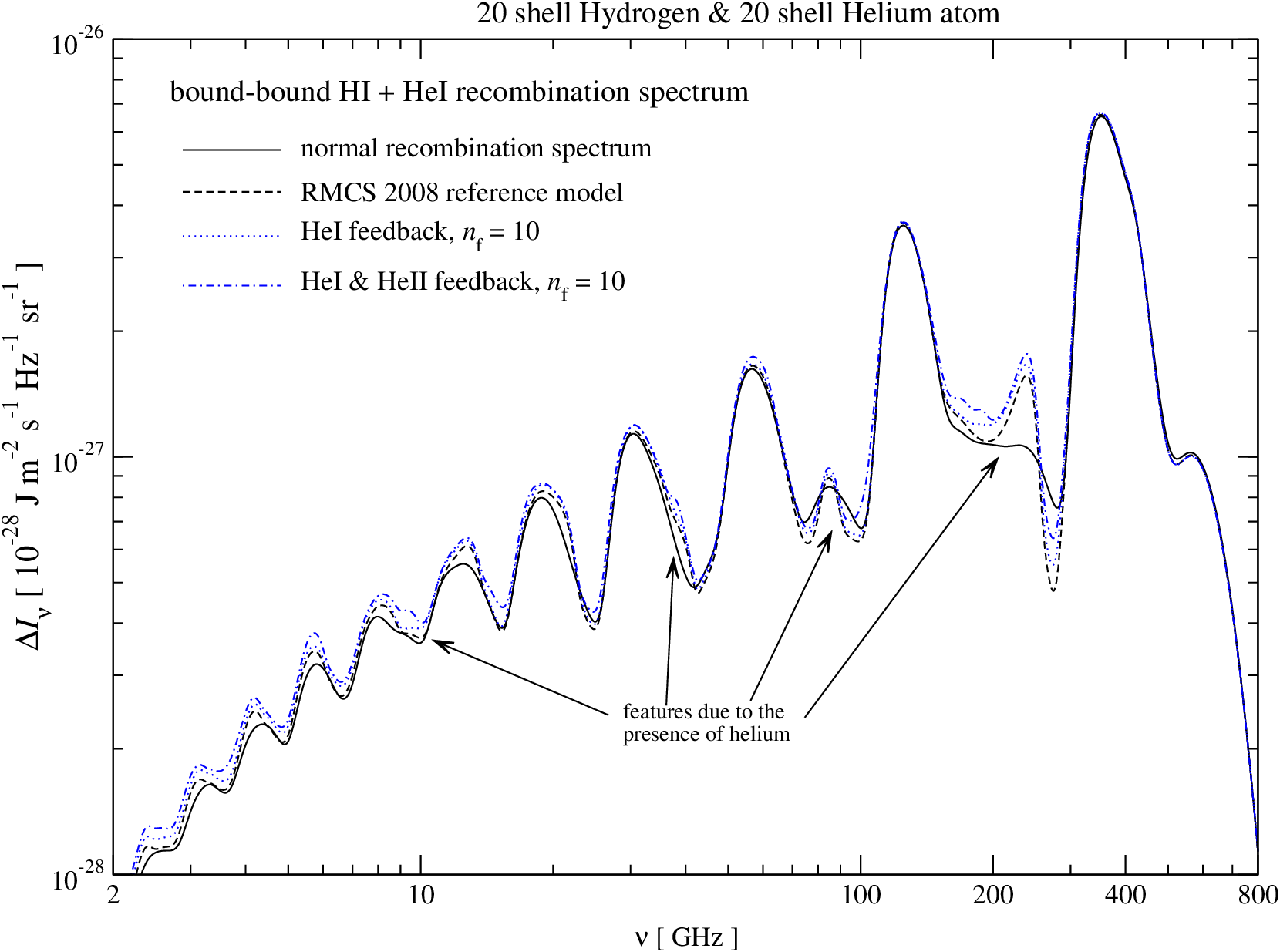}
\caption{Total bound-bound recombination spectrum from \ion{H}{i} and \ion{He}{i} at low frequencies. We only included the spectral distortions due to dipole transitions.}
\label{fig:DI.total}
\end{figure}
\subsubsection{Changes in the total \ion{H}{i}  and \ion{He}{i} bound-bound recombination spectrum}
\label{sec:spec_HI_HeI_HeII}
Finally in Fig.~\ref{fig:DI.total} we present the sum of the \ion{H}{i} and \ion{He}{i} bound-bound recombination spectrum. We only included the contributions due to dipole transitions, but neglected those from the free-bound and two-photon continua, since they are not leading to some variable signal and hence are not as interesting from an observational point of view.
One can clearly see that the feedback of helium photons leads to {\it non-trival} modifications in the recombinational radiation, with the strongest helium-related features at $\nu\sim 200 \,$GHz, $\sim 80\,$GHz, $\sim 35\,$GHz, and $\sim 10\,$GHz. However, details in the spectrum appear at practically {\it all frequencies}, and in particular the variability of the spectral distortion is altered, with significant changes in the amplitude, positions of peaks and new features appearing. 
All these traces in the recombination spectrum are only due to the presence of helium in the early Universe, and therefore may render it possible to determine the primordial helium abundance using the spectrum of the CMB. Most interestingly, the amount of helium-related photons is increased when accounting for all the feedback processes. This should make it slightly easier to use the spectrum of the CMB for measurements of the helium abundance.

\begin{table}
\caption{Total number (density) of photons, $N_{\gamma}$, produced by different species. Here $N_{\rm H}$ and $N_{\rm He}$ denote the total number (density) of hydrogen and helium nuclei. Contributions marked with 'rec' are from the normal recombinational epoch. Those marked with 'pre-rec I' are the additional photons computed for our reference model \citep{Jose2008}, while those marked with 'pre-rec II' give the additional photons when accounting for the full \ion{He}{i} feedback.
Those marked with 'pre-rec III' denote photons that are produced when including \ion{He}{ii} Lyman $\alpha$ feedback.
We also give the numbers for the whole bound-bound and free-bound spectrum, as well as the total difference (bound-bound $+$ free-bound) in the number of photons. We included 20 shells for \ion{H}{i} and 20 shells for \ion{He}{i} in the computations. For \ion{He}{ii} we used the results for 100 shells.}
\label{tab:Ng_HeII}
\centering
\begin{tabular}{@{}lrrrrr}
\hline
\hline
Atom & $N_{\gamma}/N_{\rm H}$ & $N_{\gamma}/N_{\rm He}$ \\
\hline
\ion{H}{i} (rec) & 4.51 &  -- \\
\ion{H}{i} (pre-rec I) & 0.12 & 1.5 \\
\ion{H}{i} (pre-rec II) & 0.12 &  1.5 \\
\ion{H}{i} (pre-rec III) & 0.1 &  $1.3^\ast$ \\
\ion{H}{i} (all) & 4.85 & -- \\
\hline
\hline
\ion{He}{i} (rec) & 0.29 &  3.69 \\
\ion{He}{i} (pre-rec I) & -0.06 & $-0.73^\dagger$ \\
\ion{He}{i} (pre-rec II) & -0.04 &  -0.45 \\
\ion{He}{i} (pre-rec III) & 0.04 &  $0.53^\ddagger$ \\
\ion{He}{i} (all) & 0.24 & 3.04 \\
\hline
\hline
\ion{He}{ii} (rec) & 0.37 &  4.69 \\
\ion{He}{ii} (pre-rec III) & -0.04 &  $-0.55^{\star}$ \\
\ion{He}{ii} (all) & 0.33 & 4.14 \\
\hline
\hline
total number  (rec)& 5.17 &  -- \\
total difference (pre-rec I)& 0.06 &  0.77 \\
total difference (pre-rec II)& 0.08 &  1.1 \\
total difference (pre-rec III)& 0.10 &  1.28 \\
total number (all) & 5.42 &  -- \\
\hline
\hline
total difference (pre-rec I-III)& 0.25 &  3.2 \\
\hline
\hline
\end{tabular}
\begin{minipage}{\columnwidth}
\vspace{1mm}
{\footnotesize
$^\ast$ This number includes all surviving high frequency photons from \ion{He}{i} and those from the \ion{H}{i} Lyman continuum.

$^\dagger$ About $-0.45$ is due to the removal of high frequency photons in the \ion{H}{i} Lyman continuum. The rest is due to changes in the low frequency photon production as a result of the speed-up of \ion{He}{i} recombination.

$^{\ddagger}$ This number is only an estimate, since some of the $0.55$ ionizing \ion{He}{i} photons directly escaped in the $\ion{He}{i}\,\HeIlevel{1}{1}{S}{0}$ continuum. We accounted for these photons in the computations, but did not check their number exactly. The difference in the \ion{He}{i} bound-bound spectrum was $\sim 0.25\,\gamma$ per helium nucleus. To this we added $0.55\times 50\%\sim 0.28$ of high-$n$ free bound transitions resulting in the quoted number.

$^{\star}$ We assumed that only the \ion{He}{ii} Lyman $\alpha$ line is re-processed.}
\end{minipage}
\end{table}
\subsection{Estimates for the total number of feedback induced photons from helium and hydrogen}
\label{sec:number_HeII}
Using the spectral distortions \changeI{discussed} in the previous section, one can now compute the total number of photons emitted by the different atoms.
In Table~\ref{tab:Ng_HeII} we summarize the results of these computations.
In total one finds about $3.2$ additional photons per helium atom when including all the feedback processes.
This corresponds to an increase in the total number of helium-related photons by $\sim 37\%$ in comparison to the standard calculation which assumes independent recombination histories. 
In comparison to the reference model \citep{Jose2008}, the net gain is about $2.38$ additional photons per helium atom, so that the feedback induced number of photons increased by a factor of $3.2/0.77\sim 4.2$.

Furthermore, per helium atom one has a total of $\sim 11.5\,\gamma$ related to the presence of helium in the Universe, or about $0.91\,\gamma$ per hydrogen atom, resulting in a total contribution of helium-related photons to the recombination spectrum of $\sim 17\%$.
In the standard calculation which assumes independent recombination histories helium contributes about $13\%$. 
However, here one should mention that this average increase in the helium-related number of photons can be exceeded several times at particular frequencies (see Fig.~\ref{fig:DI.total} and Sect.~\ref{sec:spec_HI_HeI_HeII}).

%

%
Since for our computations in total $0.9+0.55\sim 1.45$ quanta per helium atom were re-processed, the above results also imply that per feedback photon about $3.2/1.45 \sim 2.2$ additional photons were produced.
The production only due to primary \ion{He}{i} photons is about\footnote{Note that here we also included the small reduction (by $\sim 0.3\,\gamma$ per helium nucleus) of the total number of \ion{He}{i} photons, which is due to the changes in the dynamics of helium recombination caused by the \ion{H}{i} continuum opacity. As explained in Sect.~\ref{sec:HeI_total_num} per primary \ion{He}{i} feedback photon \ion{H}{i} produces $\sim 2.6$ extra photons per helium nucleus.} $1.9/0.9\sim 2.1$ additional photons, while the production due to \ion{He}{ii} photons is about $1.3/0.55\sim 2.3$ additional photons per helium nucleus.
Although from naive estimates due to the double re-processing of \ion{He}{ii} photon one would expect the efficiency of photon production to be larger in the latter case, due to the fact that the hydrogen and \ion{He}{i} 1s-continua are not completely optically thick (see Sect.~\ref{sec:direct}) the numbers remain comparable.

With these values one can also provide an {\it optimistic} upper limit on the total amount of helium-related photons. Assuming that every high frequency photons emitted by \ion{He}{ii} lead to $\sim 2.3$ additional photons, using the numbers given in Sect.~\ref{sec:feed_num_HeII} we expect $2.3\times (0.98+0.5)\sim 3.4$ additional photons due to \ion{He}{ii} feedback alone. This is another factor of $\sim2.6$ times more than we already obtained here. 
Also adding those photons that may be  (see Sect.~\ref{sec:HeI_total_num} and \ref{sec:add_corr_HeI}) produced due to primary \ion{He}{i} feedback ($\sim 2.6-0.3\sim 2.3\,\gamma$ per helium nucleus) one can therefore expect a total addition of $\sim 5.7\,\gamma$ per helium atom, when including all the feedbacks.
This implies a total of $\sim 14\,\gamma$ related to the presence of helium in the Universe, or about $1.1\,\gamma$ per hydrogen atom, resulting in a total contribution of helium-related photons to the recombination spectrum of $\sim 20\%$.
In comparison to the computation that does not include any of the feedback processes one therefore could expect a factor of $14/8.4\sim 1.7$ times more helium-related photons.

\section{Summary and conclusions}
\label{sec:discussion}
\label{sec:disc_con}
In this paper we considered the {\it re-processing} of high frequency photons emitted by helium during the epoch of cosmological recombination. 
We investigated both the possible changes in the {\it cosmological recombination spectrum} and the {\it cosmological ionization history} of the Universe, taking several previously neglected physical processes into account.
In particular we studied
\begin{itemize}

\item[(i)] the feedback of $\ion{He}{i}$ photons on \ion{He}{i} recombination 

\item[(ii)] the feedback of {\it primary} $\ion{He}{i}$ photons on \ion{H}{i} recombination

\item[(iii)] the feedback of {\it primary} $\ion{He}{ii}$ photons on \ion{He}{i} recombination

\item[(iv)] the feedback of {\it secondary} $\ion{He}{i}$ photons on \ion{H}{i} recombination.

\end{itemize}

The process (i) was already considered earlier \citet{Switzer2007I}, but the processes (ii)-(iv) have not been studied at full depth so far.
We find that process (i) leads to no important change in the cosmological recombination spectrum, but a small modification in the cosmological ionization history by $\Delta N_{\rm e}/N_{\rm e}\sim +0.17\%$ at $z\sim 2300$. At low redshift our correction seems to be smaller (see Fig.~\ref{fig:tNe}) than the one presented in \citet{Switzer2007II}.
However, it is clear that the difference will not be very important for the analysis of future CMB data. Nevertheless, one should check the reason for this difference in a detailed code comparison, which is currently under discussion among different groups.

Furthermore, we find that the processes (ii)-(iv) affect the ionization history at a negligible level, but could increase the total helium induced contribution to the cosmological recombination spectrum by $\sim 40\%-70\%$ in comparison to the standard calculation which assumes independent recombination histories with no feedback between the different atomic species (see Sect.~\ref{sec:number_HeII}).

The first result is related to the fact that the feedback processes (ii)-(iv) all occur in the {\it pre-recombinational} epochs of the considered atomic species, where they are still very close to full equilibrium with the free electrons and protons, so that perturbations in the ionization state are restored very fast.
On the other hand, the increase in the total number of helium-related photons occurs because the recombination of electrons that were liberated by the feedback process can be captured into highly excited atomic levels with the possibility to release {\it several} photons per {\it one} ionizing photon in the subsequent cascade to the ground state. 
Here in particular the {\it double re-processing} of \ion{He}{ii} photons (see Sect.~\ref{sec:double}) is interesting, for which {\it primary} \ion{He}{i} photons in principle feed back on both \ion{He}{i} and at later times on \ion{H}{i}. Therefore one naively expects the feedback induced number of photons from \ion{He}{ii} to be twice the one coming from \ion{He}{i} feedback.
However, for \ion{He}{ii} feedback the efficiency of low frequency photon production is not as large (see Sect.~\ref{sec:direct}), because the hydrogen and neutral helium 1s continua are not completely optically thick at the time of feedback. Therefore the final number of produced photons per ionizing \ion{He}{ii} photon remains comparable to the one for \ion{He}{i} feedback (see Sect.~\ref{sec:number_HeII}).

As a result of all the feedback processes included here a total of $\sim12\,\gamma$ per helium nucleus are produced during cosmological recombination, implying a total contribution of $\sim 17\%$ to the cosmological recombination spectrum.
In total about $\sim 3.2$ {\it additional} helium-related photons per helium nucleus were released in comparison the standard calculation which neglects feedback.
This is an important addition to cosmological recombination spectrum and may render it easier to determine the primordial abundance of helium using differential measurements of the CMB energy spectrum \citep[e.g. as discussed in][]{Sunyaev2009}.
Here most interestingly, the feedback of \ion{He}{ii} photons on \ion{He}{i} leads to the appearance of several additional rather {\it narrow spectral features} in the \ion{He}{i} recombination spectrum at low frequencies (cf. Fig.\ref{fig:DI.total}). The contribution of these features to the spectral distortion at some given frequency can exceed the average level of $17\%$ several times. In particular the bands around $\nu\sim 10 \,$GHz, $\sim 35\,$GHz, $\sim 80\,$GHz, and $\sim 200\,$GHz seem to be affected strongly (see Fig.\ref{fig:DI.total}).
In addition, one should point out that today these photons are the only witnesses of the feedback process at high redshift, and observing them in principle offers a way to check our understanding of the recombination physics.

Since for the feedback of \ion{He}{ii} photons here we only accounted for those from the \ion{He}{ii} Lyman $\alpha$ resonance, which is only $\sim 1/3$ of the total number of primary \ion{He}{ii} photons potentially available for feedback (see Sect.~\ref{sec:feed_num_HeII} for more detailed estimates), one can still expect the total number of helium-related photons to increase by another $\sim 10\%-20\%$ to a total of $\sim14\,\gamma$ per helium nucleus (see Sect.~\ref{sec:number_HeII} for estimates). However, we leave a detailed investigation of the \ion{He}{ii} feedback problem including the {\it two-photon continuum}, {\it Balmer continuum}, and the effect of {\it electron scattering} for some future work (see Sect.~\ref{sec:HeIII_feedback} for a more detailed overview of the possible processes related to this problem). 
Also in the \ion{He}{i} feedback problem so far we neglected the effect of the $\ion{He}{i}\,\HeIlevel{2}{1}{S}{0}-\HeIlevel{1}{1}{S}{0}$ two-photon continuum and the ionizing photons from levels with $n>10$. However, their effect on the cosmological recombination spectrum is expected to be smaller (see Sect.~\ref{sec:add_corr_HeI}). 

Finally, we also refined our previous computations of $\ion{He}{ii}\rightarrow \ion{He}{i}$ recombination \citep{Jose2008} checking the effects of (A) $\ion{He}{i} \;\HeIlevel{n}{1}{D}{2}-\HeIlevel{1}{1}{S}{0}$ electric quadrupole transitions; (B) higher ($n>2$) $\ion{He}{i} \;\HeIlevel{n}{3}{P}{1}-\HeIlevel{1}{1}{S}{0}$ intercombination transitions; (C) the speed-up of the higher ($n>2$) $\ion{He}{i} \;\HeIlevel{n}{1}{P}{1}-\HeIlevel{1}{1}{S}{0}$ series due to the absorption of helium photons by hydrogen; (D) possible time-dependent corrections to the $\ion{He}{i} \;\HeIlevel{2}{1}{P}{1}-\HeIlevel{1}{1}{S}{0}$ and $\ion{He}{i} \;\HeIlevel{2}{3}{P}{1}-\HeIlevel{1}{1}{S}{0}$ channels; (E) the effect of the thermodynamic factor on the $\ion{He}{i} \;\HeIlevel{2}{1}{P}{1}-\HeIlevel{1}{1}{S}{0}$ and $\ion{He}{i} \;\HeIlevel{2}{3}{P}{1}-\HeIlevel{1}{1}{S}{0}$ transitions.
These processes \changeII{affect} the ionization history but do not introduce any significant changes in the recombination spectrum.

For the processes (A)-(C) we find very good agreement with the results of  \citet{Switzer2007II}, and conclude that they can be neglected in the analysis of future CMB data.
Although processes similar to (D) and (E) lead to significant corrections during  the epoch of \ion{H}{i}  recombination \citep{Chluba2008b, Chluba2009}, here we show that for the recombination of helium they are of minor importance, mainly because the effect of the \ion{H}{i} continuum opacity on the $\ion{He}{i} \;\HeIlevel{2}{1}{P}{1}-\HeIlevel{1}{1}{S}{0}$ and $\ion{He}{i} \;\HeIlevel{2}{3}{P}{1}-\HeIlevel{1}{1}{S}{0}$ channels is much more crucial.
Therefore, they again can be neglected for theoretical computations of the CMB power spectra.

\section*{Acknowledgements}
The authors would like to thank Prof. Vainshteyn for useful discussion and help with the atomic data for helium. JC would also like to thank the MPA in Garching for hospitality during June/July 2009, where part of this work was completed.

\begin{appendix}

\section{Atomic model for helium}
\label{app:atom}
We use the same atomic model for helium as in \citet{Jose2008}, which is mostly based on energy levels and transition rates given by\footnote{For more details we refer the interested reader to \citet{Jose2008} and references therein.} \citet{Drake2007}. 
However, here we also took additional $\ion{He}{i}\;\HeIlevel{n}{3}{P}{1}-\HeIlevel{1}{1}{S}{0}$ intercombination transitions with $3\leq n \leq 10$ into account. For the $\ion{He}{i}\;\HeIlevel{3}{3}{P}{1}-\HeIlevel{1}{1}{S}{0}$ transition we used the value $A=56.1 \,\rm s^{-1}$ \citep{Laughlin1978} and like in \citet{Switzer2007I} we extrapolate to larger $n$ assuming $A\propto 1/n^3$.
Furthermore, we also added electric quadrupole transitions (E2) for the sequence $\ion{He}{i}\;\HeIlevel{n}{1}{D}{2}-\HeIlevel{1}{1}{S}{0}$ with $n\leq 10$ using the absorption oscillator strength\footnote{These are related to the quadrupole transition rate by \citep[e.g.][]{Sobelman2006} $A^{\rm E2}_{ij}=\frac{8\pi^2 e^2}{m_{\rm e} c}\frac{1}{\lambda^2_{ij}}\,\frac{g_j}{g_i}\,f^{\rm E2}_{ji}$.} given by \citet{Cann2002} for $n\leq6$ and extrapolating these values assuming $f\propto 1/n^3$ \citep[like in][]{Switzer2007I}.

\section{Escape in the continuum}
\label{app:cont}
Integrating Eq.~\eqref{eq:DNic} over $\id\nu\id\Omega$ one can directly write
\beal
\label{eq:dNgc_dt}
\int\frac{1}{c}\left.\Abl{N_{\nu}}{t}\right|^{\rm rec}_{\rm 1s}\id\nu \id\Omega
&=N_{\rm e}\,N_{\rm p}\,R^{\rm pl}_{\rm c1s}
-N^{\ion{H}{i}}_{\rm 1s}[R^{\rm pl}_{\rm 1sc}+\Delta R_{\rm 1sc}].
\end{align}
Here one has
\beal
\label{eq:DR1sc}
\Delta R_{\rm 1sc}&=
   4\pi \int \sigma_{{\rm 1s c}}(\nu)\,\Delta N_{\nu} \id\nu
\nonumber
\\
&= 4\pi [N^{\rm c}_{\rm em}-N^{\rm pl}_{\nu_{\rm c}}] \int \sigma_{{\rm 1s c}}(\nu)\,G^{\rm c}_{\nu} \id\nu
\nonumber
\\
&= 4\pi [N^{\rm c}_{\rm em}-N^{\rm pl}_{\nu_{\rm c}}] \int \frac{\sigma_{{\rm 1s c}}(\nu)}{f^{\rm c}_\nu}\,f^{\rm c}_\nu\,G^{\rm c}_{\nu} \id\nu.
\end{align}
Since in the Wien tail of the CMB one has $\int \frac{\sigma_{{\rm 1s c}}(\nu)}{f^{\rm c}_\nu}\id\nu= \int \sigma_{{\rm 1s c}}(\nu)\,\frac{N^{\rm pl}_{\nu}}{N^{\rm pl}_{\nu_{\rm c}}}\id\nu=\frac{R^{\rm pl}_{\rm 1sc}}{4\pi N^{\rm pl}_{\nu_{\rm c}}}$, one can introduce the normalized profile $\varphi_{\rm c}(\nu, z)=\frac{4\pi N^{\rm pl}_{\nu_{\rm c}}}{R^{\rm pl}_{\rm 1sc}}\frac{\sigma_{{\rm 1s c}}(\nu)}{f_\nu^{\rm c}(z)}$ and then write
\beal
\label{eq:DR1sc_b}
N^{\ion{H}{i}}_{\rm 1s}\,\Delta R_{\rm 1sc}
&= N^{\ion{H}{i}}_{\rm 1s}\, \frac{R^{\rm pl}_{\rm 1sc}}{N^{\rm pl}_{\nu_{\rm c}}} [N^{\rm c}_{\rm em}-N^{\rm pl}_{\nu_{\rm c}}] \int \varphi_{\rm c}(\nu, z)\,f^{\rm c}_\nu\,G^{\rm c}_{\nu} \id\nu.
\nonumber
\\
&= [N_{\rm e}\,N_{\rm p}\,R^{\rm pl}_{\rm c1s}-N^{\ion{H}{i}}_{\rm 1s} R^{\rm pl}_{\rm 1sc}] \int \varphi_{\rm c}(\nu, z)\,f^{\rm c}_\nu\,G^{\rm c}_{\nu} \id\nu,
\end{align}
where we used the fact that $\frac{2\nu_{\rm c}^2}{c^2}\,\tilde{f}_{\rm 1s}(T)\,R^{\rm pl}_{\rm 1sc}/N^{\rm pl}_{\nu_{\rm c}}=R^{\rm pl}_{\rm c1s}$.
Therefore one finally has
\beal
\label{eq:dNgc_dt_II}
\int\frac{1}{c}\left.\Abl{N_{\nu}}{t}\right|^{\rm rec}_{\rm 1s}\id\nu \id\Omega
&=
\left\{N_{\rm e}\,N_{\rm p}\,R^{\rm pl}_{\rm c1s}-N^{\ion{H}{i}}_{\rm 1s}\,R^{\rm pl}_{\rm 1sc}\right\}
\times P^{\rm c}_{\rm eff}
\end{align}
with $P^{\rm c}_{\rm eff}=\int_0^\infty \varphi_{\rm c}(\nu, z)[1-\,f^{\rm c}_\nu\,G^{\rm c}_{\nu}] \id\nu$.

\subsection{Quasi-stationary solution without asymmetry between continuum emission and absorption profile}
If we assume quasi-stationary conditions for the continuum and neglect the factor $1/f^{\rm c}_\nu$ in the definition of $\Theta_{\rm a}^{\rm c}$, Eq.~\eqref{app:kin_abs_em_Sol_phys_asym_F}, then one directly has $G^{\rm 
c, qs}_{\nu}=1-e^{-\tau_{\rm c}(\nu, \infty, z)}$.
Using Eq.~\eqref{eq:continuum_channel_defs_b} and Eq.~\eqref{eq:def_phi_c} with the substitution 
$\tilde{\nu}=\nu\,\frac{1+\tilde{z}}{1+z}$ one can furthermore write
\bsub
\label{eq:continuum_channel_defs_qs_1}
\beal
\label{eq:continuum_channel_defs_qs_1a}
\tau^{\rm c}_{\rm abs}(\nu, \infty, z)
&\approx \tau^{\rm esc}_{\rm c}\,\int_{\nu}^{\infty}
\varphi_{\rm c}(\tilde{\nu})\,f^{\rm c}_{\tilde{\nu}}(\tilde{z})\frac{\nu_{\rm c}}{\tilde{\nu}}\id \tilde{\nu}
\\
\label{eq:continuum_channel_defs_qs_1b}
&\approx \tau^{\rm esc}_{\rm c}\,[1-\chi_{\rm c}].
\end{align}
\esub
Here $\tau^{\rm esc}_{\rm c}=\frac{c\,N^\ion{H}{i}_{\rm 1s}}{H\,\nu_{\rm c}}\,\frac{R^{\rm pl}_{\rm 1sc}}{4\pi\,N^{\rm p}_{\nu_{\rm c}}}\approx \frac{c\,\sigma_{\rm 1sc}(\nu_{\rm c})\,N^\ion{H}{i}_{\rm 1s}}{H}\,\frac{k\Tg}{h\nu_{\rm c}}$ and $\chi_{\rm c}=\int_{\nu_{\rm c}}^\nu\,\varphi_{\rm c}(\nu')\id \nu'$. 
Also in Eq.~\eqref{eq:continuum_channel_defs_qs_1b} we have neglected the factors $\frac{\nu_{\rm c}}{\tilde{\nu}}$ and $f^{\rm c}_{\tilde{\nu}}(\tilde{z})$.
For the escape probably from the main resonances in helium and hydrogen, this would be the standard procedure to obtain the Sobolev escape probability \citep[e.g. see][]{Chluba2008b}.
Then with  $P^{\rm c}_{\rm eff}=\int_0^\infty \varphi_{\rm c}(\nu, z)[1-\,f^{\rm c}_\nu\,G^{\rm c}_{\nu}] \id\nu$ after also setting $f^{\rm c}_\nu\approx 1$ one finally has
\beal
\label{eq:Pesc_continuum_QS}
P^{\rm c, qs, no asym}_{\rm eff}\approx \frac{1-e^{-\tau^{\rm esc}_{\rm c}}}{\tau^{\rm esc}_{\rm c}}.
\end{align}
This expression has the same form as the standard Sobolev escape probability. However, here we have neglected the fact that the emission and absorption profiles in the hydrogen Lyman continuum differ by $f^{\rm c}_\nu$, which makes the absorption profile $\varphi^{\rm abs}_{\rm c}(\nu)=f^{\rm c}_\nu \varphi_{\rm c}(\nu)\sim \sigma_{\rm 1s c}(\nu)$ very broad.
This is the reason why the expression for the escape probability in the continuum scales like $P^{\rm c, qs}_{\rm eff}\approx1/[1+\tau^{\rm esc}_{\rm c}]$ instead, as already obtained in \citet{Chluba2007b}.
Below we give a short derivation of this expression in the formulation given here.

\subsection{Quasi-stationary solution including the asymmetry between continuum emission and absorption profile}
\label{app:approxII}
Here we give the derivation for the escape probability in the \ion{H}{i} Lyman continuum assuming quasi-stationarity but taking into account that the continuum emission and absorption profiles differ by $f^{\rm c}_\nu\neq 1$.
From the definition Eq.~\eqref{eq:continuum_channel_defs_b} and with $\tilde{\nu}=\nu\,\frac{1+\tilde{z}}{1+z}$ it directly follows
\beal
\label{eq:continuum_channel_defs_qs}
\tau^{\rm c}_{\rm abs}(\nu, z', z)
&\approx \tau_0\,\int_{\nu}^{\nu'}
\frac{\sigma_{\rm 1sc}(\tilde{\nu})}{\sigma_{\rm 1sc}(\nu)}\frac{\id \tilde{\nu}}{\tilde{\nu}}
\approx \tau_0\,\frac{\nu'-\nu}{\nu}
\end{align}
with $\tau_0=\frac{c\,\sigma_{\rm 1sc}(\nu)\,N^{\ion{H}{i}}_{\rm 1s}}{H}$.
Also from Eq.~\eqref{app:kin_abs_em_Sol_phys_asym_F} with $\Theta_{\rm t}^{\rm c}=1$ one directly has
\beal
\label{eq:F_continuum_qs}
F^{\rm c, qs}_\nu(z)&=\frac{\tau_0}{f^{\rm c}_\nu}\,\int^{\infty}_\nu
\frac{f^{\rm c}_\nu(z)}{f^{\rm c}_{\nu'}(z')}
\,\frac{\sigma_{\rm 1sc}(\nu')}{\sigma_{\rm 1sc}(\nu)}
\,\exp\left(-\tau_0\,\frac{\nu'-\nu}{\nu}\right)\,\frac{\id\nu'}{\nu'}
\nonumber\\
&\approx \frac{\tau_0}{f^{\rm c}_\nu}\,\int^{\infty}_\nu
\frac{\nu^3}{\nu'^3}\,
\exp\left(-\left[\tau_0+\frac{h\nu_{\rm c}}{k\Tg}\right]\,\frac{\nu'-\nu}{\nu}\right)\,
\frac{\id\nu'}{\nu'}
\nonumber\\
&\approx \frac{1}{f^{\rm c}_\nu}\,\frac{\tau^{\rm esc}_{\rm c}(\nu)}{1+\tau^{\rm esc}_{\rm c}(\nu)},
\end{align}
with $\tau^{\rm esc}_{\rm c}(\nu)=\frac{c\,\sigma_{\rm 1sc}(\nu)\,N^{\ion{H}{i}}_{\rm 1s}}{H}\,\frac{k\Tg}{h\nu}$.
Inserting this into the expression for the escape probability $P^{\rm c}_{\rm eff}=\int_0^\infty \varphi_{\rm c}(\nu, z)[1-\,f^{\rm c}_\nu\,G^{\rm c}_{\nu}] \id\nu$, and neglecting the slow (power-law) scaling of $\tau^{\rm esc}_{\rm c}(\nu)\approx \tau^{\rm esc}_{\rm c}(\nu_{\rm c})$ with frequency one then obtains $P^{\rm c, qs}_{\rm eff}\approx1/[1+\tau^{\rm esc}_{\rm c}]$.

The main difference to the approximation Eq.~\eqref{eq:Pesc_continuum_QS} is in the scaling for small $\tau^{\rm esc}_{\rm c}$. Here one has $P^{\rm c, qs}_{\rm eff}\approx 1-\tau^{\rm esc}_{\rm c}$ while from Eq.~\eqref{eq:Pesc_continuum_QS} one finds $P^{\rm c, qs, no asym}_{\rm eff}\approx 1-\frac{1}{2}\tau^{\rm esc}_{\rm c}$.
However, in the limit $\tau^{\rm esc}_{\rm c}\gg 1$ both approximations agree.

\end{appendix}


\bibliographystyle{mn2e} 
\bibliography{Lit}

\end{document}